\documentclass{aa}  
\bibpunct{(}{)}{;}{a}{}{,} % to follow the A&A style
\usepackage{graphicx}
\usepackage{txfonts}
\usepackage{hyperref}
\usepackage{multirow}
\usepackage{rotating}
\usepackage[flushleft]{threeparttable}
\usepackage{subfloat}
\usepackage{xcolor}
\usepackage{longtable}
\usepackage{tablefootnote}
\usepackage{footnote}

\renewcommand{\arraystretch}{2.0}

\def\tco{$^{13}$CO}
\def\ceo{C$^{18}$O}
\def\deg{^{\circ}}

\newcommand{\hii}{H~\textsc{ii}}
\newcommand{\msun}{$\rm M_\odot$}
\newcommand{\lsun}{$\rm L_\odot$}
\newcommand{\kms}{km~s$^{-1}$}

\newcommand{\hmole}{H$_2$}

\newcommand{\hcopone}{HCO$^+$~$J=1-0$}
\newcommand{\htcop}{H$^{13}$CO$^+$}
\newcommand{\htcopone}{H$^{13}$CO$^+$~$J=1-0$}

\newcommand{\siofive}{SiO~$J=5-4$}

\newcommand{\chtoh}{CH$_3$OH}
\newcommand{\htco}{H$_2$CO}

\newcommand{\ceoone}{C$^{18}$O~$J=1-0$}

\newcommand{\micron}{$\mu$m}

\newcommand{\tcoone}{$^{13}$CO $J=1-0$}

\newcommand{\ceotwo}{C$^{18}$O $J=2-1$}

\newcommand{\nhtcd}{$N_{\rm H_{2}}$}
\newcommand{\dustt}{$T_{\rm dust}$}
\newcommand{\nhtnd}{$n_{\rm H_{2}}$}

\newcommand{\mjth}{$M_{\rm J}^{\rm th}$}
\newcommand{\lambdajth}{$\lambda_{\rm J}^{\rm th}$}

\newcommand{\mjthclump}{$M_{\rm J, clump}^{\rm th}$}
\newcommand{\lambdajthclump}{$\lambda_{\rm J, clump}^{\rm th}$}

\newcommand{\mjtotclump}{$M_{\rm J, clump}^{\rm tot}$}
\newcommand{\lambdajtotclump}{$\lambda_{\rm J, clump}^{\rm tot}$}

\newcommand{\mjflowclump}{$M_{\rm J, clump}^{\rm com, flow}$}
\newcommand{\lambdajflowclump}{$\lambda_{\rm J, clump}^{\rm com, flow}$}

\newcommand{\mjthens}{$M_{\rm J, ens}^{\rm th}$}
\newcommand{\lambdajthens}{$\lambda_{\rm J, ens}^{\rm th}$}

\newcommand{\mach}{$\mathcal{M}$}

\definecolor{blue}{rgb}{0.0,0.0,1}
\definecolor{green}{rgb}{0.0,1,0.0}
\hypersetup{
    colorlinks,
    linkcolor=blue,
    citecolor=blue,
    urlcolor=blue
    }
\usepackage{etoolbox}
\usepackage{afterpage}

\usepackage{float}
\usepackage{perpage}

\makeatletter
\newcommand{\verytiny}{\@setfontsize{\verytiny}{7pt}{7pt}}
\makeatother

\begin{document} 
   \title{\hii~regions and high-mass starless clump candidates}

   \subtitle{II. Fragmentation and induced star formation at $\sim0.025$~pc scale: An ALMA continuum study}

   \author{S. Zhang \inst{1}
          \and
           A. Zavagno \inst{1,2}
           \and
           A. L\'opez-Sepulcre\inst{3,4}
           \and
           H. Liu  \inst{5, 6, 7} 
           \and
           F. Louvet \inst{8}          
           \and 
           M. Figueira \inst{9}
           \and
           D. Russeil \inst{1}
           \and
           Y. Wu \inst{10}
           \and
           J. Yuan \inst{11}
           \and
           T. G. S. Pillai \inst{12,13}
          }

   \institute{Aix Marseille Univ, CNRS, CNES, LAM, Marseille, France\\
              \email{sijuzhangastro@gmail.com}
         \and
             {Institut Universitaire de France (IUF)}
         \and
             {Univ. Grenoble Alpes, CNRS, Institut de Planétologie et d'Astrophysique de Grenoble (IPAG), 38000 Grenoble, France}
         \and
             {Institut de Radioastronomie Millim\'etrique (IRAM), 300 rue de la Piscine, 38406 Saint-Martin-D'H\`eres, France}
          \and
             {Department of Astronomy, Yunnan University, Kunming, 650091, China}
          \and
             {CASSACA, China-Chile Joint Center for Astronomy, Camino El Observatorio \#1515, Las Condes, Santiago, Chile}
          \and
             {Departamento de Astronom\'ia, Universidad de Concepci\'on, Av. Esteban Iturra s/n, Distrito Universitario, 160-C, Chile} 
          \and 
             {Departmento de Astronomia de Chile, Universidad de Chile, Santiago, Chile}
         \and
             {National Centre for Nuclear Research, ul. Pasteura 7, 02-093, Warszawa} 
         \and
             {Department of Astronomy, Peking University, 100871 Beijing, China}
         \and
             {National Astronomical Observatories, Chinese Academy of Sciences, 20A Datun Road, Chaoyang District, Beijing 100012, China} 
        \and
             {Institute for Astrophysical Research, 725 Commonwealth Ave, Boston University Boston, MA 02215, USA}
         \and
             {Max-Planck-Institut f{\"u}r Radioastronomie, Auf dem H{\"u}gel 69, 53121 Bonn, Germany}
             }

   \date{Received May 1, 2020; Accepted November, 24, 2020}

% \abstract{}{}{}{}{} 
% 5 {} token are mandatory
 
  \abstract
   {The ionization feedback from \hii~regions modifies the properties of high-mass starless clumps (HMSCs, of several hundred to a few thousand solar masses with a typical size of 0.1-1~pc), such as dust temperature and turbulence, on the clump scale. The question of whether the presence of \hii~regions modifies the core-scale ($\sim0.025$~pc) fragmentation and star formation in HMSCs remains to be explored.}
   {We aim to investigate the difference of 0.025~pc-scale fragmentation between candidate HMSCs that are strongly impacted by \hii~regions and less disturbed ones. We also search for evidence of mass shaping and induced star formation in the impacted candidate HMSCs.}
   {Using the ALMA 1.3~mm continuum, with a typical angular resolution of 1.3\arcsec, we imaged eight candidate HMSCs, including four impacted by \hii~regions and another four situated in the quiet environment. The less-impacted candidate HMSCs are selected on the basis of their similar mass and distance compared to the impacted ones to avoid any possible bias linked to these parameters. We carried out a comparison between the two types of candidate HMSCs. We used multi-wavelength data to analyze the interaction between \hii~regions and the impacted candidate HMSCs.}
   {A total of 51 cores were detected in eight clumps, with three to nine cores for each clump. Within our limited sample, we did not find a clear difference in the $\sim0.025$~pc-scale fragmentation between impacted and non-impacted candidate HMSCs, even though \hii~regions seem to affect the spatial distribution of the fragmented cores. Both types of candidate HMSCs present a thermal fragmentation with two-level hierarchical features at the clump thermal Jeans length \lambdajthclump\ and 0.3\lambdajthclump. The ALMA emission morphology of the impacted candidate HMSCs AGAL010.214-00.306 and AGAL018.931-00.029 sheds light on the  capacities of \hii~regions to shape gas and dust in their surroundings and possibly to trigger star formation at $\sim0.025$~pc-scale in candidate HMSCs.}
   {The fragmentation at $\sim0.025$~pc scale for both types of candidate HMSCs is likely to be thermal-dominant, meanwhile \hii~regions probably have the capacity to assist in the formation of dense structures in the impacted candidate HMSCs. Future ALMA imaging surveys covering a large number of impacted candidate HMSCs with high turbulence levels are needed to confirm the trend of fragmentation indicated in this study.}

   \keywords{Stars: formation -- Stars: massive -- ISM: \hii~regions -- ISM: structure -- Submillimeter: ISM -- Techniques: interferometric
               }

   \maketitle
%
%-------------------------------------------------------------------------------------------------------------------------------------------

\section{Introduction}
The formation process behind high-mass stars ($> 8$~\msun) is still a matter of debate due to the unsolved questions coming from both the observational and theoretical sides. Apart from the mechanisms that lead to the formation of high-mass stars, the role of the environment and its impact on the future star-formation properties are still poorly known.

Observationally, high-mass star-formation (HMSF) regions are often situated at further distances (several kpc) and deeply embedded in the molecular clouds. In these regions, the study of embedded structures smaller than 0.1~pc commonly requires high-resolution interferometric (sub)millimeter or centimeter observations.

Theoretically, the two widely discussed HMSF models, namely, monolithic collapse (MC) and competitive accretions (CA), propose a different initial mass for the cores embedded in the massive clumps \citep{zinnecker07}. The MC model proposes that massive stars form by disk accreting masses from the hosted turbulent massive cores. This picture is qualitatively similar to the scale-up version of low-mass star formation \citep{mckee03, krumholz09}. The CA model proposes that the massive clumps fragment into a number of cores with Jeans mass, around one to a few solar masses, in the thermal-dominant case. Furthermore, the cores located in the gravitational potential well center of the hosted massive clumps are expected to more easily accrete masses to form massive stars \citep{bonnell01, bonnell04}. There are a number of the Atacama Large Millimeter/submillimeter
Array (ALMA) results that have proven to be consistent with the CA model \citep{cyganowski17, fontani18}. 

One of the key methods in differentiating the mechanism of initial HMSF is to study the fragmentation of massive clumps (typical size of 0.1 to 1~pc) at their earliest evolution stage. Whether prestellar cores (typical size of 0.01-0.03~pc) with several dozen solar masses exist and whether the cores hosted in the high-mass starless clumps (HMSCs) have a mass significantly larger than the Jeans mass have been investigated by many studies. \citet{louvet19} summarized the only five high-mass prestellar core candidates that had been detected up to the publication of their paper, namely CygX-N53-MM2 \citep{duarte2014}, G11.92-0.61-MM2 \citep{cyganowski2014},  G11.11-P6-SMA1 \citep{wang2014}, G028CA9 \citep{kong2017}, and W43-MM1-\#6 \citep{nony2018, molet19}. These sources would favour the existence of high-mass prestellar cores, although a more specific check to see whether they are driving outflows is needed. Anyhow, the scarcity of high-mass prestellar cores is still significant, suggesting that the MC is not likely to be the common path leading to the formation of high-mass stars. Recent observations with ALMA and Submillimeter Array (SMA) toward 70~\micron\ quiet massive clumps ($>300$~\msun) show that most of the fragmented cores (0.02~pc scale) are low- to intermediate-mass objects \citep{li19, sanhueza19, svoboda19}. For example, \citet{sanhueza19} identified a total of 294 cores in 12 candidate HMSCs ($600$~\msun~to a few thousand \msun) and 210 of them are probably prestellar candidates, whereas only eight cores (two prestellar candidates $+$ six protostellar candidates) have masses of $>10$~\msun.

%The first bird's eye view of a gravitationally unstable accretion disk in a high mass protostellar object with a mass of 10~\msun~and an age of $<10^4$~yr has been presented by ALMA (Atacama Large Millimeter/submillimeter Array) \citep{motogi19}. Similar young massive protostellar objects with disk have also been detected in many studies, e.g., \citealt{csengeri18, sanna19, moscadelli19, zhangyichen19, zinchenko20}. These recent findings support the MC model that massive stars could form by disk accretion.

Both the MC and CA models of HMSF are expected to be merged into a consistent and unified picture with the progress of observations and theories. The evolutionary scenario proposed by \citet{motte18} suggests that the large-scale ($\sim0.1$~pc) gas reservoirs, known as starless massive dense cores (MDCs) or starless clumps, could replace the high-mass analogs of prestellar cores (about 0.01 to 0.03~pc). The phase of high-mass prestellar core could be skipped because massive reservoirs concentrate their mass into high-mass cores at the same time the stellar embryos are accreting. \citet{louvet19} tested this recent scenario by observing nine 70~\micron\ quiet MDCs (0.1~pc, $\sim100$~\msun) in NGC~6334 using ALMA, with a resolution of 0.03~pc. They reveal a lack of high-mass prestellar cores in these MDCs; therefore, this result supports the ``skipped'' prestellar phase and the scenario proposed by \citet{motte18}.

We are interested in the impact of \hii~regions on the formation of high-mass stars. In particular, we want to study how these regions impact and modify future (high-mass) star formation that occurs in their immediate surroundings. Previous studies showed that at least 30\% of HMSF in the Galaxy are located at the edges of the \hii~regions \citep{deharveng10, kendrew16, palmeirim17}. In our first paper (hereafter Paper\,I, \citealt{p1}), we investigated the feedback of \hii~regions on candidate HMSCs that are quiet from $\sim1$~\micron~to 70~\micron\ and without any star-formation signposts \citep{yuan18}. Higher resolution interferometric observations show that many of these candidate HMSCs are not absolutely starless because low- to intermediate-mass protostellar objects usually exist there \citep{Feng2016, traficante17, contoreras18, sanhueza19, svoboda19, pillai19, li19}.  Using single-dish data observed by the \textit{Herschel} and the Atacama Pathfinder EXperiment (APEX), Paper\,I explored how \hii~regions modify the candidate HMSCs properties on clump scale (0.1 to 1~pc). The results presented in Paper\,I show that more than 60\% of the candidate HMSCs are associated with \hii~regions and 30\% to 50\% of these associated HMSCs show the impacts from \hii~regions. The heating, compression, and turbulence injection by \hii~regions make the impacted candidate HMSCs have a higher dust temperature (\dustt, increment is 3 to 6~K), a higher ratio of bolometric luminosity to envelope mass ($L/M$, from $\lesssim0.5$~\lsun/\msun~ increases to $\gtrsim$~2~\lsun/\msun), a steeper profile of radial column density $N_{\rm H_{2}} \propto r^{\alpha}$ (power-law index $\alpha$ at $\sim-0.12$ decreases to $\sim-0.27$), and a higher turbulence (from supersonic to hypersonic).

We consider whether these differences, created by \hii~regions, could modify the early HMSF in massive clumps. The findings of  Paper\,I suggest that higher turbulence and steeper clump radial density profile probably favour the formation of massive fragments by limiting fragmentation (for the effect of turbulence, we refer to \citealt{maclow04}; for the effect of clump density profile, we refer to \citealt{girichidis11, palau14}). In this paper, we study the properties of fragmentation at a typical scale of $\sim0.025$~pc in the candidate HMSCs under different environments (impacted or non impacted by an \hii~region) to see whether differences exist or not. We take advantage of the candidate HMSCs studied in Paper\,I. In order to limit the distance and mass biases (similar distance and mass for the pair of impacted and non-impacted candidate HMSCs), we have compiled a relatively low number of sources for this study, as explained in Sect.~\ref{DataDescription}. 

This paper is presented as follows: Section~\ref{DataDescription} describes the sample selection including a short description of the previous studies for each source. Section~\ref{ALMAContinuum} presents the ALMA continuum data. We analyze the fragmentation in Sect.~\ref{fragmentation-sec}. In Sect. \ref{ALMAimpacted-result}, we explore the signposts of mass shaping exerted by \hii~regions on candidate HMSCs with the ALMA data. In Sects. \ref{Discussion} and \ref{Conclusion}, we discuss and summarize the results. This paper mainly presents the ALMA continuum results. The spectral analyses will be presented in a forthcoming paper. \\

\section{Sample selection} \label{DataDescription}
\subsection{Sample selection} \label{SampleSelection}
Through a process of cross-matching with star formation indicators such as infrared point sources from $\sim1$~\micron~to 70~\micron, \citet{yuan18} identified 463 candidate HMSCs in the inner Galactic plane ($\left| l \right| < 60\deg$, $\left| b \right| < 1\deg$) with 870~\micron\ images from the APEX Telescope Large Area Survey of the GALaxy (ATLASGAL, \citealt{schuller09}).  In Paper\,I, we cross-matched \hii~regions and these candidate HMSCs (90\% of them are more massive than 100~\msun) according to two main criteria: (1) The projected separation between the center of \hii~region and HMSC is less than double radii of \hii~region. (2) The velocity difference $|\Delta_{{\rm v}_{lsr}}|$ between the \hii~region and HMSC is $< 7$~\kms, which is derived from the standard deviation of the velocity difference between \hii~regions and HMSCs (see details in Section 4 of Paper I). More than 60\% of the candidate HMSCs are associated with \hii~regions and termed as \textbf{AS candidate HMSCs} or \textbf{AS}. For the candidate HMSCs faraway from \hii~regions, they are simply termed as \textbf{NA (Non-Associated) candidate HMSCs} or \textbf{NA}.

\begin{figure*}[htb]
     \centering
   \includegraphics[width=0.8\textwidth]{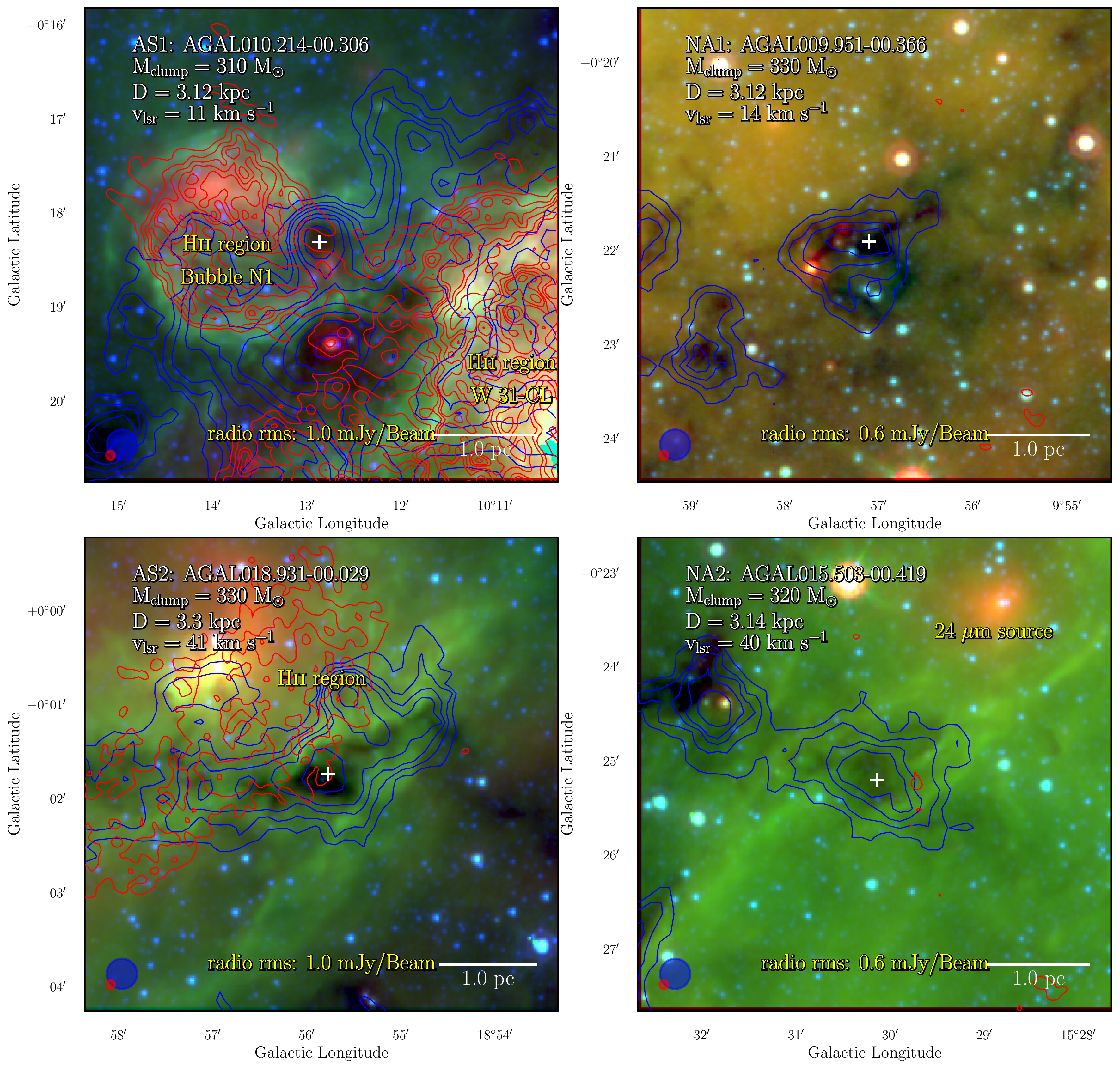}
       \caption{Environment of candidate HMSCs. Each panel shows ATLASGAL 870~\micron~contours (blue, levels are $[0.2, 0.3, 0.4, 0.5, 0.7, 0.9, 1.3, 1.8, 2.5, 4]$~Jy~Beam$^{-1}$) overlaid on RGB images constructed from \textit{Spitzer} 24~\micron~(in red), 8~\micron~(in green), and 4.5~\micron~(in blue) images. The candidate HMSCs are indicated by white crosses. Red contours indicate radio continuum emission, which are MAGPIS 20~cm for most of the regions except for NA3 and AS4 (MAGPIS 90~cm and SUMSS 35 cm, respectively; see Fig.~\ref{RadioMapplus}.). The root mean square (rms) of the radio images are indicated in the bottom center of each panel. The radio contour levels are the radio image ${\rm rms} \times [3, 4, 6, 8, 10, 15, 20, 30, 40, 60, 80]$. The beams of radio and ATLASGAL images are shown in bottom left by red and blue ellipses, respectively. The used survey data include the: Galactic Legacy Infrared Mid-Plane Survey Extraordinaire (GLIMPSE, \citealt{churchwell09}); Multiband Imaging Photometer  Galactic Plane
Survey (MIPSGAL, \citealt{carey09}); Multi-Array Galactic Plane Imaging Survey (MAGPIS, \citealt{helfand06}), and Sydney University Molonglo Sky Survey (SUMSS, \citealt{bock99}).}
        \label{RadioMap}
    \end{figure*}

\begin{figure*}[htb]
     \centering
   \includegraphics[width=0.8\textwidth]{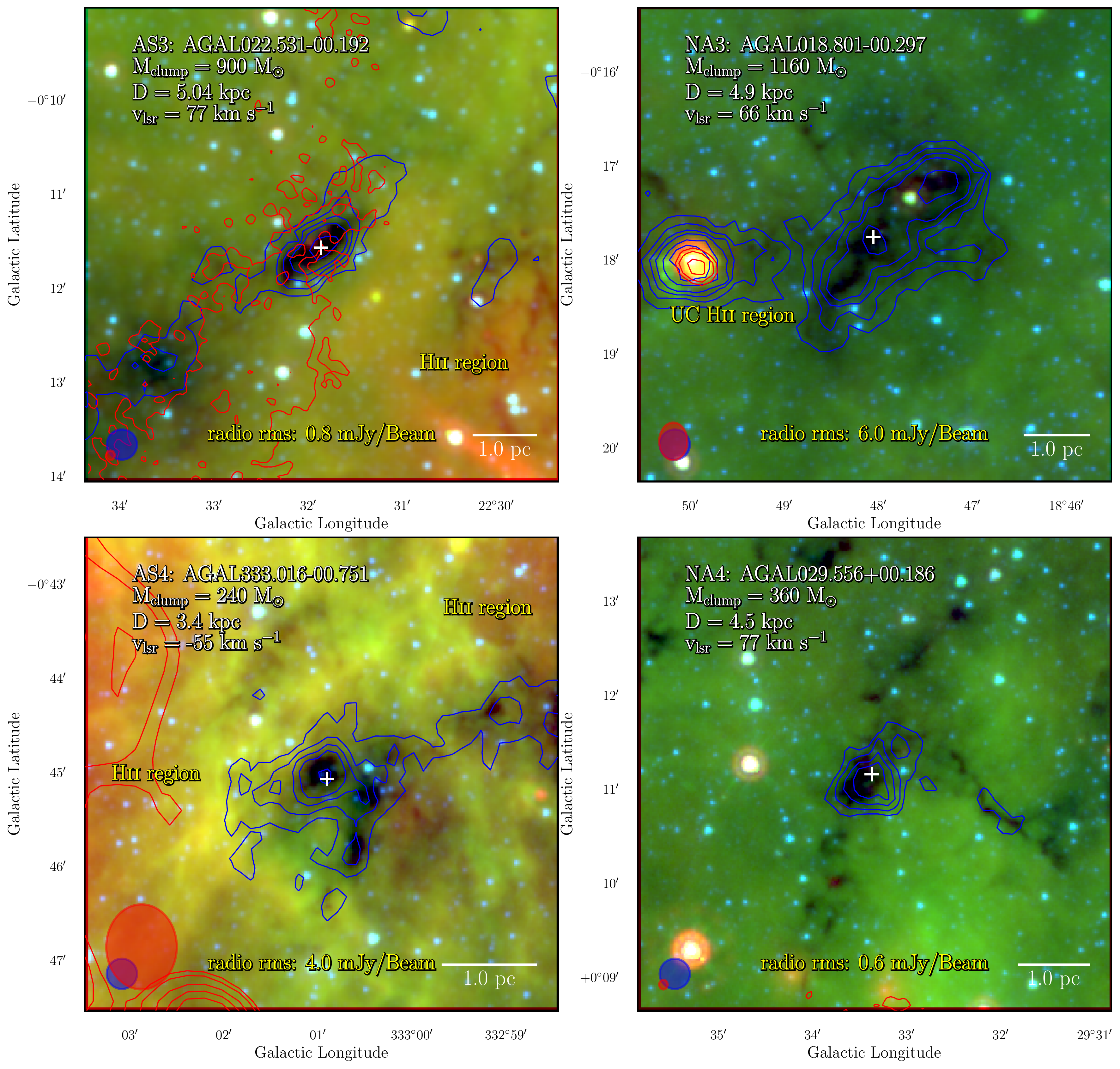}
       \caption{Environment of candidate HMSCs. The same as Fig.~\ref{RadioMap}, but for another four candidate HMSCs.}
        \label{RadioMapplus}
    \end{figure*}

We made use of the ALMA data observed in ALMA Cycle 4 by project 2016.1.01346.S\footnote{\url{http://almascience.eso.org//aq/?project\_code=2016.1.01346.S}} (PI: Thushara Pillai, project title: Galactic census of all massive starless cores within 5~kpc). This project, covering the entire inner Galactic plane, surveyed the Galactic massive starless core candidates within 5~kpc. Their selected candidates are dark from near IR to 70~\micron. We cross-matched candidate HMSCs in Paper\,I and sources for which their ALMA data are public. A total of four AS candidate HMSCs impacted by \hii~regions were extracted. To have a basis for a comparison with the non-impacted candidate HMSCs, we \textbf{tentatively} searched for additional four NA candidate HMSCs in the same ALMA observing project source pool. To limit the bias and independent variables, these NA are required to have the most similar single-dish derived clump mass ($M_{\rm clump}$), column density (\nhtcd), size (FWHM), ellipticity ($e$), and distance ($D$) compared to AS counterparts in the same ALMA project. Other single-dish derived properties, such as \dustt, $L/M$, turbulence (line width), density profile (power-law index $\alpha$), and virial ratio are not considered when searching for the possible counterparts. The reason is that the modifications of these physical properties made by the impacts of nearby \hii~regions have been proven in Paper\,I. The environment and single-dish properties of the selected candidate HMSCs are shown and listed in Figs.~\ref{RadioMap}, \ref{RadioMapplus}, and Table \ref{SingleDishPar}, respectively. The clump mass of the sample is approximately equal to the necessary clump mass to form at least one high-mass star with a mass of 8~\msun, which is 260~\msun~to 320~\msun, derived by \citet{sanhueza17,sanhueza19}, using the initial mass function of \citet{kroupa2001} and a star formation efficiency of 30\% \citep{alves2007}. \\

      \begin{table*}[htb]
      \centering % used for centering table
      \begin{threeparttable}
      \setlength{\tabcolsep}{4.5pt}
      \renewcommand{\arraystretch}{1.5}
      \tiny
      \caption{Single-dish derived properties of HMSC candidates.} % title of Table
      \label{SingleDishPar} % is used to refer this table in the text
       \begin{tabular}{c c c c r c c c c c l c} % centered columns (4 columns)
       \hline\hline % inserts double horizontal lines
             Source                  &   FWHM\tnote{\textit{(a)}}   &  $r$\tnote{\textit{(a)}}  & Distance\tnote{\textit{(b)}}    &  $M_{\rm clump}$\tnote{\textit{(c)}}    & \nhtcd\tnote{\textit{(c,d)}}             & $\Sigma$\tnote{\textit{(e)}}        &  \dustt\tnote{\textit{(c)}}      & $L/M$        &  $\alpha$\tnote{\textit{(f)}}   & $\delta {\rm v}$\tnote{\textit{(g)}}    & virial ratio    \\ 
                                  & \arcsec  &   pc   & kpc       &   \msun         &  10$^{22}$ cm$^{-2}$    &  g cm$^{-2}$  &   K        & \lsun/\msun  &           &   \kms     &                  \\
       \hline % inserts single horizontal line
   AGAL010.214-00.306 (AS1)     &  20   & 0.13 &   3.1$\pm$0.2 &    310$\pm$90    &  6.6$\pm$1.6     &   1.29      &  16.6    &    2.1$\pm$0.8   &     $-$0.21$\pm$0.18  &  3.6$\pm$0.2$^{\rm S-C^{18}O}$   &  0.8 $\pm$0.3       \\   
   AGAL009.951-00.366 (NA1)     &  31   & 0.36 &   3.1$\pm$0.2 &    330$\pm$90    &  5.9$\pm$1.5     &  0.16       &  12.0    &    0.3$\pm$0.1   &     $-$0.11$\pm$0.06  &  1.6$\pm$0.4$^{\rm S-C^{18}O}$   &  0.4$\pm$0.3       \\ 
   \hline
   AGAL018.931-00.029 (AS2)     &  42   & 0.60 &    3.3$\pm$0.2 &    330$\pm$90    &  3.2$\pm$0.8     &  0.06       &  18.2    &    4.2$\pm$1.9   &     $-$0.14$\pm$0.08  &  3.6$\pm$0.4$^{\rm F-C^{18}O}$   &  3.6$\pm$1.7        \\  
   AGAL015.503-00.419 (NA2)     &  38   & 0.51 &    3.1$\pm$0.4 &    320$\pm$110    &  3.7$\pm$0.9     &  0.08       &  14.0    &    0.8$\pm$0.3   &     $-$0.12$\pm$0.08  &  2.8$\pm$0.8$^{\rm F-C^{18}O}$   &  1.8$\pm$1.6        \\  
   \hline
   AGAL022.531-00.192 (AS3)     &  31   & 0.60 &    5.0$\pm$0.6 &    900$\pm$300    &  5.2$\pm$1.3     &  0.17       &  12.4    &    0.4$\pm$0.1   &     $-$0.15$\pm$0.19  &  5.2$\pm$1.3$^{\rm F-C^{18}O}$   &  2.5$\pm$2.0       \\  
   AGAL018.801-00.297 (NA3)     &  34   & 0.66 &    4.9$\pm$0.3 &   1160$\pm$330    &  7.0$\pm$1.8     &  0.18       &  12.3    &    0.3$\pm$0.1   &     $-$0.12$\pm$0.06  &  2.7$\pm$1.1$^{\rm F-C^{18}O}$  &  0.6$\pm$0.5        \\  
   \hline
   AGAL333.016-00.751 (AS4)     &  33   & 0.44 &    3.4$\pm$0.4 &    240$\pm$80    &  3.8$\pm$0.9     &  0.08       &  15.6    &    1.3$\pm$0.5   &     $-$0.25$\pm$0.25  &  3.8$\pm$0.7$^{\rm H^{13}CO^{+}}$   &  4.0$\pm$3.0        \\  
   AGAL029.556+00.186 (NA4)     &  31   & 0.53 &    4.5$\pm$0.3 &    360$\pm$100    &  3.3$\pm$0.8     &  0.08       &  12.8    &    0.4$\pm$0.1   &     $-$0.14$\pm$0.08  &  2.8$\pm$0.2$^{\rm F-^{13}CO}$   &  1.8$\pm$0.7        \\  
        \hline                                         
      \end{tabular}
      \begin{tablenotes}
      \item [\textit{(a)}] FWHM is measured with ATLASGAL 870~\micron~continuum by \citet{csengeri14}. $r$ is derived from FWHM and distance but deconvolved with the beam of ATLASGAL image (19.2\arcsec). See details in \citet{yuan18} and Paper\,I.
      \item [\textit{(b)}] Distance is taken from Paper\,I.
      \item [\textit{(c)}] Clump mass $M_{\rm clump}$, \hmole~column density \nhtcd, and dust temperature \dustt~are derived by fitting SED pixel-by-pixel to \textit{Herschel} 160, 250, 350, 500~\micron, and ATLASGAL 870~\micron~by \citet{yuan18} but updated with the new distance by Paper\,I.
      \item [\textit{(d)}] \nhtcd~is the beam-averaged \hmole~column density taken from SED fitting resulted maps with a beam of 36.4\arcsec.
      \item [\textit{(e)}] Surface density $\Sigma$ is derived by the clump mass divided by $\pi \times r^{2}$. The $r$ is the deconvolved radius in the third column.
      \item [\textit{(f)}] $\alpha$ is the power-law index of \hmole~column density radial profile \nhtcd$(r)\propto r^{\alpha}$, see details in Paper\,I.
      \item [\textit{(g)}] The FWHM line widths for the same pair of sources are taken from the same survey as possible as much. The abbreviation ``${\rm S-C^{18}O}$'' means SEDIGISM \ceotwo~transition (Structure, Excitation, and Dynamics of the Inner Galactic InterStellar Medium survey, \citealt{schuller17}). ``${\rm F-^{13}CO}$'' and ``${\rm F-C^{18}O}$'' represent FUGIN \tcoone~and \ceoone~transitions, respectively (FOREST Unbiased Galactic plane Imaging survey with the Nobeyama 45~m telescope, \citealt{umemoto17}). ``${\rm H^{13}CO^{+}}$'' is MALT90 \htcopone~transition (Millimetre Astronomy Legacy Team 90 GHz, \citealt{jackson13}).
      \end{tablenotes}
      \end{threeparttable}
      \end{table*}

\subsection{Single-dish derived properties and environment} \label{SingleDish}
Compared to the selected NA, the selected AS are warmer, more luminous, steeper in density profile, more turbulent, and probably less gravitationally bounded as indicated by Table~\ref{SingleDishPar}. These differences are in line with the interpretations in Paper\,I, which assert the compression and energy injection by \hii~regions play a role in modifying the clump-scale properties.

Pair 1: AS1 is located at the edge of IR dust bubble N1 \citep{deharveng10}. N1 bubble is very close to the populous star cluster W 31-CL hosted in the W 31 giant \hii~region (see Fig.~\ref{RadioMap}). The age of W 31-CL is about 0.5~$\pm$~0.5~Myr derived with near-IR data by \citet{bianchin19}. With MAGPIS 20~cm continuum \citep{helfand06} and \hii~region expansion model of \citet{tremblin14}, the estimated age of N1 is about 0.4~$\pm$~0.2~Myr with a pressure of about $(7 \pm 2) \times 10^{-10}$~dyne~cm$^{-2}$. NA1 has a similar single-dish mass and beam-averaged \nhtcd~to AS1. The FWHM of AS1 (20\arcsec) is close to the resolution of ATLASGAL (19.2\arcsec). The smallest size, the highest surface density $\Sigma$, and small $\alpha$ clearly reflect the more compact nature of AS1 compared to others.

Pair 2: AS2 is hosted in a filamentary cloud at the edge of an open \hii~region mainly created by an O8.5 star \citep{tackenberg13}. The question of whether triggered star formation is happening in this clump has been explored by the single-dish studies of \citet{tackenberg13}. Their research reveals the possible existence of gas shocked by the \hii~region. These authors did not find the features of compression in the density profile with a resolution of 37\arcsec. The appearance of the ``champagne flows'' in the \hii~region prevents us from validly applying the expansion model of \citet{tremblin14} on age estimation of the \hii~region. Alternatively, the ionized gas pressure on the interacting filament is estimated with the electron density $n_{\rm e}$ derived from 20~cm continuum integrated flux \citep{martin2005}. The pressure is $\lesssim (4.5 \pm 0.5) \times 10^{-10}$~dyne~cm$^{-2}$. 

A 24~\micron\ bright source is in the neighborhood of NA2 (see Fig.~\ref{RadioMap}). This IR bright source is probably an \hii~region considering the tight relation between 24~\micron\ hot dust emission and centimeter  free-free continuum \citep{ingallinera14, makai17}. No significant emission of MAGPIS 6 and 20~cm continuum is detected toward this 24~\micron\ bright source. Furthermore, the associated exciting OB star BD$-$15 4928 has a Gaia parallax distance of 1.6~kpc (0.6148~mas, \citealt{gaia18}). Therefore, we propose that the 24~\micron\ bright source is just a projected contamination. 

Pair 3: AS3 is associated with \hii~region G022.495-00.261 \citep{anderson14} and only very weak (1 to 3~rms) radio continuum is detected towards this region. Dust temperature \dustt~of AS3 is similar to its NA counterpart. As a result, AS3 is probably only weakly affected by the \hii~region. We note that AS3 is probably not a starless clump owing to a 24~\micron\ point source situated at an off-center position (see Fig. \ref{DarkProperties2} in Appendix). \citet{svoboda2016} identified it as a protostellar clump whereas \citet{yuan18} and ALMA project of Pillai identified it as a starless candidate. With ALMA imaging, we are able to check the protostellar properties of AS3. 

A bright MAGPIS 90~cm continuum source is located near NA3 with a separation of $\sim 2$~pc. The limited size ($<0.3$~pc) and the detection of \chtoh~maser \citep{avison16} reveal that the centimeter bright source is an UC\hii~region thus it does not have significant impacts on distant NA3.

Pair 4: AS4 is likely located in a bright rimmed structure traced by 8~\micron\ PAH\footnote{Polycyclic aromatic hydrocarbon (PAH) commonly exits in the photodissociation region (PDR) \citep{fleming10}.} emission (see Fig.~\ref{RadioMapplus}). AS4 is in the giant molecular cloud - massive star formation complex G333.0-0.5, which hosts many bubbles and pillars. The SUMSS 35~cm continuum shows that AS4 is adjacent to two interacting \hii~regions located on both western and eastern sides \citep{bock99}. NA4 is probably located in the hub of three converging filaments far away from any \hii~regions. NA4 is more massive than AS4 by a factor of 1.5, whereas their deconvolved surface density is quite similar. \\

\section{ALMA dust continuum} \label{ALMAContinuum}
\subsection{ALMA band 6 data} \label{ArchivalBand6}
We made use of the ALMA band 6 data (225~GHz, equal to 1.33~mm) of 12~m main array and 7~m Atacama Compact Array (ACA) in the ALMA Cycle 4 project 2016.1.01346.S. Table~\ref{ALMAObsPar} lists the main ALMA observing parameters. The sources were observed between March 2017 and June 2018 for the 12~m array, and between May and August 2017 for the 7~m array. The typical precipitable water vapor (PWV) is similar for all sources, except for the main array observation of AS3, which is twice as large. The time on source for each clump is 0.81 minutes for the main array and 3.53 to 6.05 minutes for the ACA.   \\

%--------------------------------------------------- One column table
      \begin{table*}[htb]
      \centering % used for centering table
      \begin{threeparttable}
      \setlength{\tabcolsep}{4.5pt}
      \renewcommand{\arraystretch}{1.5}
      \verytiny
      \caption{Information about the ALMA observations in this work.} % title of Table
      \label{ALMAObsPar} % is used to refer this table in the text
       \begin{tabular}{l l c c c c l l l l c c} % centered columns (4 columns)
       \hline\hline % inserts double horizontal lines
             Source  &  rms\tnote{\textit{(a, c)}}   & mass rms\tnote{\textit{(a, d)}}   &   Beam\tnote{\textit{(a)}}        & Resolution\tnote{\textit{(a)}}        & Config. & Baseline\tnote{\textit{(b)}}    & Antennas\tnote{\textit{(b)}} &  Time\tnote{\textit{(b, e)}}   &  Observation period\tnote{\textit{(b)}}        & PWV\tnote{\textit{(b)}} & Main array calibrators\tnote{\textit{(f)}}   \\ % table heading
                     &  mJy~Beam$^{-1}$                 &  \msun~Beam$^{-1}$   &  $\arcsec~\times~\arcsec$  & pc $\times$~pc       &       &   m         &      &    minute      &  MM/YYYY                   & mm     & Flux \& Phase\\
       \hline % inserts single horizontal line
        \multirow{2}*{AS1}  &   C: 0.55  & C: 0.15 & C: 1.80$\times$0.99 & C: 0.027$\times$0.015 &\multirow{2}*{C43-2} & M: 15--314   & M: 46     & M: 0.81  &  M: 05/2018        &   M: 1.0  & \multirow{2}*{Titan \& J1832-2039}\\   % 
                            &   A: 2.0   & A: 0.55 & A: 6.76$\times$3.90 & A: 0.102$\times$0.059 &                     & A: 9--49     & A: 8--11  & A: 6.05  &  A: 07\&05/2017    &   A: 0.9 & \\
        \hline
        \multirow{2}*{NA1}  &   C: 0.20 &  C: 0.09 & C: 1.98$\times$0.98 & C: 0.030$\times$0.015 &\multirow{2}*{C43-2} & M: 15--314   & M: 46     & M: 0.81   &  M: 05/2018          &   M: 1.0  & \multirow{2}*{Titan \& J1832-2039}\\   % 
                           &    A: 1.0  &  A: 0.45 & A: 6.98$\times$3.96 & A: 0.106$\times$0.060 &                     & A: 9--49     & A: 8--11  & A: 6.05   & A: 07\&05/2017       &   A: 0.9  &                                   \\   
       \hline
        \multirow{2}*{AS2} &  C: 0.16  &  C: 0.04  & C: 1.46$\times$1.02  & C: 0.023$\times$0.016  &\multirow{2}*{C43-2} & M: 15--314   & M: 47      & M: 0.81  & M:  05/2018         &   M: 1.1  & \multirow{2}*{Titan \& J1832-1035} \\ %
                           &  A: 1.7   &  A: 0.43  & A: 6.39$\times$4.73  & A: 0.102$\times$0.076  &                     & A: 9--48     & A: 8--10   & A: 5.04  & A: 08/2017          &   A: 0.9  &    \\
        \hline
        \multirow{2}*{NA2} &   C: 0.15  & C: 0.05  & C: 1.44$\times$1.02  & C: 0.022$\times$0.015  &\multirow{2}*{C43-2} & M: 15--314   & M: 47     & M: 0.81   & M: 05/2018          &   M: 1.1  & \multirow{2}*{Titan \& J1832-1035} \\  % 
                           &  A: 1.2    & A: 0.40  & A: 6.35$\times$4.70  &  A: 0.097$\times$0.072 &                     & A: 9--48     & A: 8--10  & A: 5.04   & A: 08/2017          &   A: 0.9  &                                     \\
       \hline
        \multirow{2}*{AS3} &   C: 0.15  &  C: 0.16 & C: 1.29$\times$1.04  & C: 0.031$\times$0.025   &\multirow{2}*{C43-2} & M: 15--314  & M: 46     & M: 0.81   &  M: 06/2018  & M: 1.1 & \multirow{2}*{Titan \& J1832-1035}\\   % 
                           &   A: 1.5   &  A: 1.6  & A: 7.48$\times$4.07  & A: 0.183$\times$0.099   &                     & A: 9--49    & A: 9--12  & A: 4.03  &  A: 07/2017  & A: 0.8   &                    \\
        \hline
        \multirow{2}*{NA3} &   C: 0.15  & C: 0.15  & C: 1.48$\times$1.04  & C: 0.035$\times$0.025   &\multirow{2}*{C43-2} & M: 15--314  & M: 47     & M: 0.81   &  M: 05/2018  & M: 1.1  & \multirow{2}*{Titan \& J1832-1035}\\   % 
                          &    A: 1.2   &   A: 1.2 & A: 6.36$\times$4.75  & A: 0.151$\times$0.113   &                     & A: 9--48    & A: 8--10  & A: 5.04   &  A: 08/2017  & A: 0.9   &                    \\
        \hline  
        \multirow{2}*{AS4} &   C: 0.20  &  C: 0.07 & C: 1.63$\times$1.31  & C: 0.027$\times$0.022   &\multirow{2}*{C40-2} & M: 15--287  & M: 45     & M: 1.01  &  M: 03/2017  & M: 2.7    & \multirow{2}*{Titan \& J1603-4904}\\   % 
                         &     A: 1.0   &  A: 0.35 & A: 6.19$\times$4.55  & A: 0.102$\times$0.075   &                     & A: 8--44    & A: 10--12 & A: 3.53  &  A: 05/2017  & A: 1.8    &                    \\
        \hline
        \multirow{2}*{NA4} &   C: 0.16  &  C: 0.13 & C: 1.36$\times$1.06  & C: 0.030$\times$0.023   &\multirow{2}*{C43-2} & M: 15--314   & M: 46      & M: 0.81  &   M: 06/2018    & M: 1.1      & \multirow{2}*{Titan \& J1832-1035} \\
                         &     A: 1.2   &  A: 0.98 & A: 7.18$\times$4.26  & A: 0.157$\times$0.093   &                     & A: 9--49     & A: 9--12   & A: 4.03 &   A: 07/2017    &  A: 0.8      &                    \\%     
        \hline                                         
      \end{tabular}
      \begin{tablenotes}
      \item [\textit{(a)}] ``C'' and ``A'' in the second to fifth columns represent the information derived from cleaned continuum images of 12~m + 7~m combined data and 7~m data, respectively.
      \item [\textit{(b)}] ``M'' and ``A'' in the seventh to eleventh columns represent the observational information of 12~m main array and 7~m ACA, respectively.
      \item [\textit{(c)}] The rms noise in the cleaned continuum images, see Sect.~\ref{ALMAContinuumReduction}.
      \item [\textit{(d)}] The rms mass sensitivity is estimated with the rms noise in cleaned continuum images, using Equation~\ref{MassEquation}.
      \item [\textit{(e)}] Time on source. It is estimated by the CASA Analysis Utilities task ``TimeOnSource'', see details in \url{https://casaguides.nrao.edu/index.php?title=TimeOnSource}.
      \item [\textit{(f)}] The 7~m ACA calibrators are different for different observations. \textbf{Flux calibrators:} Titan, Neptune, and J1733-1304 for AS1 and NA1; Neptune and J1733-1304 for AS2, NA2, AS3, NA3, and NA4; Titan, Callisto, and Ganymede for AS4. \textbf{Phase calibrators:} J1833-210B and J1832-2039 for AS1 and NA1; J1833-210B and J1733-1304 for AS2, NA2, and NA3; J1851+0035 for AS3 and NA4; J1650-5044 for AS4.
      \end{tablenotes}
      \end{threeparttable}
      \end{table*}

\subsection{ALMA data reduction} \label{ALMAContinuumReduction}
The data reductions were conducted using the same versions of Common Astronomy Software Applications (CASA, \citealt{mcmullin2007}) as those used in the QA2 pipeline products\footnote{\url{https://almascience.eso.org/processing/science-pipeline}}. The band 6 receiver is equipped with lower and upper sidebands centering at around 217.83~GHz and 232.00~GHz, respectively. The total used continuum bandwidth is around 7.2~GHz. 

 The 12~m main array and 7~m ACA data are combined by CASA task CONCAT in order to create the 12~m $+$ 7~m array combined continuum images (referred to as combined images or combined continuum, hereafter). The line-contaminated single polarization (XX) continuum is regridded to a channel width of 9~MHz in the frequency axis to accelerate the data process. The CASA task TCLEAN is then performed on the line-free channels to finally obtain the cleaned images. The pixel size ``cell'' in the cleaned image is set to be 0.3\arcsec, which is about one fifth of the synthesized beam $\simeq$~1.3\arcsec. A deconvolver using multi-term (multi-scale) multi-frequency synthesis is applied. The Briggs weighting with a robust of 0.5 is chosen. The threshold for stopping clean process is set to be 0.4~mJy~Beam$^{-1}$, about 2 to 3~rms noise. Self-calibrations are not applied because the strongest emission is just around a few dozen rms. The rms measured in the 7~m + 12~m combined cleaned continuum images is around 0.15~mJy~Beam$^{-1}$ (for NA2, AS3, and NA3) to 0.55~mJy~Beam$^{-1}$ (AS1), with a median value of 0.16~mJy~Beam$^{-1}$. The imaged field is down to 20\% power point of the primary beam. The spatial resolution for each source is about 0.018~pc ($\sim3800$~AU for NA2) to 0.029~pc ($\sim6100$~AU for NA3) with a median value of $\sim0.025$~pc.

Considering the larger synthesized beam ($\sim5.5$\arcsec) of the 7~m ACA observations, we created 7~m ACA continuum images (written as 7~m images or 7~m continuum hereafter) to more clearly show potential larger structures. The 7~m images are processed in a way similar to the combined images. The pixel size and threshold for stopping clean process in task TCLEAN are set to be 1\arcsec\ and 4~mJy~Beam$^{-1}$, respectively. Other TCLEAN parameters are similar to those for combined images. The rms noise in the final cleaned 7~m images is around 1.5~mJy~Beam$^{-1}$.

The information about ALMA observations is listed in Table~\ref{ALMAObsPar}. At 225~GHz, 12~m array and 7~m ACA have a primary beam of 25.2\arcsec\ and 44.6\arcsec, sensitive to the structure smaller than 11\arcsec\ and 19\arcsec, respectively. Interferometric observations lose the emission from larger-scale structures. Assuming a dust emissivity spectral index $\beta \simeq 1.5$, we estimate the missed 1.3~mm flux in 7~m $+$ 12~m array observations by comparing with single-dish ATLASGAL 870~\micron\ flux in the ALMA imaged field. Our 7~m + 12~m combined ALMA data set recovers 10--20\% of the single-dish integrated flux (minimum is 10\% for NA2, NA3, and AS4, maximum is 19\% for NA4). This value range is similar to other ALMA observations toward HMSC candidates such as \citet{contoreras18, beuthermagnetic18, sanhueza19}, which resulted in a value from 10\% to 30\%. Most of the large-scale emission detected in single-dish telescope is lost by interferometer, which implies that most of the masses in the earliest high-mass clump are more likely to be in large-scale diffuse envelope rather than in compact cores. \\

\subsection{Core extraction and physical properties} \label{Core-extracted-sec}
We extracted the cores in the cleaned 7~m + 12~m combined continuum images with the technique of Astrodendrograms\footnote{\url{https://dendrograms.readthedocs.io/en/stable/}} \citep{rosolowsky08}. Astrodendrogram algorithm uses a tree diagram to describe the hierarchical structures over a range of scales. The cores we extracted correspond to the ``leaves'' in the dendrogram, which means that there are no substructures for these leaves. We set the minimum emission to be considered in the dendrogam as 3~rms and the steps to differentiate the leaves as 1~rms. The minimum size considered as an independent leaf is set to be half of the synthesized beam. Table~\ref{ALMACoresBasics} lists the primary beam corrected properties of the cores identified with Astrodendrogram. 

Assuming that the dust continuum emission is optically thin, core mass is estimated with \citep{hildebrand83}:
\begin{equation}
\centering
M_{\rm core} = R_{\rm gd} \frac{F_{\nu}D^{2}}{\kappa_{\nu}{B_{\nu}(T_{\rm dust})}}, 
\label{MassEquation}
\end{equation}
where $F_{\nu}$, $\kappa_{\nu}$, and $B_{\nu}(T_{\rm dust})$ are the measured integrated flux corrected by primary beam, dust opacity per gram, and the Planck function at the frequency $\nu$. The gas-to-dust mass ratio $R_{\rm gd}$ is assumed to be 100 in this work. The $\kappa_{\nu}$ is set to be 0.9~cm$^{2}$~g$^{-1}$, corresponding to the opacity of the dust grains with thin ice mantles at a gas density of $\sim10^{6}$~cm$^{-3}$ \citep{Ossenkopf94}. 

The uncertainties of $F_{\nu}$ are derived by multiplying image rms noise with the area of the core. Distance uncertainties are taken from Table \ref{SingleDishPar}, which stem from the uncertainties in the distance estimator developed by \citet{reid16}. The combined uncertainties of ${R_{\rm gd}}/{\kappa_{\nu}}$ are around 30\% according to the estimations of \citet{sanhueza19}. 

The uncertainties of \dustt\ should be carefully treated. A unified \dustt\ for all cores embedded in one HMSC, which normally comes from single-dish clump-scale observations, is not solid due to \dustt\ gradients of the clump and different evolutionary stages of the cores. Dust temperature is expected to decline towards the center of starless clump \citep{guzman15, svoboda20, p1}. The temperature could drop to $\lesssim 10$~K for the starless center \citep{lin20}. Besides, protostellar cores or clumps are warmer than starless counterparts. The high-mass clumps which have mid-IR emission but for which the embedded \hii~region has not developed yet, have an average \dustt~of $18.6\pm0.2$~K, about 2~K higher than that of the high-mass clumps quiet in mid-IR \citep{guzman15}. In the absence of \dustt\ measurement at high resolution ($\sim1$\arcsec), we simply set three \dustt\ values, which are single-dish derived \dustt\ and \dustt~$\pm 3$~K. A difference of 3~K is estimated from \citet{guzman15} and the \dustt\ difference between candidate HMSCs close to and faraway from \hii~regions in Paper\,I. The core mass for lower and higher \dustt\ is estimated ($M_{\rm core}^{\rm cold}$ and $M_{\rm core}^{\rm warm}$). The differences between $M_{\rm core}$ and $M_{\rm core}^{\rm cold}$ or $M_{\rm core}^{\rm warm}$ could reach a level of about 30--50\%.

\begin{figure*}[p]
     \centering
      \tiny
      \includegraphics[width=0.85\textwidth]{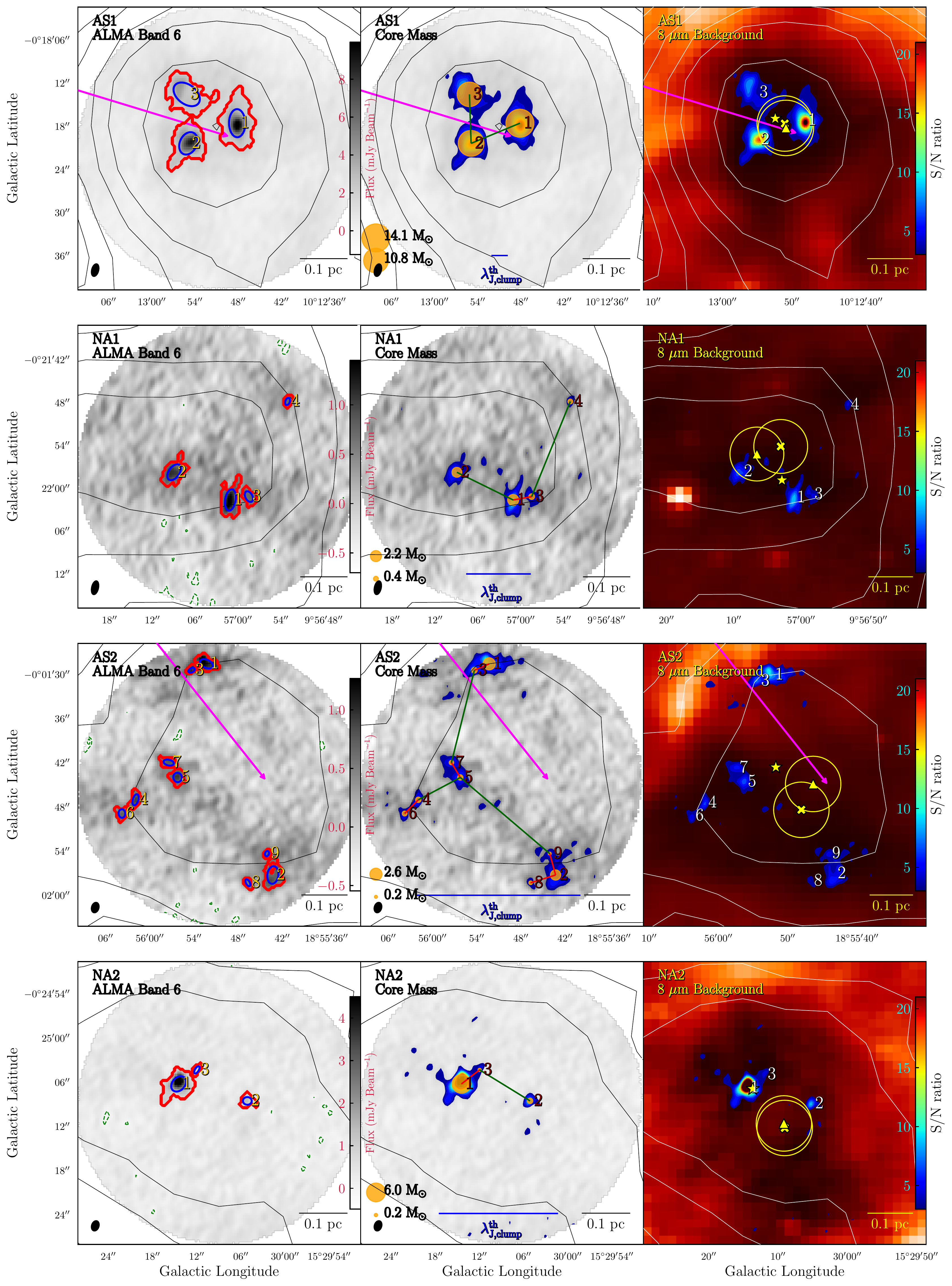}
       \caption{7~m + 12~m combined 1.3~mm continuum. The presented images are uncorrected for the primary beam for displaying a homogeneous noise background. Four rows of panels show AS1, NA1, AS2, and NA2, respectively. ALMA emission (grayscale) is overlaid with the color-filled contours starting from 3~rms to 21~rms with a step of 1~rms. Each row has three panels. \textit{Left panel:} The extracted cores (blue ellipses) and their Astrodendrogram ``leaves'' (red contours). Green dashed contours show the negative emission starting from $-3$~rms with a step of $-1$~rms. ATLASGAL emission contours (black) are the same as Fig.~\ref{RadioMap}. The synthesized beam (black ellipse) is shown in the bottom left. \textit{Central panel:} Core mass is shown by the orange circle whose area is in proportion to mass. The numbers close to cores are core indexes ranking from the highest to the lowest mass. The highest and the lowest masses are indicated in the bottom left. MST short and long separations are shown in red and green lines, respectively, see more in Fig.~\ref{FragDistribution}. Blue lines indicate \lambdajthclump. Magenta arrows show the impacted direction. \textit{Right panel:} ALMA 1.3~mm color contours overlaid on GLIMPSE 8~\micron\ background. See Sect.~\ref{ALMAimpacted-result-morphology} for the meanings of yellow stars, crosses, triangles, and associated circles.}
        \label{ALMAemission1}
    \end{figure*}

\begin{figure*}[h]
      \centering
   \includegraphics[width=0.85\textwidth]{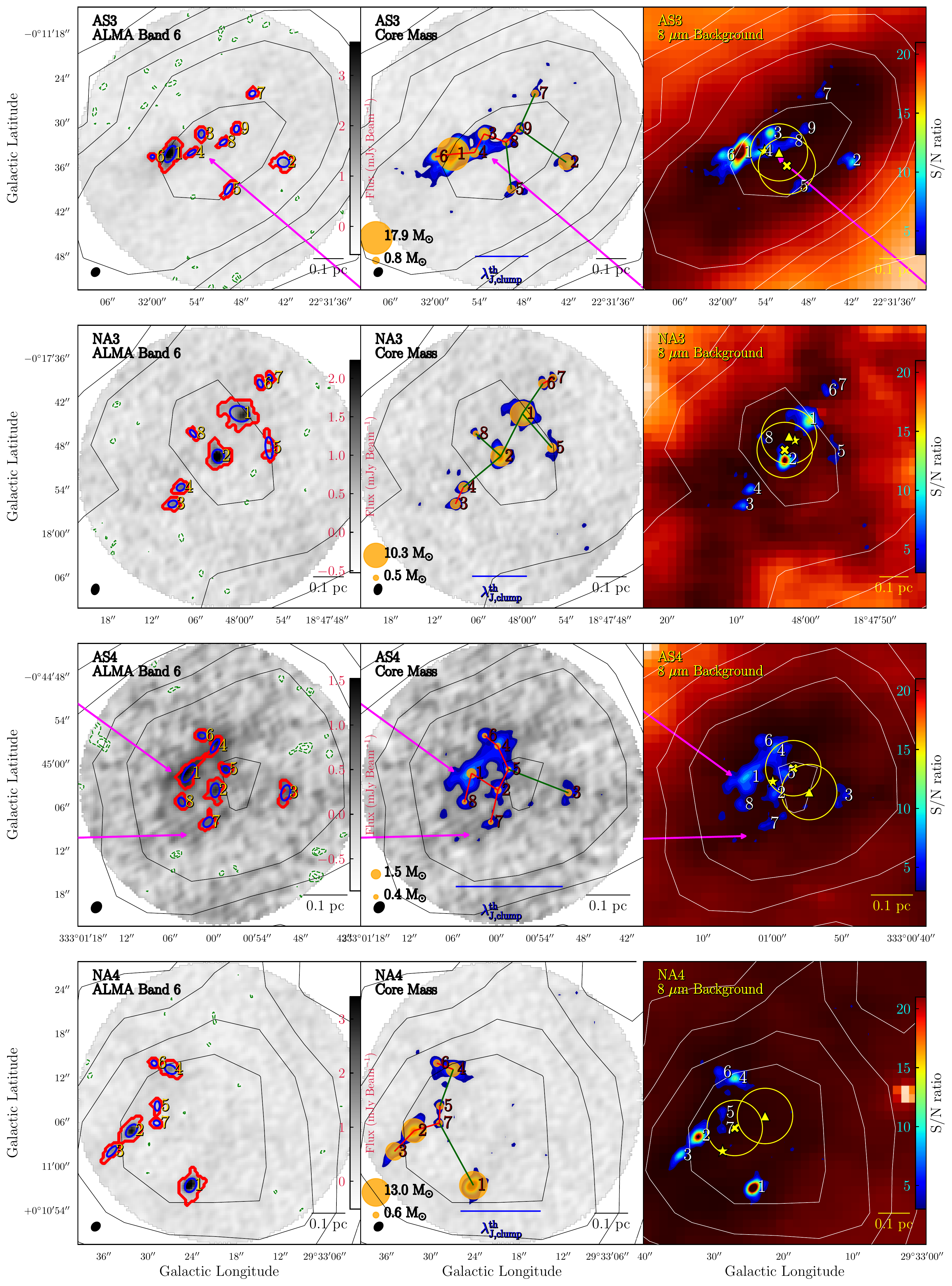}
       \caption{7~m + 12~m combined 1.3~mm continuum. The meanings of lines and marks are similar to those in  Fig.~\ref{ALMAemission1} but for AS3, NA3, AS4, and NA4. }
        \label{ALMAemission2}
    \end{figure*}

With spherical assumption, we calculate the core mass surface density $\Sigma_{\rm core}$ by $M_{\rm core}/\pi r^{2}$. Further assuming a molecular weight per hydrogen molecule of ${\mu}_{\rm H_{2}} = 2.8$ \citep{Kauffmann2008}, the \hmole~number density (\nhtnd) and column density (\nhtcd) are also estimated. All resulted physical parameters are listed in Table \ref{ALMACoresPhysical} in Appendix. The identified cores and their mass are shown in the middle panels of Figs. \ref{ALMAemission1} and \ref{ALMAemission2}. \\

\section{Fragmentation at $\sim0.025$~pc scale} \label{fragmentation-sec}
\subsection{Calculation} \label{FragCal}
Fragmentation is a multi-scale process existing in the interstellar medium from $\sim10$~kpc-scale spiral arm of the galaxy to $\sim100$~AU-scale protostellar system \citep{efremov98}. Taking advantage of the ALMA spatial resolution of $\sim0.025$~pc, in this work, we explore how the pc scale candidate HMSCs fragment into $\sim0.025$~pc-scale dense cores.

In the simplest case, fragmentation is dominated by thermal support in a non-magnetic, isothermal, homogeneous, and self-gravitating clump. Without turbulence, HMSCs are expected to fragment into cores with a mass around thermal Jeans mass (\mjth) and a core separation around thermal Jeans length (\lambdajth) \citep{jeans02,maclow04}:

\begin{equation}  
M_{\rm J}^{\rm th} = \frac{4\pi\rho}{3} \left(\frac{\lambda_{\rm J}^{\rm th}}{2}\right)^3 = \frac{\pi^{5/2}}{6} \frac{{\sigma_{\rm th}}^3}{\sqrt{G^3\rho}},
\label{equ-mj}
\end{equation}
\begin{equation}    
\lambda_{\rm J}^{\rm th} = \sigma_{\rm th} \left( \frac{\pi}{G\rho} \right) ^{1/2},
\label{equ-lambdaj}
\end{equation}
where $\rho$ is the mass density and $\sigma_{\rm th}$ is the thermal velocity dispersion: 
\begin{equation}  
\sigma_{\rm th} = \left( \frac{k_{\rm B} T}{\mu m_{\rm H}} \right) ^{1/2}.
\label{equ-thermal}
\end{equation}
The mean molecular weight per free particle $\mu$ is set to be 2.37 because $\sigma_{\rm th}$ is governed by \hmole~and He (see Appendix~\ref{Appendix-math}). The sound speed $c_{\mathrm{s}} = \sigma_{\mathrm{th}}$ in this case.

In the case that takes into account both thermal and non-thermal motions, $\sigma_{\rm th}$ is replaced by the total velocity dispersion $\sigma_{\rm {tot}}$ which is derived from the non-thermal velocity dispersion of the observed lines $\sigma_{\rm nth,line}$ and the sound speed $c_{\mathrm{s}}$ by
\begin{equation}
    \sigma_{\mathrm {tot}} = (\sigma_{\mathrm {nth,line}}^2 + c_{\mathrm{s}}^2)^{1/2},
\end{equation}
where $\sigma_{\rm nth,line} = {(\sigma ^2 - \sigma_{\rm th,line}^2)}^{1/2}$ \citep{palau2015, li19}. Here, $\sigma$ is derived from the observed line width $\delta {\rm v}$ in Table~\ref{SingleDishPar} by $\sigma = \delta {\mathrm v} /{(8 \mathrm{ln2})}^{1/2}$. The $\sigma_{\rm th,line}$ is the thermal velocity dispersion of the observed line $\sigma_{\rm th,line} = \left( \frac{k_{\rm B} T}{\mu_{\rm mole} m_{\rm H}} \right) ^{1/2}$. For $\mu_{\rm mole}$, the weight for different molecules, is 29 for \tco\ and 30 for \ceo\ and \htcop.

The corresponding Jeans parameters are \citep{palau2015}:
\begin{equation}  
M_{\rm J}^{\rm tot} = \frac{\pi^{5/2}}{6} \frac{{\sigma_{\rm tot}}^3}{\sqrt{G^3\rho}},\\
\lambda_{\rm J}^{\rm tot} = \sigma_{\rm tot} \left(\frac{\pi}{G\rho} \right) ^{1/2}.
\label{equ-jeans-tot}
\end{equation}
More specifically, considering the cases that the compression of large scale supersonic flow creates density enhancement by a factor of Mach number square, ${\mathcal {M}}^2$, the corresponding Jeans parameters are \citep{maclow04, palau2015}:
\begin{equation}  
M_{\rm J}^{\rm com, flow}  = \frac{\pi^{5/2}}{6} \frac{\sigma_{\rm nth,line}^3}{\sqrt{G^3\rho_{\rm eff}}},\\
\lambda_{\rm J}^{\rm com, flow} = \sigma_{\rm nth,line} \left(\frac{\pi}{G\rho_{\rm eff}} \right) ^{1/2},
\label{equ-jeans-flow}
\end{equation}
where $\rho_{\rm eff}$ is the effective density $\rho_{\rm eff} = \rho \mathcal{M} ^2$ and \mach~is equal to $\mathrm {v_{flow}} / c_{\mathrm s} = \sqrt{3} \sigma_{\rm nth,line} / c_{\rm s}$.

 The comparison between $M_{\rm core}$ and Jeans mass~is probably more straightforward than that between core separations $S_{\rm core}$ and Jeans length. The actual separations are longer than or equal to those separations derived from the images owing to the  projection effect, which is the main uncertainty when using separations to analyze the fragmentation. On average, the projected separations are shorter than the actual separations by a factor of $2/\pi$ (see math derivations in Appendix~\ref{Appendix-math}). A unified clump-scale \dustt\ derived from the SED fitting for each core embedded in the clump is a loose approximation for $M_{\rm core}$. Some cores have already presented molecular emission, such as the outflow tracer \siofive\ (see Sect.~\ref{FragmentationDiversity}), indicating star-forming activities. As a result, these cores are actually warmer and the core mass derived at the clump-scale \dustt~is an upper limit. The warm core mass $M_{\rm core}^{\rm warm}$ is probably more appropriate for these cores.  Both core mass and separation estimations have pros and cons and therefore we analyze the fragmentation by combining both \mjth\ and \lambdajth\ to attain a more consistent conclusion.

We derived the nearest neighbouring separations for all cores in each clump using the minimum spanning tree (MST) method. It works by connecting a set of points with a set of straight lines and ending up with a minimum sum of line lengths. The MST is frequently used in the research of the spatial distribution of star-forming objects, such as YSOs \citep{panwar19, pandeymnras20, pandeyapj20} and dense cores \citep{dib19,sanhueza19}. In particular,  \citet{clarke19} have demonstrated that MST is able to robustly detect the characteristic fragmentation scales. We plot the core MST for each clump in the middle panels of Figs.~\ref{ALMAemission1} and \ref{ALMAemission2}.  \\

%The fragmentation in some \textcolor{red}{candidate HMSCs} presents a hierarchical nature at $\gtrsim 0.1$~pc and $\lesssim 0.05$~pc in  Figs.~\ref{ALMAemission1} and \ref{ALMAemission2}. First, we define an ALMA emission morphology, \textbf{ensemble}, which describes the structure that hosts several very close ALMA cores in the iso-intensity contour.  To differentiate fragmentation scales, we highlight the corresponding MST with orange lines that connect cores in the iso-contour ($>2.5$~rms) of ALMA continuum in Figs.~\ref{ALMAemission1} and \ref{ALMAemission2}. 

\subsection{General trend shown by combined images} \label{FragTrend}
\begin{figure}[htb]
     \centering
   \includegraphics[width=0.45\textwidth]{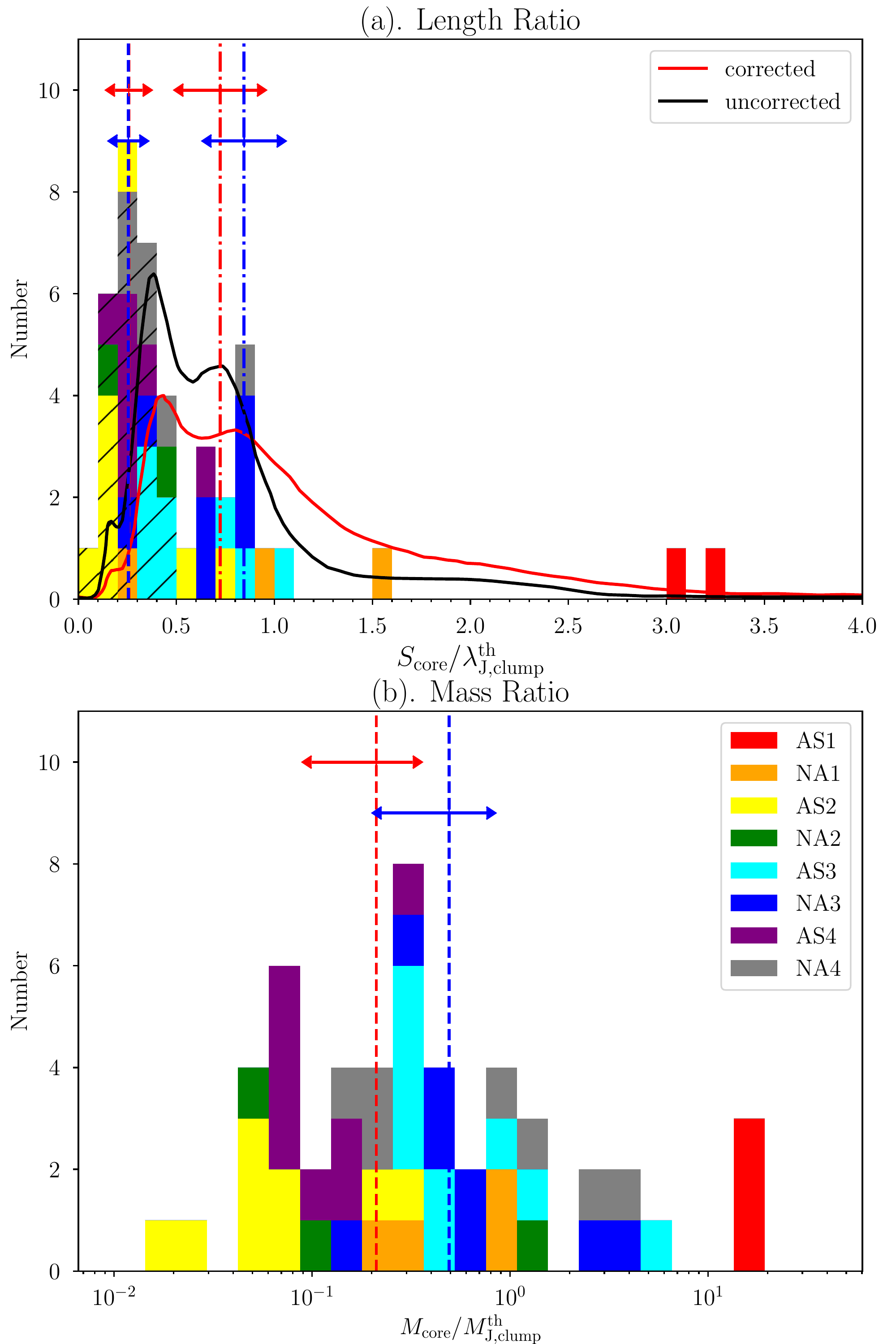}
       \caption{Core separations and mass scaled by clump thermal Jeans parameters. Histograms with different colors represent different candidate HMSCs. \textit{Panel (a)} shows the scaled core separations $S_{\rm core}/{\lambda_{\rm J,clump}^{\rm th}}$. The hatched histograms indicate short separations that two cores are in the same ensembles (see Sect.~\ref{FragTrend7m}). Black and red solid lines represent the distributions derived from \citet{svoboda19} by multiplying their probability density functions (PDFs) with the total number of cores in our cases. Black and red lines show the projected nearest neighbor separations and the projection-corrected nearest neighbor separations, respectively, see details in \citet{svoboda19}. Blue and red dash lines are median values of short separations for NA and AS, respectively. Similarly, dash-dotted lines indicate median values of long separations. \textit{Panel (b)} shows the scaled mass $M_{\rm core}/{M_{\rm J,clump}^{\rm th}}$. The red and blue lines indicate median values for AS and NA, respectively.}
        \label{FragDistribution}
    \end{figure}

We show the core MST separations and mass scaled by clump thermal Jeans parameters in Fig.~\ref{FragDistribution}. Most of the scaled core mass ($M_{\rm core}/{M_{\rm J,clump}^{\rm th}}$, the ratio of core mass to clump thermal Jeans mass) are less than one~except for the most massive cores and AS1, suggesting the dominant role of the thermal motions in the fragmentation. Similar trend is also implied by the scaled core MST separations ($S_{\rm core}/{\lambda_{\rm J,clump}^{\rm th}}$, the ratio of core separation to clump thermal Jeans length), which are less than one for most sources except for AS1.

The distribution of $S_{\rm core}/{\lambda_{\rm J,clump}^{\rm th}}$ presents a bimodal profile with $S_{\rm core}/{\lambda_{\rm J,clump}^{\rm th}}$ peaked at $\sim0.3$ and $\sim0.8$ in our limited sample. \citet{svoboda19} found a similar bimodal distribution peaking at around 0.3 and 1 when surveying twelve candidate HMSCs (400~\msun~to 3600~\msun\ with a median mass of 800~\msun) by ALMA Band 6 observations. The physical scale they studied (0.015~pc in 0.5 pc-scale clump) is similar to our cases. Their bimodal profiles that are corrected and uncorrected for projection effects are shown in Fig.~\ref{FragDistribution}. By checking with 7~m + 12~m combined images in Figs.~\ref{ALMAemission1} and \ref{ALMAemission2}, the reason for the bimodal profile could be the hierarchical fragmentation at two scales $\gtrsim 0.1$~pc and $\lesssim 0.05$~pc. The most significant case is AS2, which shows four groups of cores with a separation of 0.1-0.2~pc and each group contains two to three cores with separations of $\lesssim 0.05$~pc, corresponding to the double peaks in the bimodal profile. \\

\subsection{Hierarchical fragmentation shown by 7~m data} 
\label{FragTrend7m}
\begin{figure*}[htbp]
     \centering
   \includegraphics[width=0.73\textwidth]{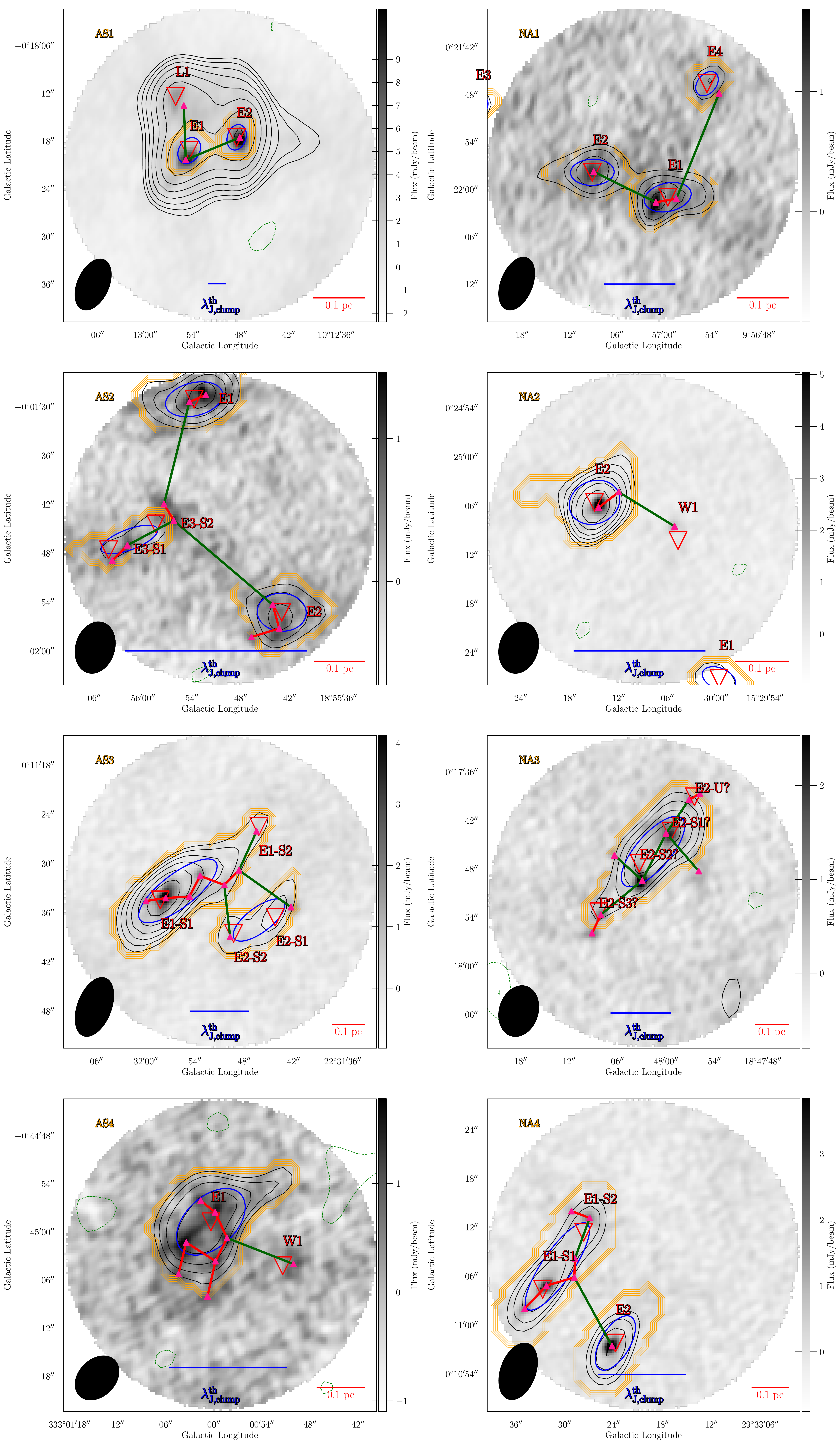}
       \caption{Hierarchical fragmentation. Grayscale shows the primary beam uncorrected 7~m + 12~m combined images, overlaid with primary beam uncorrected 7~m ACA continuum in black (positive) and green (negative) contours with levels of $\pm$ 7~m image $ {\rm rms} \times [3, 3^{1.25}, 3^{1.50}, 3^{1.75}, 3^{2.0}, 3^{2.25}, 3^{2.50}, 3^{2.75}, 3^{3.0}]$. Blue ellipses and yellow contours show the ``leaves'' and their shapes derived by Astrodendro. Large and small triangles mark the positions of candidate ensembles and cores, respectively. Green and red lines are MST core separations the same as those in Figs.~\ref{ALMAemission1} and \ref{ALMAemission2}. Blue lines indicate the clump thermal Jeans length, \lambdajthclump.}
        \label{Subfragmentation}
    \end{figure*}

To confirm the hierarchical nature of fragmentation, the 7~m~+~12~m combined continuum overlaid with 7~m ACA continuum contours are presented in Fig.~\ref{Subfragmentation}. It is obvious that a number of very close cores are hosted in the same 7~m iso-contours, indicating the sub-fragmentation from structures outlined by iso-contours to cores. Here we term these 7~m ACA emission structures as ``ensembles''. We first identify ``leaves'' within Astrodendro in 7~m images uncorrected for the primary beam. The criteria of Astrodendro are looser than those used in the combined images (minimum emission $= 2.5$~rms, step $= 1$~rms, and minimum size $= $ half of the 7~m ACA synthesized beam). The information of all identified 7~m image ``leaves'' could be found in Figs.~\ref{Emission7m1}, \ref{Emission7m2}, and Table~\ref{ACALeavesTable} in Appendix. These 7~m image ``leaves'' are possible candidate ensembles, however, we classify the ``ensembles'' in a more rigorous way here:

First, if the identified 7~m ``leaves'' have a major-to-minor ratio less than 2.5, they are directly selected as ensembles. The ensemble mass and density are calculated in a manner that
is similar to that in Equation~\ref{MassEquation}.

Next, for the identified 7~m ``leaves'', which present a prolonged shape (the major-to-minor ratio $> 2.5$), they are probably made of more than two individual candidate ensembles if the ensemble peaks are too similar to be differentiated by Astrodendro. We carefully check the 7~m ACA continuum contours to search for the peak and subpeak for these prolonged ``leaves'' and then simply assume that these peaks and subpeaks are the positions of the candidate ensembles. Taking into account the roughly equal peaks, we directly use the density of the ``leaves'' as the estimated density of each ensemble.

Lastly, we note that some 7~m ACA emission structures are not identified in Astrodendero due to their low signal-to-noise ratio (S/N), whereas they host one or few low-mass cores detected in the 7~m + 12~m combined images, such as the candidate ensemble W1 in NA2. We mark these weak structures by eye and assume that they are also candidate ensembles. The mass and density are not calculated due to their low S/N. \\

All the identified ensembles, their density, and thermal Jeans parameters are listed in Table~\ref{Jeansparameters_clumps_ensembles}. With regard to the calculation of the ensemble thermal Jeans length, \lambdajthens, we can see that AS3 and NA3 are special because a number of their cores are nearly equally spaced in a prolonged leaf, invoking the cylindrical fragmentation. For an isothermal gas ``cylinder'', it could fragment into equally spaced cores under gravitational instability. The typical separation and mass in the case of cylindrical thermal fragmentation is: 
\begin{equation}
\lambda_{\rm J,cylinder}^{\rm th} = 22\sigma_{\rm th} \left( 4\pi G\rho_{c} \right) ^{-1/2}, 
M_{\rm J,cylinder}^{\rm th} = \frac{22}{{\pi}^{1/2}} \frac{{\sigma_{\rm th}}^3}{\sqrt{G^3\rho_{\rm c}}},
\label{equ-lambdaj-cylinder}
\end{equation}
where $\rho_{c}$ is the density at the center of the ``cylinder'' in virial equilibrium \citep{chandrasekhar1953, jackson2010, wang2014}. We corrected the \lambdajthens~of AS3 main leaf according to Equation~\ref{equ-lambdaj-cylinder}. The $\rho_{c}$ is assumed to be the minimum density of the cores embedded in the filament, which is about one magnitude larger than the filament average density, a result that is similar to other methods of estimations in studies of filament fragmentation, such as in \citet{jackson2010, wang2014, lu2018}.

Ensemble identification is not clear for NA3. The identified 7~m leaf has a major-to-minor axis ratio of $>2.5$ but the presented three subpeaks \textit{E2-S1?}, \textit{E2-S2?}, and \textit{E2-S3?} cannot cover all cores (see Fig.~\ref{Subfragmentation}). NA3 is the most distant source in our sample (4.9~kpc compared to 3.1--3.4~kpc of other sources) except for AS3. In addition, the major axis of 7~m image synthesized beam roughly aligns with the filament. This evidence may shed light on the possibility that its longer core separations correspond to separations between ensembles rather than to core separations in the same ensembles. To cover all possibilities, we considered two cases for NA3: (1). The identified 7~m image ``leaf'' is merged by few ensembles and, thus, it represents the fragmentation from clump to ensemble. (2). The ``leaf'' is just one filamentary ensemble and, therefore, it represents the fragmentation from ensemble (filament) to core. In the second case, we apply the calculation of \lambdajthens\, that is corrected with Equation~\ref{equ-lambdaj-cylinder}.

%The One problematic source is NA3 owing to its large major to minor ratio and unclear subpeaks. We tentatively mark the three subpeaks for NA3. The candidate ensembles and their properties are listed in Table xx. We estimate the ensemble thermal Jeans length \lambdajthens~with the properties of identified ``leaves'' and Equ.~\ref{equ-lambdaj}. For the candidate ensembles in the same ``leaves'', we simply use the ``leaves'' density as the ensemble density considering their size and column density are similar (because in iso-contour). The ensemble separation scaled by clump Jeans length \lambdajthclump~and separation of cores embedded in ensembles scaled by ensemble Jeans length are shown in Fig.~\ref{EnsembleSubfragmentation}.

%\textcolor{red}{It shows a clearly filamentary structure. }

% Jeans tables 
      \begin{table*}[htb]
      \centering % used for centering table
      \begin{threeparttable}
      \setlength{\tabcolsep}{3.5pt}
      \renewcommand{\arraystretch}{1.5}
      \tiny
      \caption{Jeans parameters for clumps and ensembles.} % title of Table
      \label{Jeansparameters_clumps_ensembles} % is used to refer this table in the text
       \begin{tabular}{c c c c c c c c c c c c l l l} % centered columns (4 columns)
       \hline\hline % inserts double horizontal lines
     Source  &  \lambdajthclump & $\frac{\lambda_{\rm J,clump}^{\rm tot}}{\lambda_{\rm J,clump}^{\rm th}}$ & $\frac{\lambda_{\rm J,clump}^{\rm com,flow}}{\lambda_{\rm J,clump}^{\rm th}}$ &\mjthclump  & $\frac{M_{\rm J,clump}^{\rm tot}}{M_{\rm J,clump}^{\rm th}}$ & $\frac{M_{\rm J,clump}^{\rm com,flow}}{M_{\rm J,clump}^{\rm th}}$ &   Ens.\tnote{\textit{(a)}}  & Subfrag.\tnote{\textit{(b)}}  & Ens. \nhtnd\tnote{\textit{(c)}} & $\overline{M_{\rm ens}}$\tnote{\textit{(c)}} & \lambdajthens & \mjthens & Cores\tnote{\textit{(d)}} & 7~m Leaves\tnote{\textit{(e)}}   \\ % table heading
           &  pc                                     &  & &  \msun      & & &           &  &  10$^6$cm$^{-3}$     &  \msun &  pc         &  \msun   &  ALMA$n$ & ACA$n$          \\
       \hline % inserts single horizontal line
        \multirow{3}*{AS1}  & \multirow{3}*{0.034$\pm$0.007}&\multirow{3}*{6.8} & \multirow{3}*{0.58} & \multirow{3}*{0.8$\pm$0.3} & \multirow{3}*{311}& \multirow{3}*{26} & E1  & N & 3.6 & \multirow{3}*{$\lesssim7$} & 0.013$\pm$0.003  & 0.3$\pm$0.1    &  2  &  1 \\   %
                            &                               & & &                              & & & E2  & N & 4.1 &  & 0.012$\pm$0.002  & 0.3$\pm$0.1    &  1  & 2   \\
                            &                               & & &                              & & & W1  & N & $-$ &  & $-$  &   $-$    &  3   &  $-$       \\
       \hline
        \multirow{4}*{NA1}  &   \multirow{4}*{0.14$\pm$0.03} &\multirow{4}*{2.7} & \multirow{4}*{0.58} & \multirow{4}*{2.2$\pm$0.8} & \multirow{4}*{19} & \multirow{4}*{3.5} & E1  & Y & 0.50 & \multirow{4}*{$\sim 2.5$}  & 0.030$\pm$0.006  & 0.5$\pm$0.2    &   1, 3 & 1  \\   %
                           &                                & & &                             & & & E2  & N & 0.52 & & 0.029$\pm$0.006  & 0.5$\pm$0.2    &  2    &  2 \\    
                           &                                & & &                             & & & E3  & N & 0.98 & & 0.021$\pm$0.005  & 0.3$\pm$0.1    &  $-$  &  3  \\  
                           &                                & & &                             & & & E4  & N & 0.66 & & 0.026$\pm$0.006  & 0.4$\pm$0.2    &  4    &  4  \\  
       \hline
        \multirow{4}*{AS2} &   \multirow{4}*{0.36$\pm$0.07} &\multirow{4}*{5.7} & \multirow{4}*{0.58} & \multirow{4}*{8.6$\pm$2.9}  & \multirow{4}*{181}& \multirow{4}*{18}& E1  & Y  & 0.51 & \multirow{4}*{$\sim 3.5$} & 0.036$\pm$0.007   & 0.9$\pm$0.3    &  1, 3  &     1   \\   % 
                           &                                & & &                             & & & E2    & Y & 0.34 & & 0.044$\pm$0.010   & 1.1$\pm$0.4    &  2, 8, 9  &   2     \\   % 
                           &                                & & &                             & & & E3-S1 & Y & 0.27 & & 0.049$\pm$0.012   & 1.2$\pm$0.5    &  4, 6  &   3     \\   % 
                           &                                & & &                             & & & E3-S2 & Y & 0.27 & & 0.049$\pm$0.012   & 1.2$\pm$0.5    &  5, 7  &   3     \\   % 
        \hline
        \multirow{3}*{NA2} &   \multirow{3}*{0.25$\pm$0.07} &\multirow{3}*{5.7} & \multirow{3}*{0.58} & \multirow{3}*{4.6$\pm$1.8}  &\multirow{3}*{182} & \multirow{3}*{18} & E1   & Y & 0.53  & \multirow{3}*{$\lesssim 4.9$} & 0.031$\pm$0.007   &  0.6$\pm$0.2   & 1, 3   &    1    \\   %   % 
                           &                                & & &                             & & & E2  & N  & 0.65  & & 0.028$\pm$0.007   &  0.5$\pm$0.2   & $-$   &    2    \\   % 
                           &                                & & &                             & & & W1  & N  & $-$   & & $-$ &     &  2  &    $-$    \\   % 
       \hline
        \multirow{4}*{AS3} &   \multirow{4}*{0.18$\pm$0.05} &\multirow{4}*{10.7} & \multirow{4}*{0.58} & \multirow{4}*{2.9$\pm$1.1} &\multirow{4}*{1216} & \multirow{4}*{65} & E1-S1 & Y & 0.41 &  \multirow{4}*{$\lesssim 13$}& 0.034$\pm$0.007 &  0.6$\pm$0.2   & 1, 3, 4, 6, 8, 9   &  1  \\   % 
                           &                                & & &                            & & & E1-S2 & N & 0.41 & &0.034$\pm$0.007 &  0.6$\pm$0.2   & 7   &      1  \\   % 
                           &                                & & &                            & & & E2-S1 & N & 0.25 & &0.043$\pm$0.010 &  0.7$\pm$0.3   & 2   &      2  \\   %      
                           &                                & & &                            & & & E2-S2 & N & 0.25 & &0.043$\pm$0.010 &  0.7$\pm$0.2   & 5   &      2  \\   %   
      \hline
        \multirow{4}*{NA3} &   \multirow{4}*{0.18$\pm$0.04}  &\multirow{4}*{6.8} & \multirow{4}*{0.58} &\multirow{4}*{2.9$\pm$1.0}  & \multirow{4}*{320}& \multirow{4}*{26} & E2-S1? & $-$ & \multirow{4}*{0.65}& \multirow{4}*{$-$} &\multirow{4}*{0.043$\pm$0.01} &  \multirow{4}*{0.7$\pm$0.2}  &  1?  &  \multirow{4}*{2}       \\   % 
                           &                                 & & &                            & & & E2-S2?  & $-$ & & &  &     &  2?  &        \\   % 
                           &                                 & & &                            & & & E2-S3?  & $-$ & & &  &     &  3, 4?  &        \\
                           &                                 & & &                            & & & E2-U?  & $-$ & & &  &     &  6, 7?  &        \\                           
        \hline  
        \multirow{2}*{AS4} &   \multirow{2}*{0.24$\pm$0.07}  &\multirow{2}*{9.4} & \multirow{2}*{0.58} & \multirow{2}*{5.0$\pm$2.0}  & \multirow{2}*{819}& \multirow{2}*{50} & E1 & Y  & 0.2 & \multirow{2}*{$\sim 5$} & 0.054$\pm$0.012  & 1.1$\pm$0.4    & 1, 2, 4, 5, 6, 7, 8   &    1     \\   %  
                           &                                 & & &                             & & & W1 & N & $-$ & &$-$              &  $-$           & 3                     &    $-$    \\   % 
        \hline
        \multirow{3}*{NA4} &   \multirow{3}*{0.24$\pm$0.05}  &\multirow{3}*{6.3} & \multirow{3}*{0.58} & \multirow{3}*{4.0$\pm$1.4} & \multirow{3}*{251}& \multirow{3}*{22} & E1-S1 & Y & 0.37 & \multirow{3}*{$\sim 15$} & 0.036$\pm$0.007 & 0.6$\pm$0.2   & 2,3,5,7   & 1       \\   % 
                           &                                   & & &                            & & & E1-S2 & Y & 0.37 & & 0.036$\pm$0.007 & 0.6$\pm$0.2   & 4, 6      &    1    \\   % 
                           &                                   & & &                            & & & E1-S3 & N & 0.49 & & 0.031$\pm$0.007 & 0.5$\pm$0.2   &  1        &     2   \\   %                 
        \hline                                         
      \end{tabular}
      \begin{tablenotes}
      
      \item [\textit{(a)}] Candidate ensembles. The nomenclature: (1) E$n$. Candidate ensemble shown by 7~m image leaf ACA$n$ in Table~\ref{ACALeavesTable} of Appendix. (2) E$n$-S$m$. Subpeak $m$ of 7~m image leaf ACA$n$ in Table~\ref{ACALeavesTable}. (3) W$n$. Weak-emission ensemble. See details about three types of ensembles in Sect.~\ref{FragTrend7m}. \textit{E2-S1?}, \textit{E2-S2?}, and \textit{E2-S3?} are uncertain subpeaks for NA3, as explained in Sect.~\ref{FragTrend7m}. \textit{E2-U?} is just a potential candidate because of the existence of two close cores ALMA 6 and 7, but this candidate is highly uncertain.
      \item [\textit{(b)}] Whether sub-fragmentation is detected.
      \item [\textit{(c)}] Estimated ensemble number density and average mass, see details in Sect.~\ref{FragTrend7m}.
      \item [\textit{(d)}] Embedded cores identified in combined images. The ``?'' following the core index for NA3 means the association between cores and ensembles is uncertain, see Sect.~\ref{FragTrend7m}.
      \item [\textit{(e)}] 7~m image ``leaves'', listed in Table~\ref{ACALeavesTable}, correspond to candidate ensembles.
      \end{tablenotes}
      \end{threeparttable}
      \end{table*}

      \begin{figure*}[ptbh]
      \centering
   \includegraphics[width=0.95\textwidth]{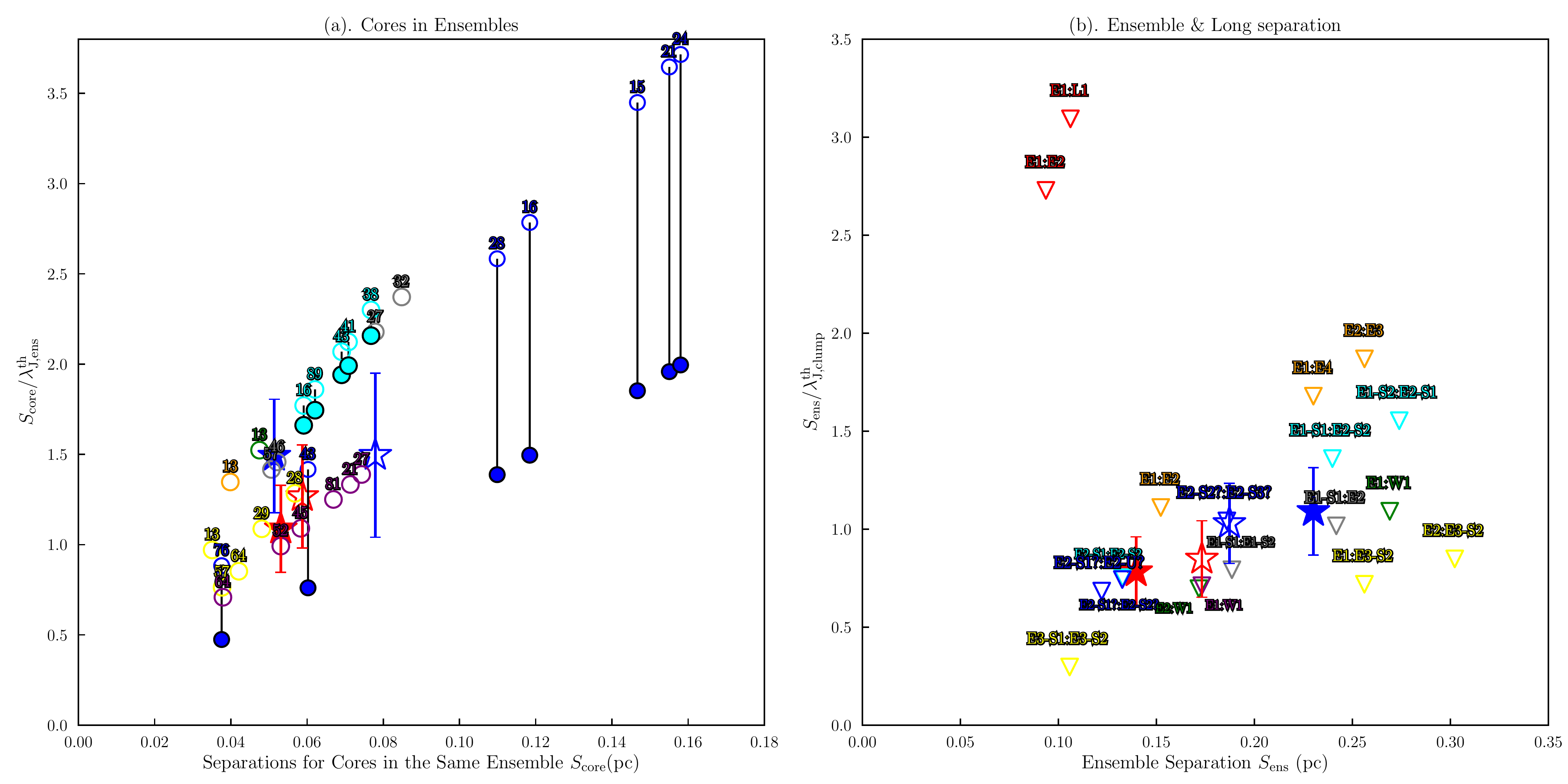}
       \caption{Thermal dominant fragmentation. Red and blue filled stars represent the median values for AS and NA, except for AS3 and NA3, while the stars with blank facecolor represent the median values for all AS and NA. The labels near the data points mark the ALMA cores (in the same ensembles) or ensembles making up the separations. \textit{Panel (a)} shows the separations for the cores in the same ensembles, scaled by ensemble thermal Jeans length \lambdajthens. The color-coding for different clumps is similar to that of Fig.~\ref{FragDistribution}. The circles with cyan and blue facecolors are the recalculations in the case of cylindrical fragmentation for AS3 and NA3, respectively. \textit{Panel (b)} shows the ensemble MST separations scaled by clump thermal Jeans length \lambdajthclump.  }
        \label{EnsembleSubfragmentation}
    \end{figure*}

Figure~\ref{EnsembleSubfragmentation} (a) shows the core separations scaled by ensemble thermal Jeans length ($S_{\rm core}/{\lambda_{\rm J,ens}^{\rm th}}$) for the cores in the same ensembles.  The median $S_{\rm core}/{\lambda_{\rm J,ens}^{\rm th}}$ is around 1 to 1.5. The $S_{\rm core}/{\lambda_{\rm J,ens}^{\rm th}}$ for AS3 and NA3 is significantly reduced from $>2$ to $<2$ in the case of cylindrical fragmentation. A maximum factor of 1.5\lambdajth\ means that there is a non-thermal velocity dispersion, $\sigma_{\rm nth}$, playing a role equivalent to thermal dispersion, $\sigma_{\rm th}$. Even if the ensemble-scale ($<0.1$~pc) line measurement is not presented in this continuum study, other studies toward embedded structures with similar scale in HMSCs reveal a line dispersion of $\gtrsim0.5$--2~\kms, which is twice larger than $\sigma_{\rm th}$ at least \citep{li19}. Another potential bias is that the mass and density of most massive 7~m ``leaves'' are likely overestimated due to the existing star-formation activities (see Sect.~\ref{FragmentationDiversity}). A dust temperature improvement of 3~K in ensemble mass calculation could reduce the derived mass and density by one third, leading to a new thermal Jeans length $ \lambda_{\rm J, ens}^{\rm th, warm}>\sqrt{3/2}$\lambdajthens. Therefore, the thermal motions are more effective in sub-fragmentation from ensembles to cores although other effects such as turbulence and magnetic field may play a limited assistant role in the sub-fragmentation process.

Figure~\ref{EnsembleSubfragmentation} (b) shows ensemble separations scaled by clump thermal Jeans length ($S_{\rm ens}/{\lambda_{\rm J,clump}^{\rm th}}$). The median value of $S_{\rm ens}/{\lambda_{\rm J,clump}^{\rm th}}$ is around 1. Nearly all $S_{\rm ens}/{\lambda_{\rm J,clump}^{\rm th}}$ are less than 1.5 except for AS1 and NA1, confirming the dominance of the thermal motions in fragmentation from clumps to ensembles for most candidate HMSCs. The mass of NA1 ensembles is in low-mass range (1--3~\msun), similarly to \mjthclump. The turbulent fragmentation model is inappropriate for most HMSCs because both ensemble mass and separations are significantly smaller than \mjtotclump\ and \lambdajtotclump. Taking the compression of supersonic flow into account, the resulting \lambdajflowclump, which is even smaller than \lambdajthclump, gives a more contradictory picture than the one given by the thermal fragmentation.\\

As a result, the early-stage fragmentation in the high-mass clumps probably takes place in a hierarchical fashion: clumps fragment into ensembles with separations around clump thermal Jeans length, \lambdajthclump, and then some of these ensembles continuously sub-fragment into cores with separations slightly larger than ensemble Jeans length, \lambdajthens.\\

\subsection{Considering possible differences between AS and NA}
In this limited statistic, there is no clear overall trend that fragmentation between AS and NA has a significant difference. The $S_{\rm core}/{\lambda_{\rm J,clump}^{\rm th}}$, $S_{\rm ens}/{\lambda_{\rm J,clump}^{\rm th}}$,  $S_{\rm core}/{\lambda_{\rm J,ens}^{\rm th}}$ in Figs.~\ref{FragDistribution} and~\ref{EnsembleSubfragmentation} only show a slight difference between AS and NA. The median values of $M_{\rm core}/{M_{\rm J,clump}^{\rm th}}$ in Fig.~\ref{FragDistribution} indicate that the cores in NA are probably more massive than AS, but we note that AS2 has much more low-mass cores than NA2. The nine cores of AS2 in the low-mass end could be the reason for which the cores in type AS have a lower median $M_{\rm core}/{M_{\rm J,clump}^{\rm th}}$ when compared to those in NA. 

\begin{figure}[htb]
     \centering
   \includegraphics[width=0.49\textwidth]{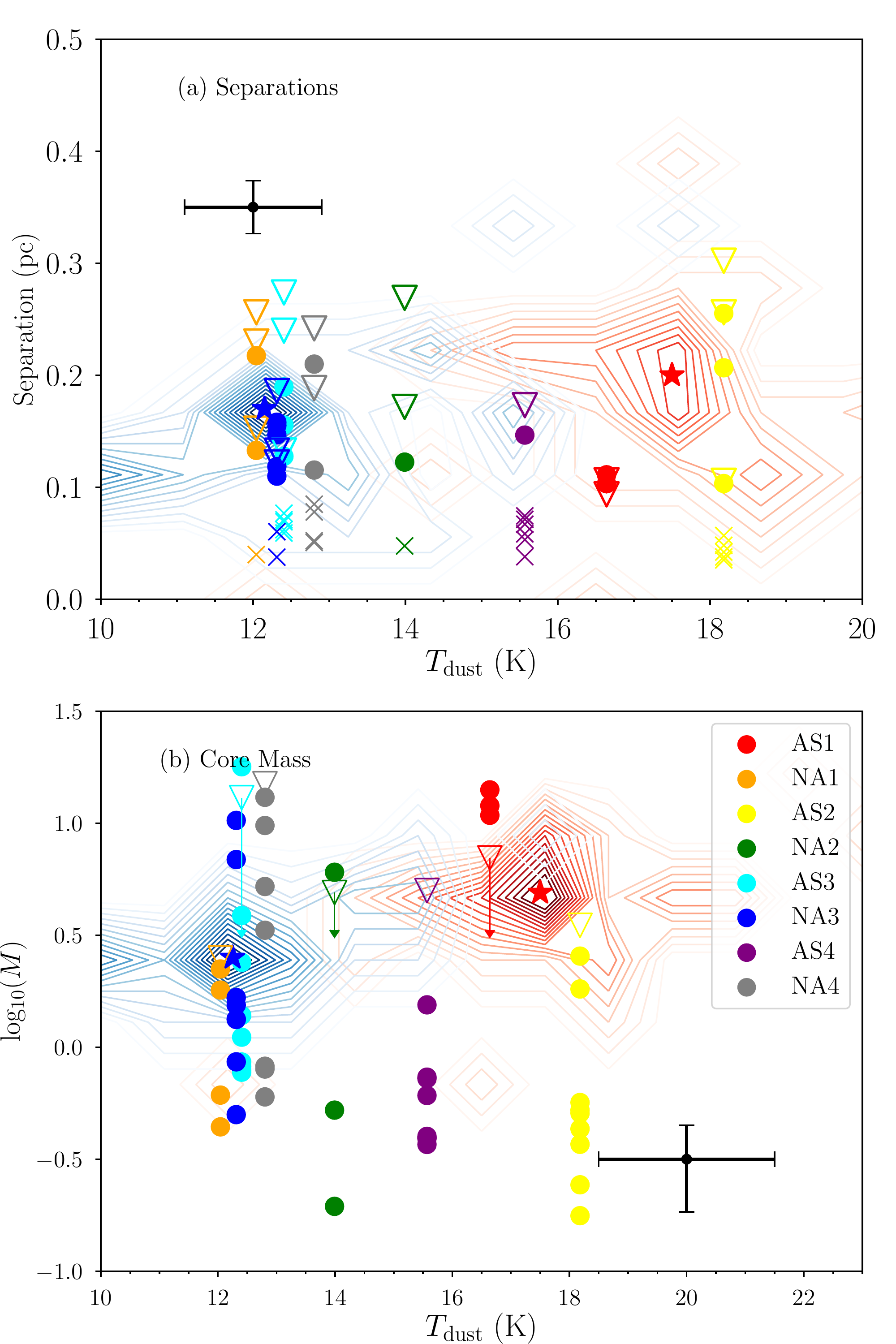}
       \caption{Expected \lambdajthclump\ and \mjthclump\ for 58 candidate HMSCs located in the 3--5~kpc distance range. Red and blue contours show the distributions of \mjthclump\ and \lambdajthclump\ for all strongly impacted AS (31 S-type AS, see details in Paper\,I) and all NA (27), respectively. Contours with a heavier color mean a larger number of points. Red and blue stars mark the peaks of corresponding distributions. \textit{Panel (a)} shows the distributions of \lambdajthclump. The triangles, crosses, and circles represent ensemble MST separations, core short MST separations, and core long MST separations for the ALMA sources in this paper, respectively. \textit{Panel (b)} shows the distributions of \mjthclump. Circles and triangles represent core mass and average ensemble mass for the ALMA sources in this paper, respectively.}
        \label{FragmentationOverall}
    \end{figure}

For thermal fragmentation, a higher \dustt\ results in a larger \mjth\ and \lambdajth\ if density $\rho$ is roughly kept constant. A higher \dustt\ (3--6~K) and a similar $\rho$ for AS, as compared to NA, have been revealed with the single-dish results in Paper\,I. Taking advantage of 58 candidate HMSCs located in the 3--5~kpc range given in Paper\,I, the expected \mjthclump\ and \lambdajthclump\ for strongly impacted AS (S-type AS, 31 in total, see Paper\,I) and non-impacted NA HMSCs (27 in total) are shown in Fig.~\ref{FragmentationOverall}. The peak of \mjthclump\ distribution shifts from about 2~\msun\ for NA to 6~\msun\ for AS. If most AS and NA HMSCs follow thermal fragmentation, the fragmented cores in AS are expected to be more massive than NA on average. The study of eight candidate HMSCs in this paper is a pilot and tentative exploration due to the very limited sample. An ALMA survey covering several dozens of AS and NA candidates observed with the similar spatial resolution is crucial to resolving the trend suggested in this work.\\

\subsection{Diversity of fragmentation} \label{FragmentationDiversity}
In this section, we discuss in greater detail more the fragmentation of the candidate HMSCs, presenting diverse properties besides the general hierarchical fragmentation found. A scaled separation $S_{\rm core}/{\lambda_{\rm J,clump}^{\rm th}}$ of $\sim3$ suggests that cores in AS1 probably result from not only thermal motions. Meanwhile, massive cores probably exist in some candidate HMSCs. The most massive cores in AS1, AS3, NA3, and NA4 have a scaled mass $M_{\rm core}/$\mjthclump\ of 20, 6, 3, and 3, respectively. As mentioned in Sect.~\ref{Core-extracted-sec}, the evolutionary stage of cores is an important source of uncertainty when estimating $M_{\rm core}$ because a large scaled mass could be due to an underestimated \dustt.\\

\textbf{Notable case AS1:} AS1 fragments into three massive cores with mass from 10-17~\msun\ and a separation of $\simeq 0.1$~pc. Both \mjthclump\ and \lambdajthclump\ show that its fragmentation cannot be explained by thermal motions alone. The turbulent Jeans parameters in total support case (\mjtotclump\ and \lambdajtotclump) are much larger than the observed parameters, whereas taking compressible flow into account gives a closer \mjflowclump\, ($\sim20$~\msun) with a more contradict \lambdajflowclump\, ($\sim1/2$\lambdajthclump). It implies that turbulence alone is equally insufficient to fully explain AS1 fragmentation. Introducing more physical factors, such as magnetic field support, and considering a different relative significance for each factor in the fragmentation process could probably lead to the building of a more suitable scenario to explain this process \citep{tang19}.

A potential bias is that the worse rms mass sensitivity (0.15~\msun~Beam$^{-1}$) of AS1 combined image compared to that of other candidate HMSCs would make us leave out more low-mass cores. We image AS1 continuum with only 12~m array dataset, resulting in an image rms of 0.2--0.3~mJy Beam$^{-1}$, which is better than combined image of AS1 (0.55~mJy Beam$^{-1}$) and close to combined images of other candidate HMSCs. Using the same Astrodendro parameter, the number of identified cores in AS1 12~m image is the same as the one in the combined image. Therefore, we propose that the number of missed cores in AS1 is likely at a level similar to other candidate HMSCs. 

High-velocity components ($>20$~\kms) of outflow tracer \siofive\ \citep{Louvet2016, matsushita19} are detected within our data set toward cores ALMA1 and ALMA2, indicating the stars are forming and still accreting mass there. Meanwhile, thermal methanol lines \chtoh~4(2, 2)-3(1, 2)-E at 218.44 GHz (E$_u$ = 45.46~K), which probably trace the shocked gas \citep{Zapata2011, Johnston2014}, are also detected toward both two cores. When taking the warm mass $M_{\rm core}^{\rm warm}$ as core mass, the scaled mass $M_{\rm core}^{\rm warm}/$\mjthclump\ is still around 8-13. Combining it with the mass accretion and very early evolutionary stage of AS1 (70~\micron\ dark), AS1 cores probably have the potential to form high-mass stars. The highest clump surface density compared to other candidate HMSCs, 1.3~g~cm$^{-2}$, also meets the high-mass star formation density criterion of 1~g~cm$^{-2}$ suggested by \citet{krumholz08}. \\

\textbf{Star formation and warm gas indicators:} A check on whether massive cores show star formation signposts is beneficial to understand the overestimated mass and whether high-mass prestellar cores exist. We will present a complete spectral study of this ALMA data set in a forthcoming paper. Here we only describe a few cases of significant spectral detection in our regridded spectra.

Recent studies show that \htco\ mainly traces denser and warmer gas ($\gtrsim30$~K, $\sim10^{5}$~cm$^{-3}$), such as warm gas in OMC-1 \citep{tang18} and cometary globule (bright-rimmed cloud impacted by \hii~regions, \citealt{mookerjea19}). The detection of \htco\ is also related to shock chemistry corresponding to outflow \citep{contoreras18, gieser19}. The most massive ensemble E1 in AS2 (see Fig.~\ref{Subfragmentation}), containing cores ALMA1 and ALMA3, is just in front of the PDR traced by PAH 8~\micron\ emission, showing the gas and dust distribution shaped by the \hii~region. This structure will be discussed in more detailed in Sects.~\ref{shaping-result} and \ref{shaping-dis}. The small organic molecular thermometer \htco\ is only detected toward this ensemble while there is no detection for other ensembles in AS2, suggesting that ALMA1 and ALMA3 are in a warmer environment. \htco\ is also detected in the most massive cores of NA2 (ALMA1, $M_{\rm core}\simeq6$~\msun), NA3 (ALMA1, $M_{\rm core}\simeq10.3$~\msun; ALMA2, $M_{\rm core}\simeq6.9$~\msun), indicating shocked gas or warmer gas, or both. As a result, the mass of these massive cores could be overestimated.

There are high-velocity components ($\gtrsim10$~\kms) of \siofive\, detected toward the most massive core of NA4 (ALMA1, $M_{\rm core}\sim13$~\msun) meanwhile c-H$_2$COCH$_2$ (Ethylene Oxide, hot core tracer, see \citealt{ikeda01}) is detected towards ALMA2 ($M_{\rm core}\sim10$~\msun), showing that massive cores of NA4 are already at the protostellar stage. 

\begin{figure}[htb]
     \centering
   \includegraphics[width=0.48\textwidth]{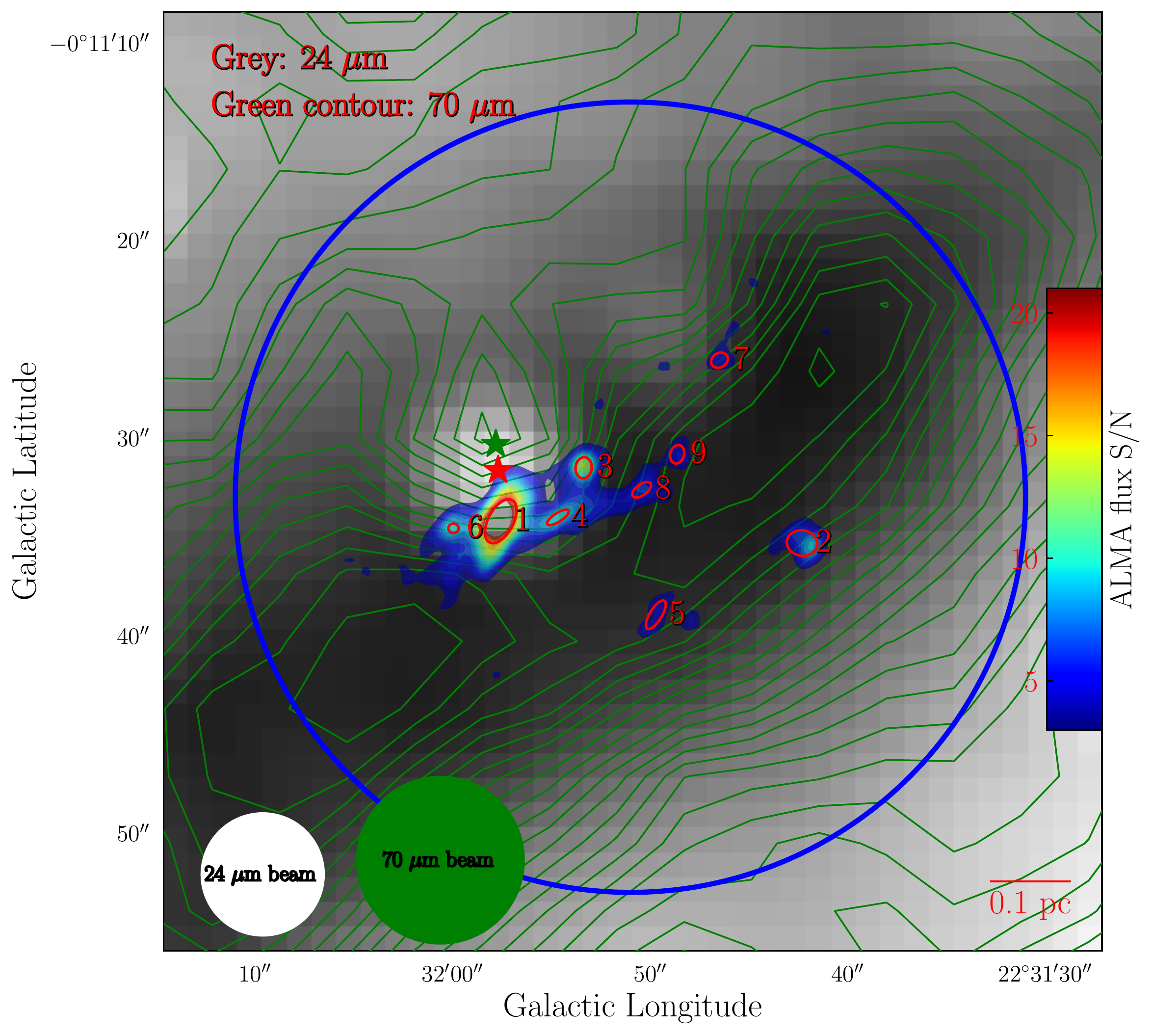}
       \caption{AS3 24~\micron\ and 70~\micron\ emission. Green contours, star, and circle indicate the 70~\micron\ emission, its peak and beam, respectively. Background shows 24~\micron\ emission with a peak at the red star. The beam of 24~\micron\ image is shown in white circle. Color-filled contours show 7~m + 12~m combined 1.3~mm emission with levels similar to those in Fig.~\ref{ALMAemission2}.}
        \label{AS3-viewer}
    \end{figure}

AS3 presents a filamentary structure in ALMA 1.3~mm; \htco\ is detected towards ALMA1, 3, 4, 8, and 9, indicating that the shocked gas or warm gas is present along the filament. Meanwhile, \siofive\ is detected towards ALMA1. AS3 has been identified as a 70~\micron\ quiet clump with single-dish (sub)mm images by \cite{yuan18, traficante18, urquhart18}, and the ALMA project of Pillai but a comparison between ALMA image, MIPSGAL 24~\micron\ (Beam$\sim6$\arcsec, \citealt{carey09}), and \textit{Herschel} 70~\micron\ (Beam$\sim8.5$\arcsec, \citealt{molinari10}) reveals that ALMA1 ($M_{\rm core}/$\mjthclump$\sim6$) is probably associated with an IR source at 24~\micron\ (20.984~mJy, FWHM$\simeq6.24$\arcsec, \citealt{gutermuth15}) and 70~\micron\ (278~mJy, FWHM$\simeq$10.1\arcsec, \citealt{molinari16}) with an offset of 2.5\arcsec\ and 3.8\arcsec, respectively, as shown in Fig.~\ref{AS3-viewer}.  The possible reason for ignoring the far IR source for these authors could be the significant offset between the IR source and the single-dish (sub)mm emission peak (see Fig.~\ref{DarkProperties2} in Appendix).

To sum up, the fragmentation in most of the candidate HMSCs in our sample is thermal-dominated except for AS1 which shows that only thermal motions or turbulence cannot fully explain its fragmentation. The most massive cores ($M_{\rm core}>$ 8~\msun) in our candidate HMSCs commonly present star-forming signposts, implying that there is no high-mass prestellar core in our sample. Besides, we see that a large part ($\gtrsim60$\%) of cores with $M_{\rm core}>$ \mjthclump\, may already show star-forming signposts. One possible reason could be the evolutionary bias that the protostellar cores accrete more mass with evolution; another reason is that the \dustt\ of these star-forming cores is likely to be inaccurate that is, in this case, underestimated and leads to an overestimation of the core mass.\\

\section{Mass distribution shaping revealed by ALMA} \label{ALMAimpacted-result}
In the work discussed up through the last section, we did not find a clear difference of fragmentation between AS and NA at $\sim0.025$~pc scale in our limited sample. However, this does not mean that there is no implication of the impacts of \hii~regions on dense ensembles or cores. In this section, we use multi wavelength data to search for signs of impacts from \hii~regions and compare these with ALMA 1.3~mm emission.\\

\subsection{Morphology of cores and clumps} \label{ALMAimpacted-result-morphology}
Compared to millimeter interferometric observation, which is only sensitive to compact emission, single-dish observation could detect more diffuse and larger-scale emission from clump envelope. ALMA observations recover 10--20\% of the corresponding single-dish flux (see Sect.~\ref{ALMAContinuumReduction}), showing that most masses of 70~\micron\ quiet high-mass clumps exist in the envelope rather than in the compact structures. Due to its lower density, the clump envelope is expected to be more easily modified by the compression of \hii~regions. \citet{zhang19} studied a sample of eight impacted high-mass protocluster clumps with a mass and resolution similar to the ones in our study. These authors proposed that the affected clumps present a morphology for which the embedded compact structures observed by interferometer are expected to be closer to the interface of interaction because the clump envelope is compressed more distinctly towards the interface.

The ATLASGAL flux peak, flux weighted clump center, and the ATLAGSAL pointing error are shown in the right panels of Figs.~\ref{ALMAemission1} and \ref{ALMAemission2} with yellow crosses, triangles, and circles (radius $= 4\arcsec$, \citealt{schuller09}), respectively. The uncertainty in pixel flux makes the flux peak taken from the sole pixel not reliable as the flux weighted center. AS2 and AS4, which are deeply impacted by the \hii~regions, present an asymmetrical morphology as shown by the systematic offset between flux weighted clump center and mass weighted core center (yellow stars in the right panels of Figs.~\ref{ALMAemission1} and \ref{ALMAemission2}). The reason why deeply impacted AS1 does not present a clear such offset could be the density improvement due to the clump evolution (the densest clump in eight candidate HMSCs) overwhelming the density structure modified by the compression of \hii~region. AS3 is an exceptional case where the compact structures are faraway from the interaction interface. AS3 is only weakly affected as described in Sect.~\ref{SingleDish} therefore no significant compression of the clump envelope is seen. The 3D distribution of the envelope may bias the conclusion by projection effect, which could be found for the irregular center positions for some NA sources.  

\begin{figure*}[htb]
     \centering
   \includegraphics[width=0.85\textwidth]{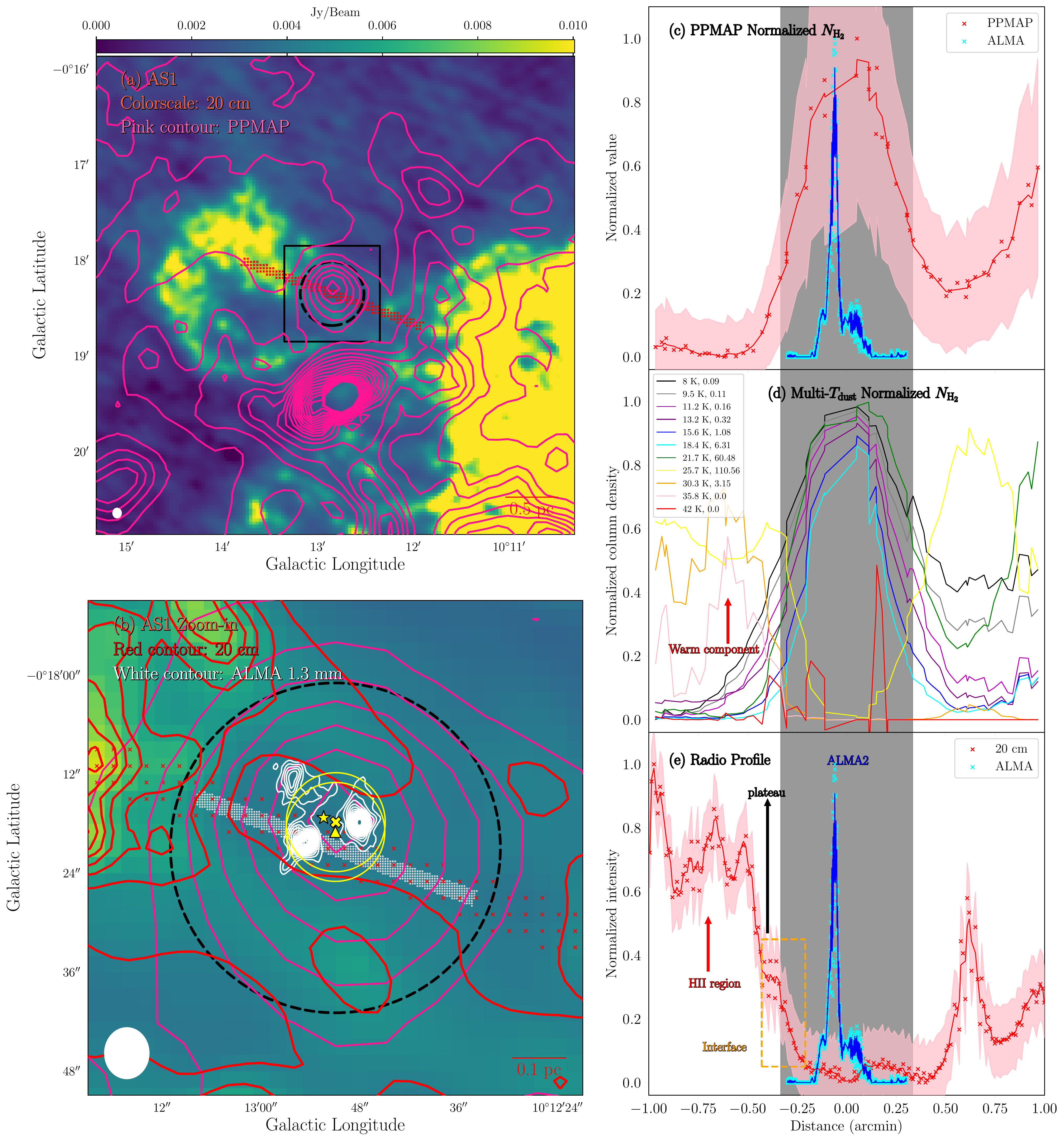}
       \caption{Impacted candidate HMSC AS1. \textit{Panel (a)} shows the large-scale 20~cm continuum overlaid with the Hi-GAL PPMAP \nhtcd\ contours starting from $10^{22}$~cm$^{-2}$ with a step of $10^{22}$~cm$^{-2}$. The red bar indicates the pixels used to create intensity profile. \textit{Panel (b)} is a closer view of the black box in \textit{Panel (a)}, overlaid with 7~m +12 m combined 1.3~mm emission contours (white) starting from 3~rms with a step of 1~rms. The 20~cm continuum is indicated by red contours starting from 3~rms with a step of 1~rms. The black dashed circle indicates the field of view (20\% power point) of the 1.3~mm image. The white bar indicates the pixels used for ALMA intensity profile. The yellow star, cross, and triangle are similar to those in Figs.~\ref{ALMAemission1} and \ref{ALMAemission2}, but for the PPMAP \nhtcd. Yellow circles and white ellipse indicate the PPMAP and cm continuum beams, respectively. \textit{Panel (c)} shows the normalized PPMAP \nhtcd\ and the 7~m + 12~m combined 1.3~mm intensity profiles cut along the direction indicated in \textit{Panel (a)} and \textit{(b)}. The gray region highlights the ALMA imaging field. \textit{Panel (d)} shows the normalized profiles of \nhtcd\ with different \dustt\ derived by PPMAP. The numbers after \dustt~values in the legend are the S/N of the corresponding profiles. The S/N is set to be zero when $< 0.01$. \textit{Panel (e)} shows the 20~cm emission profile. The \hii~region, interface, and plateau are indicated by different labels.}
        \label{triggered-viewer1}
    \end{figure*}

\begin{sidewaysfigure*}[p]
     \centering
   \includegraphics[width=0.9\textwidth]{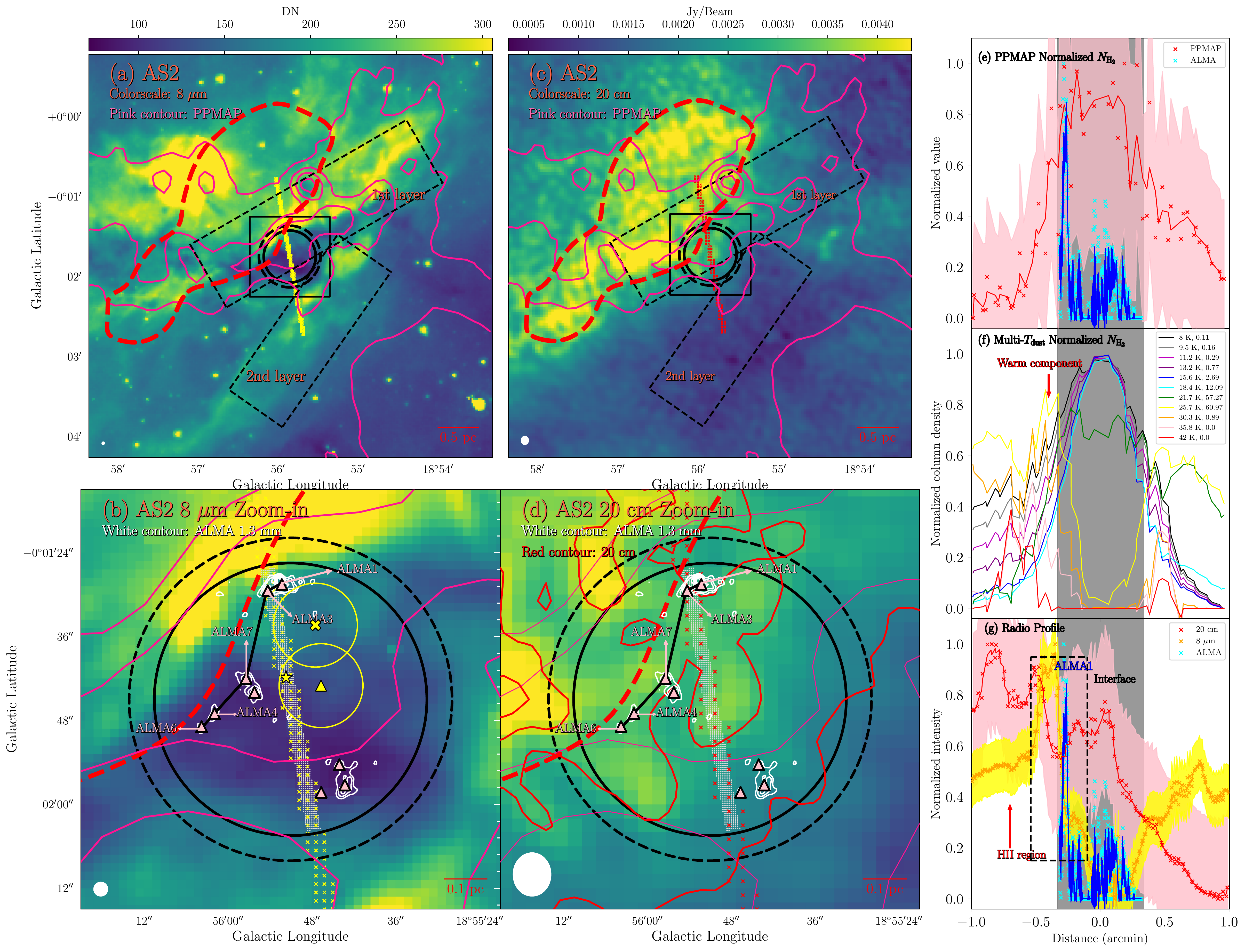}
       \caption{Impacted candidate HMSC AS2. \textit{Panel (a)} shows 8~\micron\ emission. The pink contours have the same meanings as in \textit{Panel (a)} of Fig.~\ref{triggered-viewer1}. Black dashed rectangles indicate two PDRs. The red dashed contour shows the smoothed 20~cm continuum with a level of 3.3~rms. The yellow bar marks the pixels used to create the intensity profile. \textit{Panel (b)} shows a close view of \textit{Panel (a)}, overlaid with 7~m + 12~m combined 1.3~mm continuum contours (white) starting from 3~rms with a step of 1~rms. Black solid and dashed circles represent the 7~m + 12~m combined fields of view cut at 20\% and 10\% power points of primary beam, respectively. The white bar shows the pixels used for ALMA 1.3~mm profile. The light pink triangles mark the ALMA cores. The lines connecting ALMA1, 3, 7, 4, and 6 indicate a wall-like morphology. The yellow star, cross (large), triangle, and circles are similar to Fig.~\ref{triggered-viewer1}. \textit{Panel (c)} shows 20~cm continuum. The red bar indicates the pixels used to show intensity profile. \textit{Panel (d)} shows a closer view of \textit{Panel (c)}. The red solid contours indicate the 20~cm continuum with levels starting from 2~rms with a step of 1~rms. \textit{Panels (e)} to \textit{(g)} are similar in content to Fig.~\ref{triggered-viewer1}.}
        \label{triggered-viewer2}
\end{sidewaysfigure*}

\begin{figure*}
     \centering
   \includegraphics[width=0.75\textwidth]{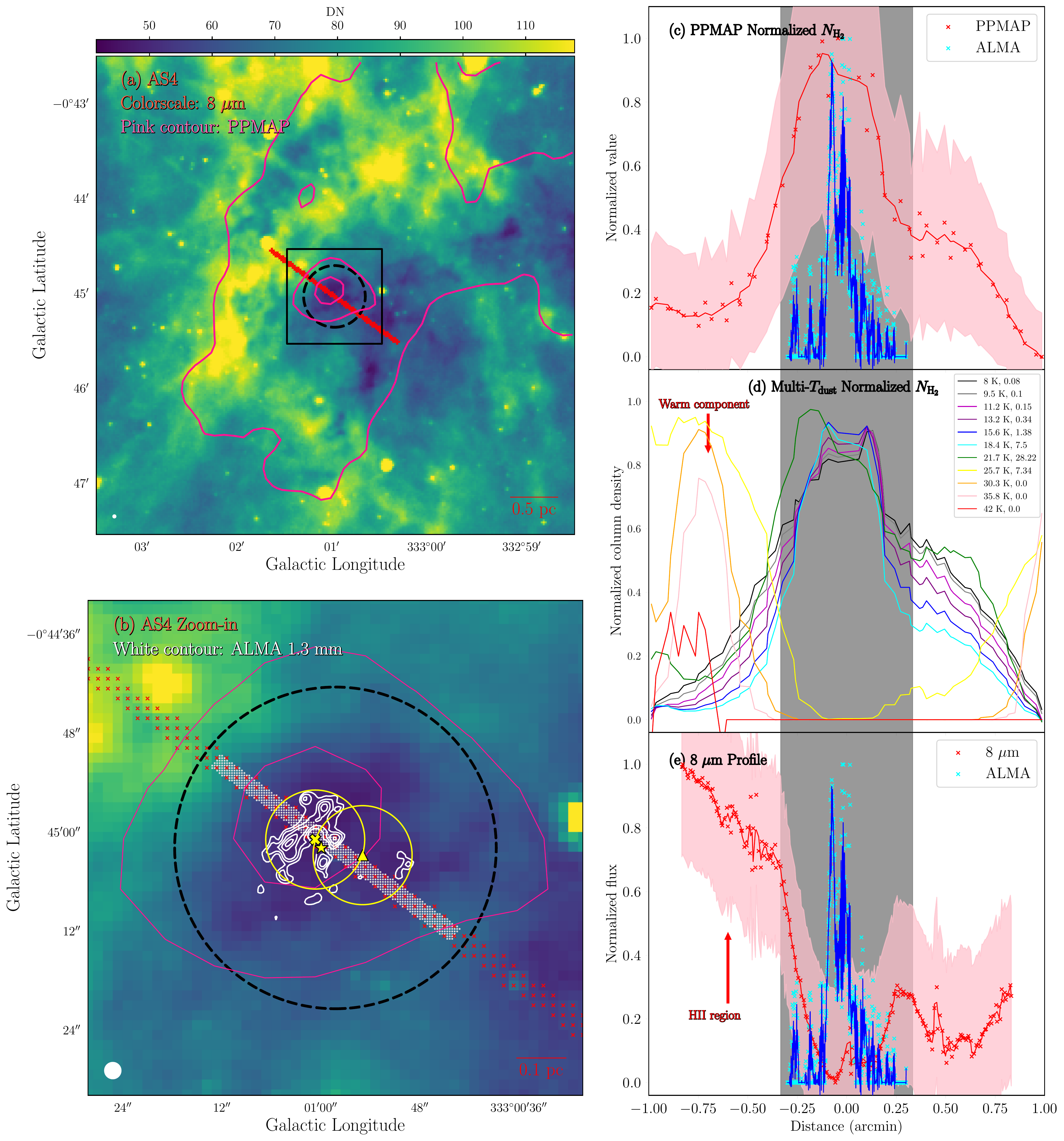}
       \caption{Impacted candidate HMSC AS4. \textit{Panel (a)} shows the 8~\micron\ emission overlaid with PPMAP \nhtcd\ contours with levels similar to Fig.~\ref{triggered-viewer1}. The red bar shows the pixels used to create the profile. \textit{Panel (b)} is a closer view of \textit{Panel (a)}, overlaid with 7~m + 12~m combined 1.3~mm contours with levels starting from 3~rms with a step of 1~rms. \textit{Panels (c)} to \textit{(e)} are similar to those in Fig.~\ref{triggered-viewer1}.}
        \label{triggered-viewer3}
    \end{figure*}

Single-dish \hmole\ column density \nhtcd\ maps are shown with pink contours in Figs.~\ref{triggered-viewer1}, \ref{triggered-viewer2}, and \ref{triggered-viewer3} for AS1, AS2, and AS4, respectively. The \nhtcd~is derived from SED fitting of Hi-GAL 70, 160, 250, 350, and 500 \micron\ emission (Herschel Infrared Galactic Plane Survey, \citealt{molinari10}) with the techniques of point process mapping (PPMAP) by \citet{marsh16}, which utilize the full instrumental point source functions to reach a spatial resolution of 12\arcsec\ \citep{marsh15, marsh17}. The Hi-GAL PPMAP column density contours suggest that AS1 and AS2 probably have an asymmetrical mass distribution while this asymmetry is not clearly seen in the column density profiles cut along the direction of the interaction (see panels named as ``PPMAP Normalized \nhtcd'' in Figs.~\ref{triggered-viewer1}, \ref{triggered-viewer2}, and \ref{triggered-viewer3}). The PPMAP column density peak and column density weighted clump center are marked in Figs.~\ref{triggered-viewer1}, \ref{triggered-viewer2}, and \ref{triggered-viewer3} and they show a similar trend that AS2 and AS4 core centers are closer to the interaction interference compared to the weighted clump center.

The PPMAP column densities are equally spaced in log space between 8~K and 50~K (see panels named as ``Multi-\dustt\ Normalized \nhtcd'' in Figs.~\ref{triggered-viewer1}, \ref{triggered-viewer2}, and \ref{triggered-viewer3}). Cold dust (\dustt~$<21$~K) normally shows a density profile similar to that of the cumulative density of all dust temperature. Warm dust (\dustt~$\gtrsim 21$~K) profile presents a characteristic dip at the clump center and a peak close to \hii~regions. Warm dust at 20--30~K is located in the PDR surrounding the \hii~regions \citep{marsh19}, thus, the asymmetrical warm dust distributions suggest stronger photodissociation, which is as expected from the incoming radiation from ionizing stars. \\

\subsection{Shaping structures in candidate HMSCs at $\sim0.025$~pc scale} 
\label{shaping-result}
Among the four AS clumps of our sample, AS1 and AS2 are probably most deeply impacted by \hii~regions, according to the clump and cm free-free continuum morphology. The lack of high-resolution cm continuum data for AS4 environment does not allow for a detailed analysis of its associated ionized gas. %Observations toward objects where induced star formation is possibly at work due to the overpressure of \hii~regions such as bright rimmed clumps (Radiation Driven Implosion mechanism, RDI) and bubbles (Collect and Collapse mechanism, C\&C), indicate an ionized gas pressure of $10^{-10}$ to $10^{-8}$~dyn~cm$^{-2}$ \citep{morgan04, liu16, marshall19, rodrguez19}.  The pressure exerted by \hii~regions on AS1 and AS2 is similar to the values observed in these objects.

\begin{figure}[htb]
     \centering
   \includegraphics[width=0.48\textwidth]{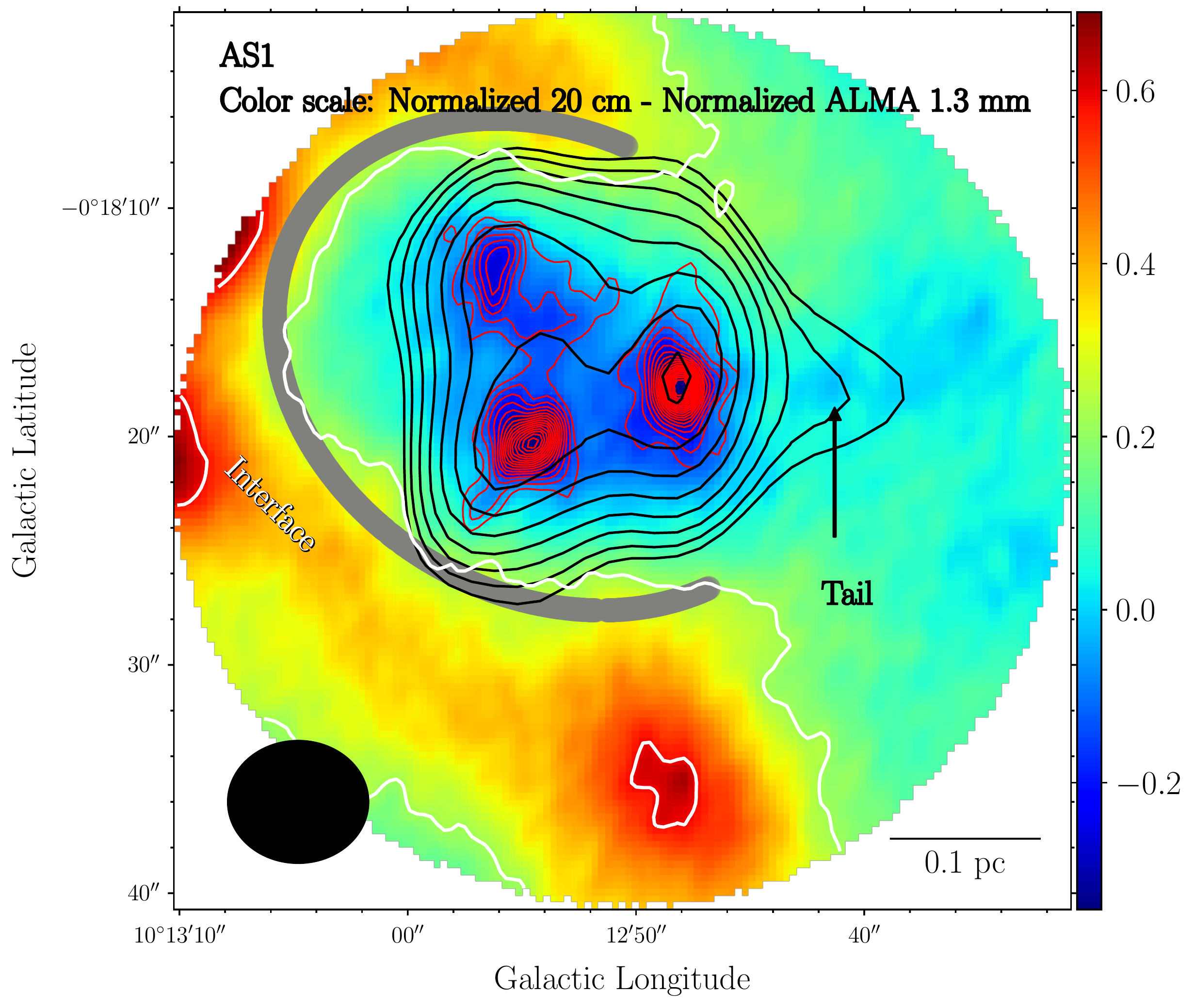}

       \caption{Normalized intensity of AS1. This map shows the value of normalized 20~cm intensity minus normalized 7~m + 12~m combined 1.3~mm continuum. White contours start from 3~rms with a step of 1~rms. The gray curve indicates the IF or the interface of interaction. Red and black contours represent the 7~m + 12~m combined and 7~m alone 1.3~mm continuum with the levels similar to Figs.~\ref{ALMAemission1} and \ref{Subfragmentation}, respectively.}
        \label{AS1-viewer}
\end{figure}

\begin{figure}
     \centering
   \includegraphics[width=0.45\textwidth]{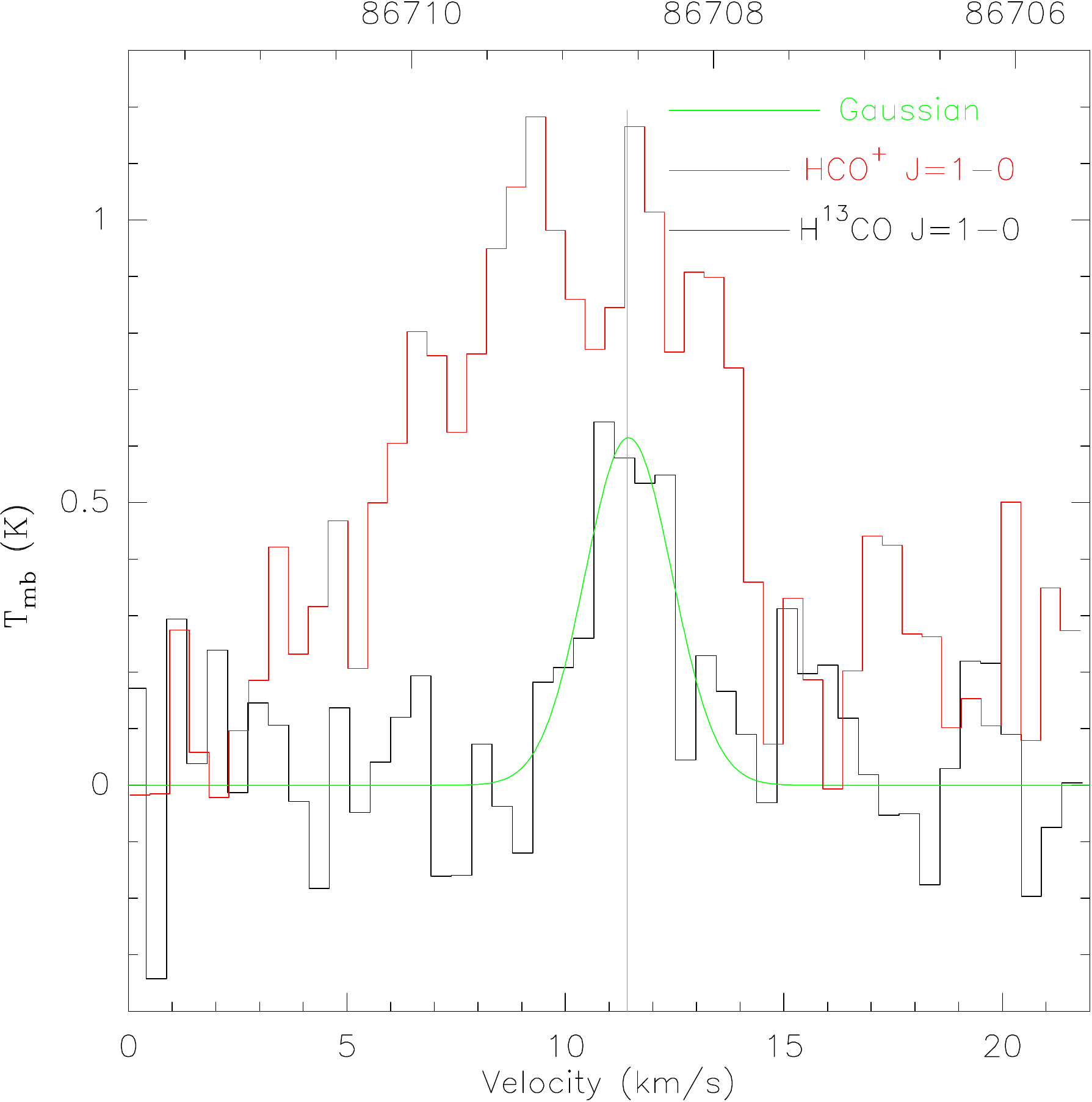}
       \caption{Single-dish spectra of AS1. Red and black spectra show the single-dish optically thick MALT90 \hcopone\ and optically thin MALT90 \htcopone\ transitions, respectively. The spatial and velocity resolutions are 38\arcsec\ and 0.11~\kms, respectively. To improve the depiction, the spectra have been regridded to 0.45~\kms. Green lines are fitted Gaussian and its center.}
        \label{AS1infall}
    \end{figure}

\textbf{Evidence for gas and dust shaping in AS1:} The 7~m + 12~m combined 1.3~mm emission of AS1 likely emerges on a dip of 20~cm continuum emission as shown in Fig.~\ref{triggered-viewer1}b. The 20~cm continuum of \hii~region is optically thin and dominated by free-free emission of the ionized gas. Therefore, this dip is not due to the extinction. To highlight the ionized gas and cold dust distribution, we created a map of normalized 20~cm continuum intensity minus normalized combined 1.3~mm intensity, as presented in the color scale in Fig.~\ref{AS1-viewer}. It clearly shows an overpressured ionized front (IF) that surrounds more than half of the compact ALMA 1.3~mm emission, meanwhile, a morphology of cometary globule with short tail following the interacting direction is clearly presented by 7~m data in Fig.~\ref{AS1-viewer} \citep{lefloch94, tremblin2012}. The 20~cm continuum profile in Fig.~\ref{triggered-viewer1}e reveals that an intensity plateau (with a width of $\sim$~three beams) of 20~cm continuum profile appears on the interface, implying that the expanding ionized gas is jammed on the clump envelope. A small fraction of ionized gas continues to leak into the envelope but stops at more compact inner regions traced by combined 1.3~mm emission. A blue shift peak and strong blue wing of MALT90 optically thick \hcopone\ (beam $\sim38\arcsec$) are detected as shown in Fig.~\ref{AS1infall}. The blue component may trace the gas shocked by \hii~regions or potential infalling envelope of AS1 \citep{mardones97, evans99, zhang16, traficante17}.

We test whether AS1 is formed through the collect and collapse (C\&C) mechanism, which describes a scenario in which the supersonic expansion drives a shock front (SF) in front of the IF and then a shell is collected between SF and IF. When the shell is too dense to be gravitational stable, the shell fragments into several clumps \citep{elmegreen1977, zavagno2007}. The gravitational fragmentation of the collected shell is expected to happen at time $t_{\rm frag} = 1.56 {C_{0.2}^{7/11}} {N_{49}^{-1/11}} {n_3^{-5/11}}$~Myr, where $C_{0.2}$ is the sound speed inside the shell in unit of 0.2~\kms, which is set as AS1 sound speed 0.24~\kms. $N_{49}$ and $n_{3}$ are the ionizing photon flux in unit of $10^{49}$~s$^{-1}$ and the initial H number density in unit of cm$^{-3}$, respectively \citep{whitworth1994, liu16}. Taking the same $N_{49}$ and $n_{3}$ as those used in the estimations of bubble age and pressure, $t_{\rm frag} \simeq 0.9$~Myr. The $t_{\rm frag}$ is larger than bubble age $\simeq 0.4$~Myr, revealing that AS1 is a pre-existing clump rather than a C\&C triggered clump. Taken as a whole, AS1, presenting a cometary globule  morphology, is likely to be a pre-existing massive clump under the feedback of bubble N1. Additional pressure from ionized gas possibly accelerates the formation of protostellar cores in AS1.

\textbf{Evidence for gas and dust shaping in AS2:} AS2 and its associated several-pc scale filament are proposed to be pre-existing structures by \citet{tackenberg13}. Whether triggered star formation is happening in AS2 is not clearly revealed by the single-dish research of \citet{tackenberg13}. Our ALMA mapping shows that the most massive core, ALMA1, is exactly located in front of the PDR traced by PAH 8~\micron~emission as shown in Fig.~\ref{triggered-viewer2}b and g. We also note that several ALMA cores are situated close to the PDR edge such as ALMA3, 4, 6, and 7. These cores and ALMA1 reside in a wall-like morphology (indicated by the connected lines in Fig.~\ref{triggered-viewer2}b and d) whose shape resembles the IF. The shape of the IF is indicated by the smoothed contour of 20~cm continuum with a level of 3.3~rms in Fig.~\ref{triggered-viewer2}. Considering the consistency between the IF shape and the wall-like structure, the rapid decline of PAH 8~\micron\ emission towards ALMA1 is not only due to the IR extinction but also the dense photodissociation region situated at the interface between the \hii~region and AS2. 

Noisy 20~cm continuum ($\rm S/N \lesssim4$) prevents us from detecting tiny structures of the ionized gas surrounding the clump as observed in Fig.~\ref{AS1-viewer}. The profile of 20~cm continuum intensity cut along the interaction direction does not present a clear plateau or rapid decline at the interface between \hii~region and clump. The smoothly decreasing intensity profile, without significant fluctuation, is explained with the leaking ionization radiation by \citet{luisi19}. Leaking ionization radiation is the possible formation mechanism for the two PDR layers indicated by 8~\micron\ emission in Fig.~\ref{triggered-viewer2}a. The ionization radiation leaked from the closer layer and then acted on further layer. These two layers have an average velocity difference of 1 to 2~\kms, suggesting the possible kinematic difference driven by the feedback of \hii~region.

 The strongest 8~\micron\ emission surrounding ALMA1 may reveal the strongest impact of the \hii~region on ALMA1, which is also suggested by the sole significant detection of warm or shocked gas tracer \htco~in ALMA1. We check the 7~m + 12~m combined 1.3~mm continuum in a slightly extended field of view by lowering the cut limit of the primary beam from 20\% to 10\% power point, as shown by the black circles in Fig.~\ref{triggered-viewer2}b and d. The rms mass sensitivity in 20\% to 10\% power point of the primary beam is around 0.2~\msun~Beam$^{-1}$ to 0.4~\msun~Beam$^{-1}$. The minimum core mass that could be identified by Astrodendro is $\sim0.6$~\msun. No new structure is found in the extended field of view, which means that the potential undetected structures in front of the most massive core, ALMA1, should be smaller than 0.6~\msun. Therefore, we propose that ALMA1 and the other cores in the wall-like structure are probably the closest to the IF. In combining all the evidence, some of the cores in the wall-like structure are likely to have a triggered origin, especially for the most massive core, ALMA1. \\

\section{Discussion} \label{Discussion}
\subsection{Fragmentation mechanism}
In our limited sample, a large population ($>70\%$) of cores are low-mass objects with a mass $<2$~\msun, meanwhile, the most massive cores ($>8$~\msun) in candidate HMSCs commonly present star formation activities. The absence of high-mass prestellar cores in candidate HMSCs is fully revealed by a series of studies with similar sampled scale of 0.02~pc to 0.03~pc, for example, \citet{sanhueza17,contoreras18, sanhueza19, svoboda19, louvet19}. It is not only core mass, but also core separations that present a hierarchical thermal fragmentation, favouring the CA model for high-mass star formation. The pc-scale candidate HMSCs fragment into several ensembles with separations of clump Jeans length \lambdajthclump\ and then some of the ensembles continue to fragment into cores with separations of ensemble Jeans length, \lambdajthens. 

 Even if the core spatial distributions present a diverse morphology, from cluster-like (AS1 and AS4) to filamentary (AS3 and NA3) or simply irregular (AS2 and NA4), a common pattern between the core and ensemble separations seems to exist within the diverse morphology. The typical ensemble separation $S_{\rm ens}$ is $\sim0.18$~pc within a range of about 0.1 to 0.3~pc meanwhile the typical core separation $S_{\rm core}$ in the ensemble is $\sim0.06$~pc within a range about 0.04 to 0.08~pc, as shown in Fig.~\ref{EnsembleSubfragmentation}. It casts light on the typical thermal Jeans length ratio of ensemble to clump \lambdajthens/\lambdajthclump\ is $\lesssim1/3$.

We collect the information from previous 7~m + 12~m combined band 6 observations toward 31 high-mass clumps at 3--5~kpc \citep{contoreras18, sanhueza19,svoboda19}, which are quiet from $\sim1$~\micron\ to 70~\micron\ and have a mass from 240~\msun\ to 5200~\msun\ with a median of 760~\msun. The resolution is around 0.01 to 0.03~pc. The sensitivity and coverage of these observations are all different. \citet{contoreras18} and \citet{sanhueza19} are mosaic observations covering full clump with a sensitivity of $\sim0.1$~mJy~Beam$^{-1}$ whereas \citet{svoboda19} and our data are single pointing observations with a sensitivity of $\sim0.05$~mJy~Beam$^{-1}$ and $\sim0.15$~mJy~Beam$^{-1}$, respectively. This data collection is statistically significant for obtaining a universal fragmentation mechanism for candidate HMSCs. Notably, nearly all collected candidate HMSCs are NA besides the AS sample in this paper and G341.039-00.114 from \citet{sanhueza19} because these authors usually set an upper limitation on \dustt, causing a bias of preference selection of NA. 

\begin{figure}[htb]
     \centering
   \includegraphics[width=0.49\textwidth]{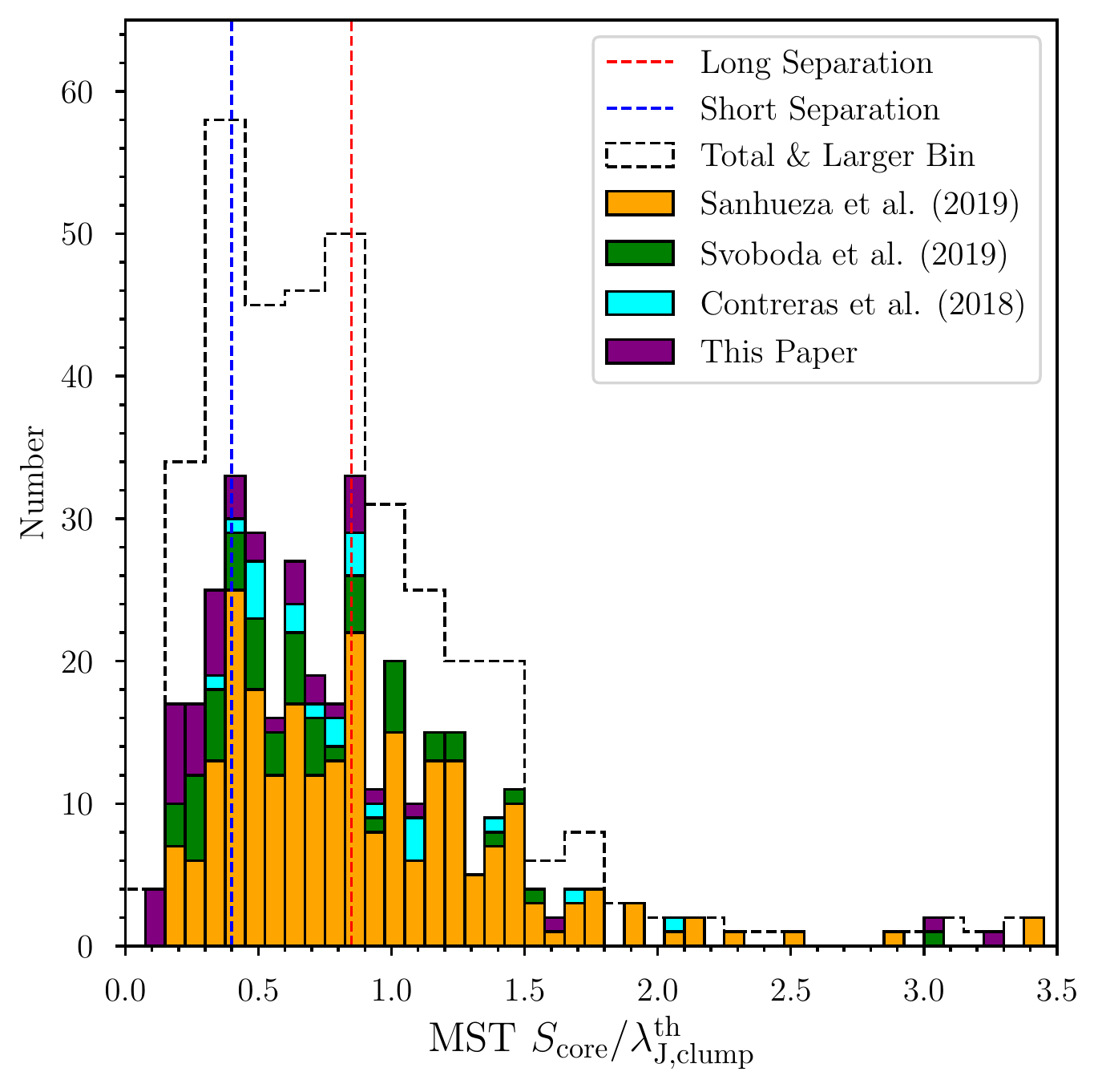}
       \caption{Statistics of core MST separations for 31 candidate HMSCs. The data have been taken from \citet{contoreras18, svoboda19,sanhueza19}, and from the sources in this paper. Clump properties for the candidate HMSCs in the single pointing mapping of \citet{svoboda19} are taken from Paper\,I when possible, to improve the consistency of the data. Therefore, the number distributions for the sources of \citet{svoboda19} here are slightly different from the raw ones. There are a total of 363 separations. The size of the bin is set to be 0.075 (color-filled histogram) and 0.15 (black-edge histogram) to avoid bias. Red and blue lines mark the two peaks of the distribution.}
        \label{DistributionSample}
    \end{figure}

Figure~\ref{DistributionSample} shows the number distribution for total 363 MST core separations scaled by clump thermal Jeans length \lambdajthclump\ of the collected sources. More than 70\% of MST separations are less than \lambdajthclump. The maximum and sub-maximum are peaked at around 0.38\lambdajthclump\ and 0.9\lambdajthclump, respectively. Firstly, the assumption of the unimodal distribution is excluded for Fig.~\ref{DistributionSample}, by applying Hartigan’s Dip test \citep{diptest1985}. Then we test multimodal distributions by ACR methods (see details about ACR in \citealt{acr2016}) using the ``modetest'' function in the R package ``multimode'' \footnote{See more details on the ``multimode'' package in \citet{multimode2018}.}. The test's results support the assumption that the bimodal distribution is the most likely distribution in the scaled separation range of 0 to 1 in Fig.~\ref{DistributionSample} and therefore prove the reliability of maximum and sub-maximum marked in Fig.~\ref{DistributionSample}. Similar bimodal fragmentation nature presented in Fig.~\ref{DistributionSample} for the samples of \citet{svoboda19}, \citet{sanhueza19}, and this paper reveals that different coverage and sensitivity of ALMA observations do not largely impact on the result of the bimodal fragmentation nature. Although some low-mass cores could be lost in our single-pointing, slightly more noisy observation, the bimodal nature of fragmentation revealed here is reliable.

As a result, most of the NA clumps fragment in a hierarchical and thermal way, consistent with several studies of the high-mass clumps at a later evolutionary stage. \citet{klaassen18} analyzed nine hypercompact \hii~regions with ALMA at a resolution of $\lesssim 0.01$~pc and found that the embedded cores are in line with the consequences of thermal fragmentation. This result coincides with that revealed in the CORE program \citep{beuther18}, which surveyed 20 young high-mass star-forming regions containing high-mass protostellar objects (HMPOs) and revealed that turbulent fragmentation is less effective than thermal fragmentation at a resolution of $\lesssim 0.01$~pc. 

With the assumption that most of the candidate HMSCs are dominated by thermal fragmentation, candidate HMSCs with a similar density $\rho$ but higher \dustt\ are expected to fragment to cores with a higher mass and longer separation. In our limited sample, it is only AS1 that obviously deviates from thermal fragmentation. The core separations and mass are at least three times larger than the determined thermal values. Compared to other candidate HMSCs, AS1 is probably the most evolved considering that its clump surface density $\Sigma\sim 1$~g~cm$^{-2}$ is one magnitude larger than others ($\sim 0.1$~g~cm$^{-2}$). A higher core mass is probably due to evolutionary effects. An observational comparison between high-mass prestellar clumps and high-mass protostellar clumps with ALMA at a resolution of 0.02~pc shows that the fraction of total mass in cores to clump mass increases from $\lesssim0.1$ to $\gtrsim0.2$ \citep{neupane20}, agreeing with the simulations of \citet{molinari19}. Notably, core separations in AS1 are similar to that in NA1 or other candidate HMSCs, at a scale of $\lesssim0.1$~pc, therefore the plausible conclusion of multi-factors dominated fragmentation in AS1 derived from equations in Sect.~\ref{FragCal} is probably biased. AS1 contracts more significantly due to the pressure of \hii~regions or the evolution, increasing clump $\Sigma$ to reduce the derived \lambdajthclump, making it quite different from \lambdajthclump\ at an earlier stage. Even if \dustt\ also increases with feedback of \hii~region or evolution, the change fraction is only $\sim30$\% to 50\%, much smaller than the increase in $\Sigma$. It is possible that AS1 had a thermal fragmentation when observed at an earlier stage.

\subsection{Transition of fragmentation mechanism}
Fragmentation is a process that is determined by a multitude of factors. The relative importance between gravity, thermal motion, magnetic field, and turbulence can lead to a different fragmentation \citep{tang19}. The dynamics of 70~\micron\ quiet massive clumps are believed to change from turbulent to gravity-dominant when they approach a critical $\Sigma$ of $\sim0.1$~g~cm$^{-2}$ \citep{traficante20}. One problem is that the turbulence level of AS is obviously increased by the feedback of \hii~regions, as shown in Paper\,I. Therefore $\Sigma$ needs to be higher for AS in order to keep thermal fragmentation effective.  AS2 and AS4 still present a thermal fragmentation when the turbulence level increases by less than 60\% (traced by line width, see Table~\ref{SingleDishPar}), compared to their counterparts NA2 and NA4, which implies that turbulent fragmentation at $\sim0.025$~pc scale requires huge energy injection from \hii~regions. 

\citet{rebolledo20} likely found the transition from thermal fragmentation to turbulent fragmentation under the effect of \hii~regions by comparing a pair of pc-scale massive star forming clumps in the Carina region, one is severely impacted by the \hii~region and the other is less disturbed. Mach number $\mathcal{M}$ derived from ALMA + ATLASGAL combined images with the techniques of densities probability distributions (N-PDF) are 1.37 (supersonic) and 0.81 (subsonic) for impacted one and less disturbed one, respectively. The 0.02~pc-scale fragmented cores in the more impacted clump are less numerous and five times more massive than the cores in the less disturbed clump. The differences are explained with the different fragmentation mechanisms (turbulent and thermal) by \citet{rebolledo20}. 

Notably, our previous single-dish spectral analyses in Paper\,I suggest that most of candidate HMSCs tend to be supersonic while the impacted ones prefer a hypersonic status ($\mathcal{M}\gtrsim5$). At this stage, it is not clear whether this turbulence increment could change a fragmentation from thermal-dominated to turbulent-dominated. The fragmentation at 0.01~pc scale in the most massive star-forming clump ($\mathcal{M}\sim4$) impacted by the \hii~region RCW~120 \citep{kirsanova19, zavagno2020} is investigated by \citet{figueira18} and they suggested that turbulent fragmentation is dominant due to the \hii~region feedback. Even for the cases where the fragmentation at around 0.02~pc scale in the clumps is still thermal-dominated under the vigorous feedback of \hii~regions, the fragments could be more massive or less numerous \citep{liu17}.

Magnetic fields make the fragmentation more complex. The presence of filamentary structure in candidate HMSCs such as AS3 may arise from a highly magnetised environment according to \citet{fontani16, fontani18}. Some studies have revealed the important role of the  magnetic field in the evolution of clumps impacted by \hii~regions, such as in bright rimmed clouds \citep{soam17, soam18}. \citet{eswaraiah19} studied the $B$ fields of massive clumps located at the waist of the bipolar \hii~region S201. These authors found that the $B$ fields of clumps are compressed and enhanced by the impact of the  \hii~region, leading to a status characterized by the magnetic pressure dominating over the turbulent and thermal pressure. \citet{busquet16} explored the fragmentation of twin filament hubs at a scale of 0.03~pc. These two hubs are similar in mass, density, turbulence, and even temperature (2~K difference). The hub more impacted by the \hii~region shows a lower fragmentation level compared to the less impacted one. \citet{busquet16} explained it as a stronger magnetic field suppressing the fragmentation in the more impacted hub. \\

To summarize, the fragmentation of candidate HMSCs at $\sim0.025$~pc scale is dominated by thermal fragmentation. Turbulent or magnetic fragmentation (or both) may happen only in some AS HMSCs that are highly impacted by \hii~regions. A specific follow-up ALMA survey towards AS HMSCs with broad single-dish line widths, which imply an extremely turbulent status, is needed to answer whether \hii~regions drive a turbulent fragmentation at $\sim0.025$~pc scale. Besides, ALMA observations of polarized dust emission are also needed to ascertain the role of magnetic field in the fragmentation of AS HMSCs \citep{beuthermagnetic18, dallolio19, cortes2019}.

\subsection{Considering induced star formation in HMSCs at $\sim0.025$~pc scale} \label{shaping-dis}
AS1 and AS2 clearly show how the \hii~regions shape the morphology of dense material distribution in pre-existing candidate HMSCs. The cometary globule morphology of ALMA dense emission of AS1 that is located just at the edge of the \hii~region probably reveals the classical triggered star-formation mechanism, with radiation driven implosion (RDI) at work here. The RDI presents a picture where the UV photons ionize the surface of a pre-existing clump to form an overpressured ionized layer and then help the clump to collapse \citep{lefloch94, kessel03, miao09}. The young associated bubble N1 ($\sim 0.4 \pm 0.2$~Myr) likely came across AS1 during the expansion and then the ionized gas of the bubble interacted with the recently ionized clump surface to form the ionized layer observed in Fig.~\ref{AS1-viewer}. Additional pressure from the ionized layer probably promotes the collapse. IR-quiet AS1 is expected to have an age of 0.05--0.1~Myr according to the statistical summary of the starless MDCs by \citet{motte18}. We assume that the IF of the bubble just propagated beyond half of AS1 with an averaged bubble expanding velocity of $\lesssim2$~\kms, which is estimated from bubble size divided by age. The interaction time between the \hii~region and AS1 is around 0.04--0.13~Myr. This roughly agrees with the expected age of AS1. Following all the evidence found here, we suggest that the cores in AS1 probably have a triggered origin.

AS2 presents an impacted morphology different from AS1. \hii~region is shaping a pre-existing 2--3 pc scale filament. The alignment between IF and dense gas from several-pc-scale filament, one-pc clumps to 0.02~pc-scale cores, reveals the multi-scale shaping effect of the  \hii~region.  All cores in AS2 are low-mass cores ($\lesssim2$~\msun). \citet{tackenberg13} found that AS2 single-dish \htcop\ spectra have two velocity components with a shift of $\sim2$~\kms\ (see Fig.~11 in their paper) and suggested that these two components are emission from pre-existing clump gas and shocked gas, respectively. The intensity ratio between the two components indicates that the shock does not propagate beyond the single-dish density peak of AS2. The stopped shock is explained by \citet{tackenberg13} with the resistance of the highest density region. With ALMA images, the shock probably has propagated beyond the wall-like structure and stopped at the further ensemble of cores shown in Fig.~\ref{triggered-viewer2}. We propose that some of the cores in wall-like structure probably have an induced origin. Follow-up ALMA spectral mapping towards these cores would help us to determine the evolutionary stage of each core and, thus, to demonstrate their triggered origin. For example, if the cores in the wall-like structure are more evolved than other less impacted cores such as ALMA 2, 8, and 9, which may indicate that the impacted cores formed earlier under the impacts of \hii~regions.

In short, ALMA data of AS1 and AS2 reveal that \hii~regions are able to modify the dense structures at $\sim0.025$~pc scale in candidate HMSCs, which may help the formation of stars in this kind of early stage clumps.

\section{Conclusion} \label{Conclusion}
Our study is based on an ALMA continuum investigation of the ionization feedback on $\sim0.025$~pc-scale fragmentation in candidate high-mass starless clumps (HMSCs), in which we explored four pairs of candidate HMSCs, including four clumps affected by \hii~regions (referred to as AS HMSCs) and another four that were located in quiet environment (referred to as NA HMSCs). The AS and NA in each pair are required to be similar in mass and distance to avoid any possible bias related to these quantities. Using the ALMA 1.3~mm continuum, with a resolution of $\sim0.025$~pc, we drew the following conclusions:
\begin{enumerate}
\item  We did not find a clear difference in the fragmented ALMA core mass and separations between AS HMSCs and NA HMSCs in our limited sample, although \hii~regions seem to affect the morphology of the cores' spatial distribution.
\item At $\sim0.025$~pc scale, thermal Jeans fragmentation with hierarchical nature dominates the fragmentation of most of the parsec-scale candidate HMSCs. The candidate HMSCs fragment into ensembles with separations of clump Jeans length, \lambdajthclump, and then some of the ensembles continue to thermally fragment into cores with separations of about one third of the \lambdajthclump.
\item AGAL010.214-00.306 (AS1), which is affected by the \hii~region, is the only candidate HMSC in our eight sources where other factors beside thermal motions could play an important role in the fragmentation process. The fragmented core mass (10-17~\msun) and separations ($\sim0.1$~pc) are at least three times larger than the thermal-determined values. 
\item \hii~regions are able to modify the dense material distributions embedded in candidate HMSCs, which probably helps the formation of next-generation stars. AS1 and AGAL018.931-00.029 (AS2) are the two most-likely cases in our sample which indicate signs of impact from the \hii~region. ALMA continuum reveals that AS1 presents a cometary globular morphology meanwhile the fragmented cores in AS2 show a spatial distribution aligned with the ionization front. 
\end{enumerate}

      \begin{figure}[htb]
     \centering
   \includegraphics[width=0.49\textwidth]{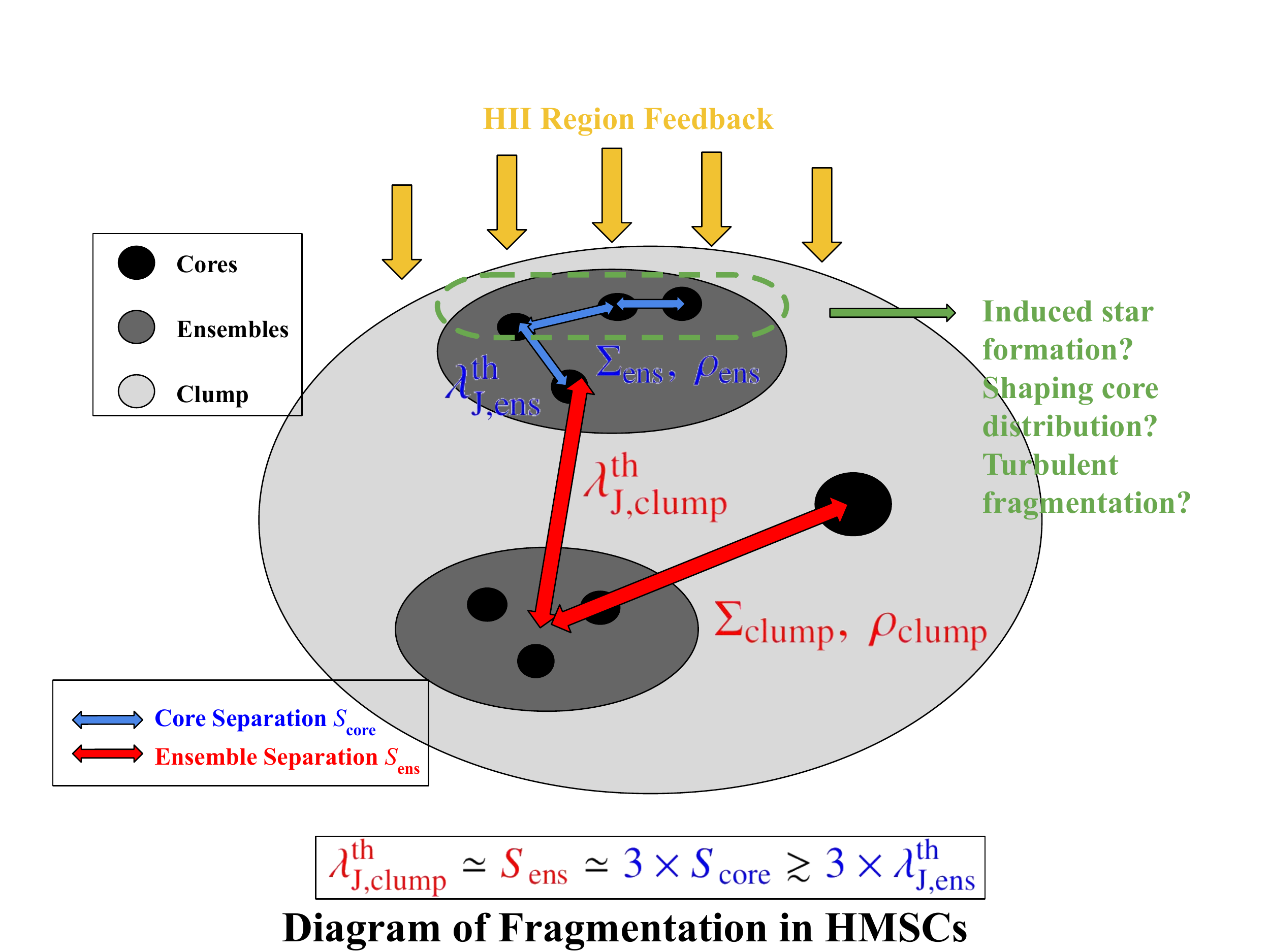}
       \caption{Proposed scenario for $\sim0.025$~pc-scale fragmentation in the impacted candidate HMSCs. $\Sigma_{\rm clump}$ and $\rho_{\rm clump}$ are the clump surface density and volume density, respectively, while $\Sigma_{\rm ens}$ and $\rho_{\rm ens}$ are the corresponding values for ensemble. Clump and ensemble thermal Jeans lengths are \lambdajthclump\ and \lambdajthens, respectively. Whether and how induced star formation and turbulent or magnetic fragmentation take place in extremely impacted HMSCs need more studies.}
        \label{Diagram}
    \end{figure}  
    
Figure~\ref{Diagram} tentatively draws a picture of hierarchical, thermal-dominated fragmentation for candidate HMSCs under the impacts of \hii~regions. Future ALMA polarized imaging toward a large sample of highly-turbulent candidate HMSCs (with the broadest lines in single-dish spectra) that are strongly impacted by \hii~regions will ascertain the role of thermal fragmentation in candidate HMSCs and shed light on whether turbulent or magnetic fragmentation could take over in extreme cases.

%\begin{equation}
%    \textcolor{red}{\lambda_{\rm J,clump}^{\rm th}} \simeq \textcolor{red}{S_{\rm ens}} \simeq \textcolor{blue}{3 \times S_{\rm core}} \gtrsim \textcolor{blue}{3 \times \lambda_{\rm J,ens}^{\rm th}}
%\end{equation}
%\textcolor{red}{$\lambda_{\rm J,clump}^{\rm th}$}\\
%\textcolor{blue}{$\lambda_{\rm J,ens}^{\rm th}$}\\
%\textcolor{red}{$\Sigma_{\rm clump},~\rho_{\rm clump}$}\\

%\textcolor{blue}{$\Sigma_{\rm ens},~\rho_{\rm ens}$}\\

%\textcolor{red}{$S_{\rm ens}$}\\

%\textcolor{blue}{$S_{\rm core}$}\\

\begin{acknowledgements}
      SZ thanks to the funding of ALMA Region Center Grenoble and IRAM.  AZ thanks the support of the Institut Universitaire de France. TGSP gratefully acknowledges support by the National Science Foundation under grant No. AST-2009842. We want to thank the anonymous referee for the constructive comments that helped to improve the quality of the paper. 
      
      \textit{Herschel} Hi-GAL data processing, map production and source catalogue generation is the result of a multi-year effort thanks to Contracts I/038/080/0 and I/029/12/0 from ASI (Agenzia Spaziale Italiana). This paper makes use of the following ALMA data: ADS/JAO.ALMA\#2016.1.01346.S ALMA is a partnership of ESO (representing its member states), NSF (USA) and NINS (Japan), together with NRC (Canada), MOST and ASIAA (Taiwan), and KASI (Republic of Korea), in cooperation with the Republic of Chile. The Joint ALMA Observatory is operated by ESO, AUI/NRAO and NAOJ.

\end{acknowledgements}

\bibliographystyle{aa} % style aa.bst
\bibliography{hmsc}

\begin{appendix} 
\section{Supplementary data}
\begin{figure*}
     \centering
   \includegraphics[width=0.75\textwidth]{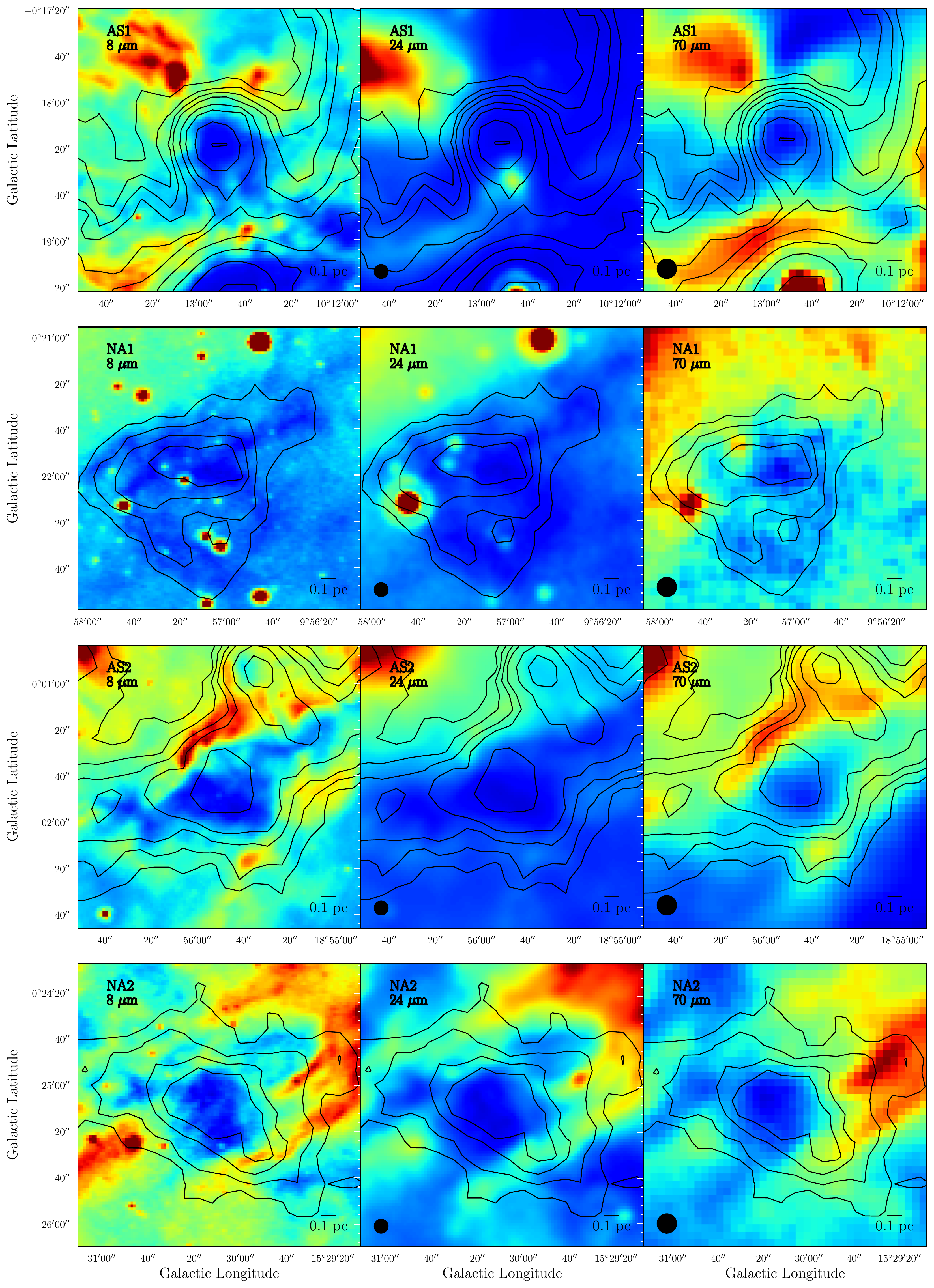}
       \caption{IR emission of candidate HMSCs' region. Each row shows 8~\micron, 24~\micron, and 70~\micron\ emission for each source. ATLASGAL 870~\micron\ emission is shown with black contours with levels the same as Fig.~\ref{RadioMap}. The black circles indicate the beam size of IR images, except for  8~\micron\ images, owing to their small beam size ($\simeq$~2\arcsec).}
        \label{DarkProperties1}
    \end{figure*}

\begin{figure*}
     \centering
   \includegraphics[width=0.75\textwidth]{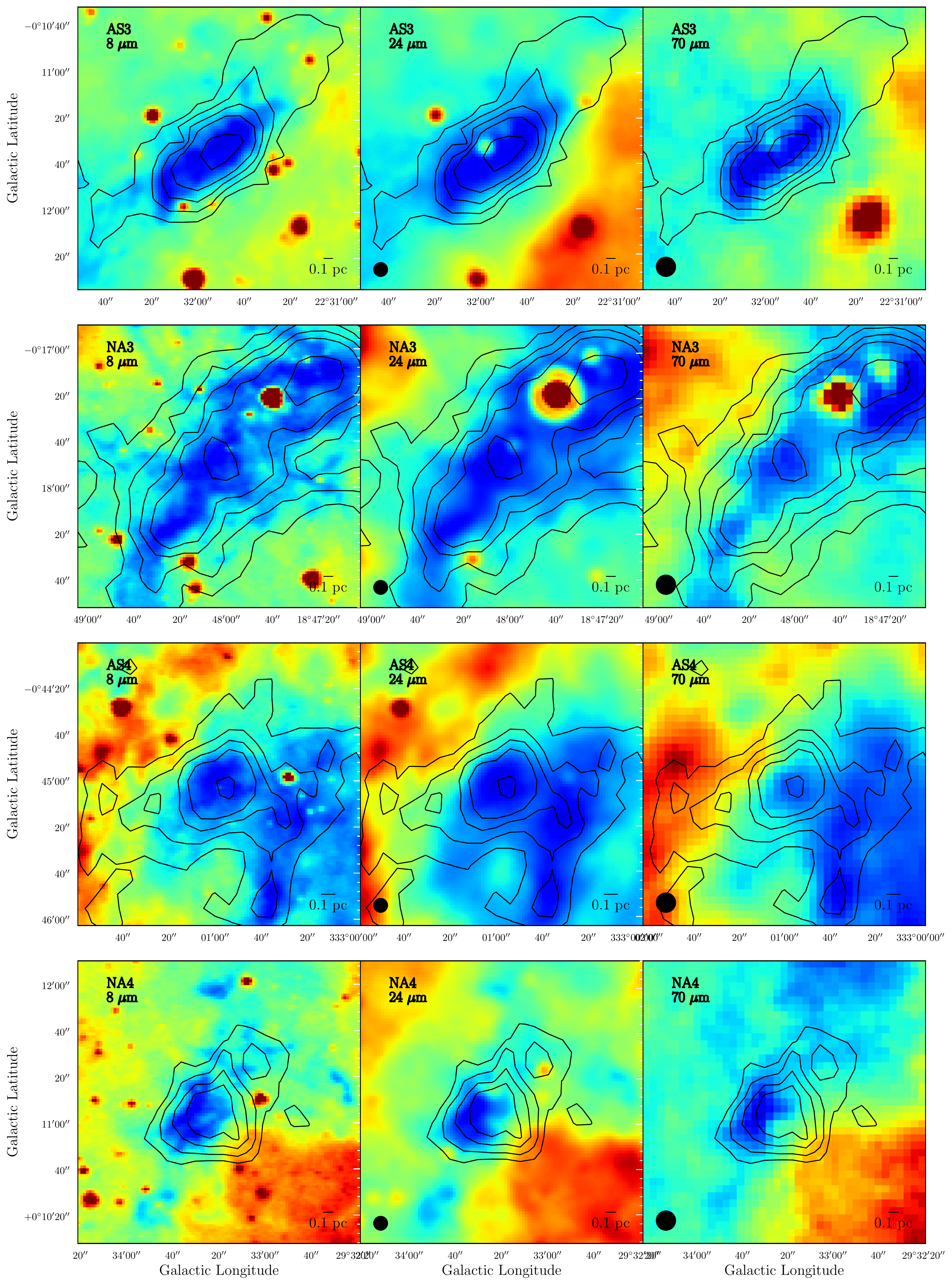}
       \caption{IR emission of candidate HMSCs' region. The meanings of each panel are similar to Fig.~\ref{DarkProperties1} but for different HMSCs.}
        \label{DarkProperties2}
    \end{figure*}

Figures~\ref{DarkProperties1} and \ref{DarkProperties2} present GLIMPSE 8~\micron, MIPSGAL 24~\micron~and Hi-GAL 70~\micron~emission of the candidate HMSCs' region. The cores identified with Astrodendro and the derived physical parameters are listed in Table~\ref{ALMACoresBasics} and \ref{ALMACoresPhysical}, respectively. 

\begin{figure*}
     \centering
   \includegraphics[width=0.75\textwidth]{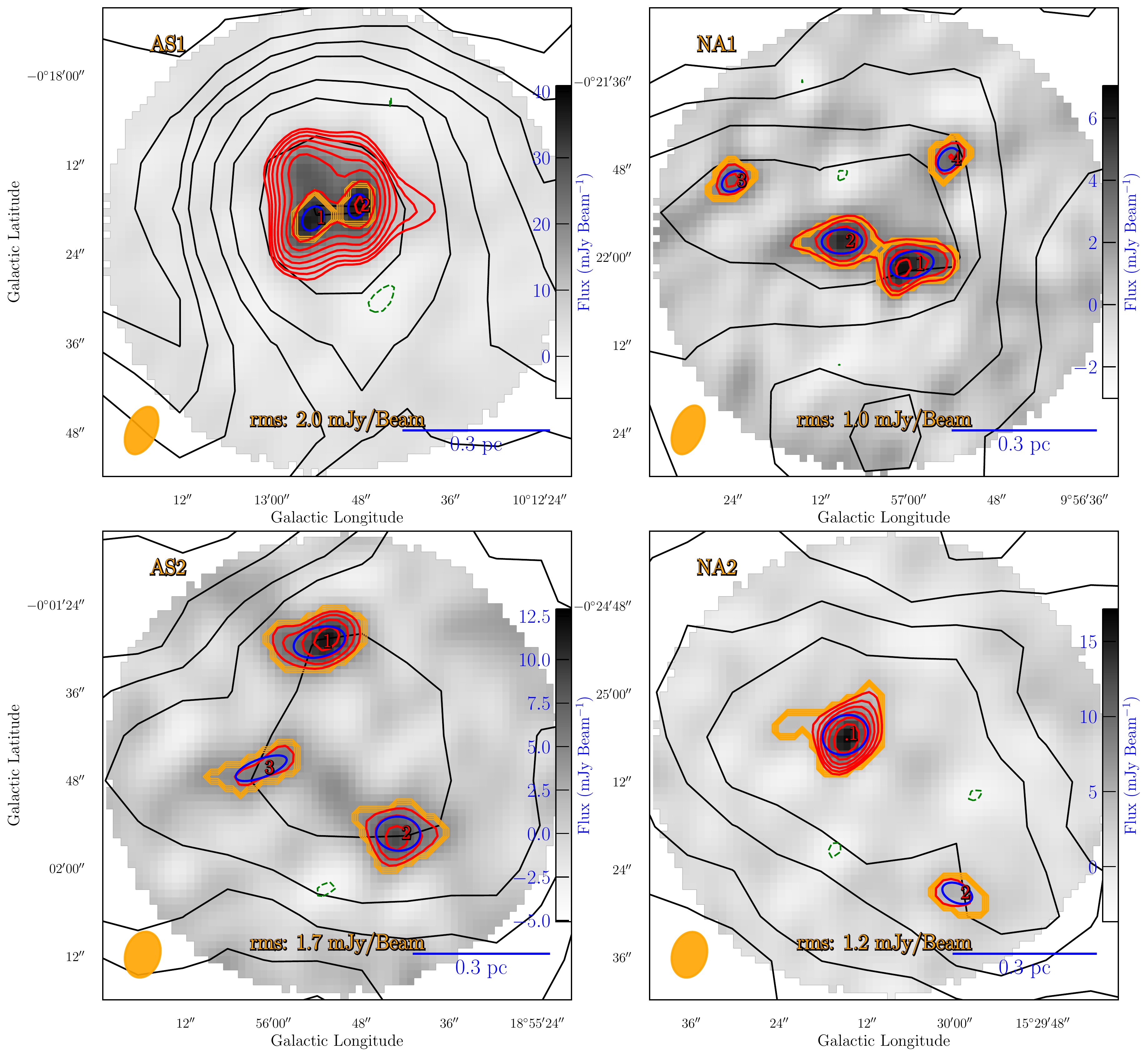}
       \caption{7~m ACA continuum images uncorrected for primary beam. ATLASGAL 870~\micron\ emission is shown with black contours with levels the same as Fig.~\ref{RadioMap}. Red (positive) and green (negative) contours show the 7~m ACA emission with levels of $\pm {\rm rms}\times[3, 3^{1.25}, 3^{1.50}, 3^{1.75}, 3^{2.0}, 3^{2.25}, 3^{2.50}, 3^{2.75}, 3^{3.0}]$. Blue ellipses and orange contours indicate the leaves and their pixels derived by Astrodendro, respectively. The red numbers mark the ranking of leaves' mass in Table~\ref{ACALeavesTable}.}
        \label{Emission7m1}
    \end{figure*}

\begin{figure*}
     \centering
   \includegraphics[width=0.75\textwidth]{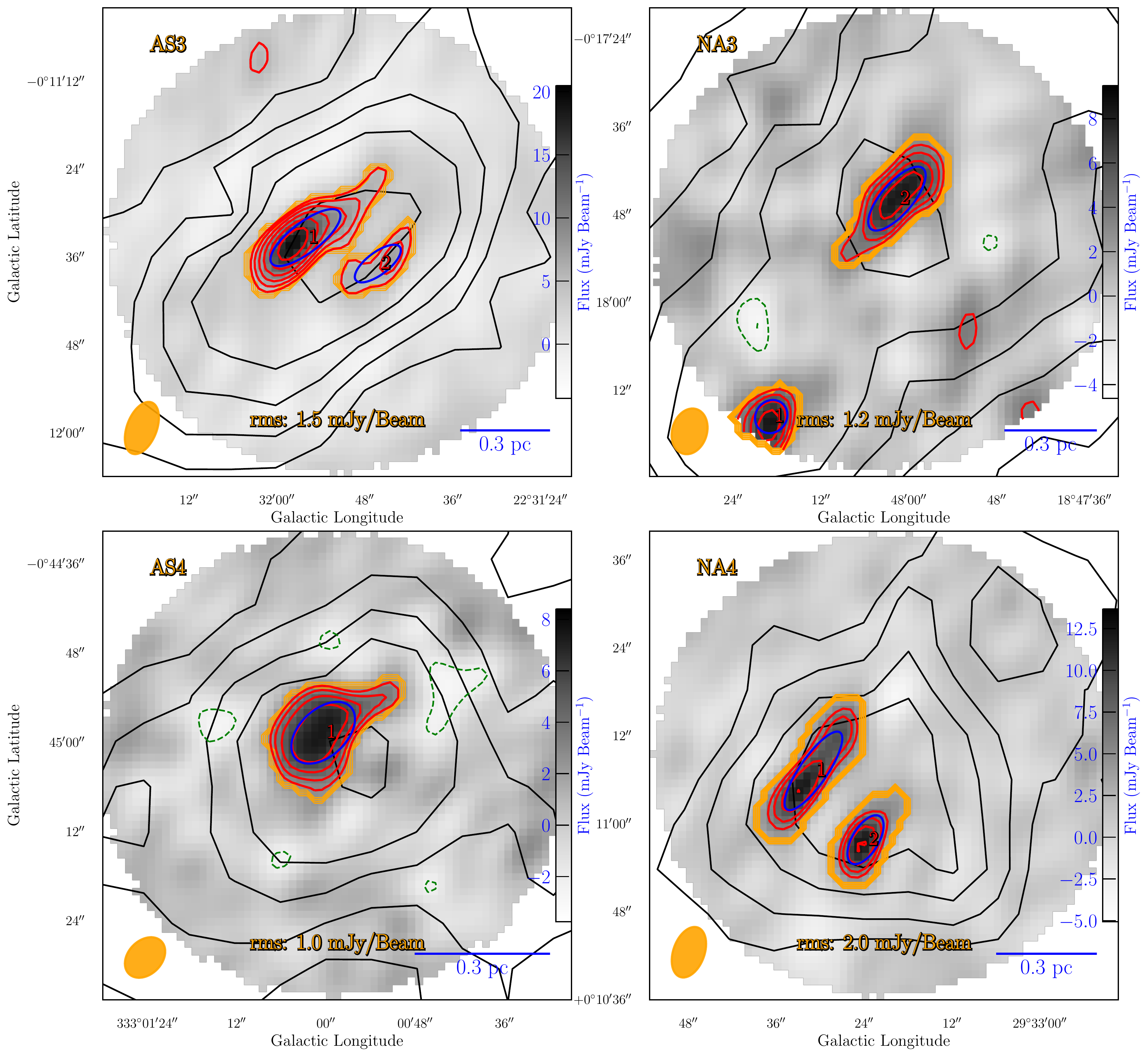}
       \caption{7~m ACA continuum images uncorrected for primary beam. The meanings of lines and marks are similar to Fig.~\ref{Emission7m1} but for another four candidate HMSCs.}
        \label{Emission7m2}
    \end{figure*}

Figures~\ref{Emission7m1} and \ref{Emission7m2} show the 7~m array continuum and Astrodendro leaves extracted in the 7~m images. The parameters of leaves are listed in Table~\ref{ACALeavesTable}.

\begin{table*}
\begin{threeparttable}
\centering
\tiny
\setlength{\tabcolsep}{5pt}
\renewcommand{\arraystretch}{1.}
\caption{Astrodendrogram results for the combined images.} \label{ALMACoresBasics} 
\begin{tabular}{rrrrrrrrrrrr}
\hline
\hline
Clump & Core\tnote{\textit{(a)}} & $l$     &    $b$     & major\tnote{\textit{(b)}} & minor\tnote{\textit{(b)}} & PA & $R$\tnote{\textit{(b)}} & $F$\tnote{\textit{(c)}} & $F_{1}^{\rm p}$\tnote{\textit{(d)}} & $F_{2}^{\rm p}$\tnote{\textit{(d)}} & S/N \\
      &      &($\deg$) &   ($\deg$) & (\arcsec)& (\arcsec)& ($\deg$) & (\arcsec) &(mJy) & (mJy)  &  (mJy Beam$^{-1}$) & \\

    \hline                                                                                                                                          
  \multirow{3}*{\shortstack{AGAL010.214-00.306 \\ (AS1)}} & ALMA1 & 10.2134 & -0.3049  & 1.66 & 0.96 & 90.8   & 1.26 &  $ 49.80 \pm 8.13 $ & $ 0.59  \pm 0.026 $ &  $13 \pm 0.59$  & 6.1 \\ 
                                    & ALMA2 & 10.2153 & -0.3057  & 1.39 & 1.00 & 64.6   & 1.18 &  $ 42.32 \pm 6.39 $ & $ 0.48  \pm 0.027 $ & $11 \pm 0.60$   & 6.6 \\ 
                                    & ALMA3 & 10.2153 & -0.3038  & 1.75 & 1.13 & 142.5  & 1.41 &  $ 38.34 \pm 8.89 $ & $ 0.28  \pm 0.033 $ &  $6.3 \pm 0.74$   & 4.3 \\ 
    \hline                                                                                                                                          
  \multirow{4}*{\shortstack{AGAL009.951-00.366 \\ (NA1)}} & ALMA1 & 9.9503  & -0.3671  & 1.29 & 0.62 & 80.7   & 0.90 &  $ 4.96  \pm 1.01 $ & $ 0.081 \pm 0.009 $ & $ 2.0    \pm   0.22 $   & 4.9 \\ 
                                    & ALMA2 & 9.9525  & -0.3660  & 1.07 & 0.63 & 54.5   & 0.82 &  $ 4.00  \pm 0.92 $ & $ 0.063 \pm 0.010 $ & $ 1.5  \pm   0.23 $   & 4.3 \\ 
                                    & ALMA3 & 9.9496  & -0.3670  & 0.68 & 0.42 & 118.0  & 0.54 &  $ 1.36  \pm 0.37 $ & $ 0.041 \pm 0.009 $ & $ 1.0 \pm   0.23 $   & 3.7 \\ 
                                    & ALMA4 & 9.9481  & -0.3633  & 0.45 & 0.27 & 70.4   & 0.35 &  $ 0.98  \pm 0.28 $ & $ 0.067 \pm 0.015 $ & $ 1.6  \pm   0.37 $   & 3.5 \\ 
    \hline                                                                                                                                          
  \multirow{9}*{\shortstack{AGAL018.931-00.029 \\ (AS2)}} & ALMA1 & 18.9312 & -0.0246  & 0.80 & 0.47 & 154.5  & 0.61 &  $ 9.12  \pm 1.30 $ & $ 0.27  \pm 0.025 $ & $ 5.1  \pm   0.47 $   & 7.0 \\ 
                                    & ALMA2 & 18.9287 & -0.0326  & 1.02 & 0.63 & 82.5   & 0.80 &  $ 6.51  \pm 1.45 $ & $ 0.12  \pm 0.019 $ &  $ 2.3  \pm   0.35 $  & 4.5 \\ 
                                    & ALMA3 & 18.9317 & -0.0248  & 0.40 & 0.24 & -150.8 & 0.31 &  $ 2.02  \pm 0.32 $ & $ 0.17  \pm 0.023 $ & $ 3.1  \pm   0.43 $   & 6.3 \\ 
                                    & ALMA4 & 18.9339 & -0.0297  & 0.67 & 0.32 & 73.3   & 0.46 &  $ 1.87  \pm 0.43 $ & $ 0.087 \pm 0.014 $ &  $ 1.6  \pm   0.26 $  & 4.3 \\ 
                                    & ALMA5 & 18.9323 & -0.0289  & 0.54 & 0.49 & 104.6  & 0.51 &  $ 1.81  \pm 0.37 $ & $ 0.064 \pm 0.010 $ &  $ 1.2  \pm   0.18 $  & 4.8 \\ 
                                    & ALMA6 & 18.9344 & -0.0302  & 0.50 & 0.38 & 93.4   & 0.44 &  $ 1.54  \pm 0.42 $ & $ 0.085 \pm 0.017 $ &  $ 1.6  \pm   0.33 $  & 3.7 \\ 
                                    & ALMA7 & 18.9326 & -0.0283  & 0.61 & 0.33 & 166.5  & 0.45 &  $ 1.32  \pm 0.29 $ & $ 0.057 \pm 0.010 $ &  $ 1.1  \pm   0.20  $  & 4.5 \\ 
                                    & ALMA8 & 18.9296 & -0.0329  & 0.44 & 0.22 & 119.1  & 0.31 &  $ 0.87  \pm 0.25 $ & $ 0.071 \pm 0.018 $ &  $ 1.3  \pm   0.33 $  & 3.5 \\ 
                                    & ALMA9 & 18.9289 & -0.0318  & 0.32 & 0.23 & 74.7   & 0.27 &  $ 0.63  \pm 0.15 $ & $ 0.066 \pm 0.014 $ &  $ 1.2  \pm   0.26 $  & 4.1 \\ 
     \hline                                                                                                                                         
  \multirow{3}*{\shortstack{AGAL015.503-00.419 \\ (NA2)}} & ALMA1 & 15.5040 & -0.4184  & 0.99 & 0.73 & 47.2   & 0.85 &  $ 16.62 \pm 1.41 $ & $ 0.43  \pm 0.009 $ & $ 7.9  \pm   0.17 $   & 11.7 \\
                                    & ALMA2 & 15.5014 & -0.4190  & 0.51 & 0.44 & 179.1  & 0.48 &  $ 1.44  \pm 0.28 $ & $ 0.073 \pm 0.009 $ & $ 1.4  \pm   0.16 $   & 5.1 \\ 
                                    & ALMA3 & 15.5033 & -0.4179  & 0.48 & 0.22 & 58.3   & 0.32 &  $ 0.54  \pm 0.14 $ & $ 0.041 \pm 0.009 $ & $ 0.76 \pm   0.17 $   & 4.0 \\ 
     \hline                                                                                                                                         
  \multirow{9}*{\shortstack{AGAL022.531-00.192 \\ (AS3)}} & ALMA1 & 22.5327 & -0.1928  & 1.03 & 0.54 & 61.8   & 0.74 &  $ 15.97 \pm 1.02 $ & $ 0.34  \pm 0.010 $ & $ 5.8  \pm   0.18 $   & 15.7 \\
                                    & ALMA2 & 22.5284 & -0.1931  & 0.66 & 0.55 & 171.6  & 0.60 &  $ 3.46  \pm 0.65 $ & $ 0.12  \pm 0.012 $ &  $ 2.0    \pm   0.20  $  & 5.3 \\ 
                                    & ALMA3 & 22.5315 & -0.1921  & 0.46 & 0.35 & 83.3   & 0.40 &  $ 2.14  \pm 0.24 $ & $ 0.13  \pm 0.009 $ &  $ 2.2  \pm   0.16 $  & 9.0 \\ 
                                    & ALMA4 & 22.5318 & -0.1928  & 0.54 & 0.19 & -147.0 & 0.32 &  $ 1.24  \pm 0.14 $ & $ 0.094 \pm 0.009 $ &  $ 1.6  \pm   0.16 $  & 8.8 \\ 
                                    & ALMA5 & 22.5305 & -0.1942  & 0.70 & 0.26 & 58.4   & 0.43 &  $ 0.99  \pm 0.26 $ & $ 0.046 \pm 0.010 $ &  $ 0.77 \pm   0.17 $  & 3.8 \\ 
                                    & ALMA6 & 22.5333 & -0.1929  & 0.23 & 0.21 & -178.6 & 0.22 &  $ 0.77  \pm 0.08 $ & $ 0.12  \pm 0.012 $ &  $ 2.0    \pm   0.20  $  & 9.1 \\ 
                                    & ALMA7 & 22.5296 & -0.1906  & 0.38 & 0.30 & -149.6 & 0.34 &  $ 0.73  \pm 0.20 $ & $ 0.055 \pm 0.012 $ &  $ 0.93 \pm   0.20  $  & 3.7 \\ 
                                    & ALMA8 & 22.5307 & -0.1924  & 0.47 & 0.23 & -144.1 & 0.33 &  $ 0.70  \pm 0.12 $ & $ 0.058 \pm 0.009 $ &  $ 0.98 \pm   0.15 $  & 5.6 \\ 
                                    & ALMA9 & 22.5302 & -0.1919  & 0.40 & 0.30 & 75.1   & 0.35 &  $ 0.69  \pm 0.17 $ & $ 0.049 \pm 0.009 $ &  $ 0.83 \pm   0.16 $  & 4.2 \\ 
     \hline                                                                                                                                         
  \multirow{8}*{\shortstack{AGAL018.801-00.297 \\ (NA3)}} & ALMA1 & 18.8000 & -0.2954  & 1.18 & 0.87 & 158.2  & 1.01 &  $ 9.62  \pm 1.63 $ & $ 0.14  \pm 0.010 $ & $ 2.7  \pm   0.19 $   & 5.9 \\ 
                                    & ALMA2 & 18.8008 & -0.2971  & 0.79 & 0.68 & 100.1  & 0.73 &  $ 6.43  \pm 0.78 $ & $ 0.19  \pm 0.008 $ & $ 3.7  \pm   0.15 $   & 8.2 \\ 
                                    & ALMA3 & 18.8025 & -0.2989  & 0.53 & 0.39 & -164.5 & 0.45 &  $ 1.56  \pm 0.32 $ & $ 0.074 \pm 0.010 $ & $ 1.4  \pm   0.19 $   & 4.9 \\ 
                                    & ALMA4 & 18.8022 & -0.2982  & 0.51 & 0.38 & -150.5 & 0.44 &  $ 1.44  \pm 0.28 $ & $ 0.073 \pm 0.009 $ & $ 1.4  \pm   0.17 $   & 5.1 \\ 
                                    & ALMA5 & 18.7989 & -0.2967  & 1.25 & 0.43 & 95.1   & 0.73 &  $ 1.25  \pm 0.36 $ & $ 0.038 \pm 0.009 $ & $ 0.74 \pm   0.18 $   & 3.5 \\ 
                                    & ALMA6 & 18.7992 & -0.2943  & 0.49 & 0.28 & 112.4  & 0.37 &  $ 1.24  \pm 0.30 $ & $ 0.075 \pm 0.014 $ & $ 1.5  \pm   0.28 $   & 4.1 \\ 
                                    & ALMA7 & 18.7988 & -0.2941  & 0.40 & 0.21 & 115.1  & 0.29 &  $ 0.80  \pm 0.22 $ & $ 0.076 \pm 0.017 $ & $ 1.5  \pm   0.32 $   & 3.7 \\ 
                                    & ALMA8 & 18.8018 & -0.2962  & 0.51 & 0.23 & 128.0  & 0.34 &  $ 0.47  \pm 0.13 $ & $ 0.041 \pm 0.009 $ & $ 0.79 \pm   0.17 $   & 3.6 \\ 
    \hline                                                                                                                                          
  \multirow{8}*{\shortstack{AGAL333.016-00.751 \\ (AS4)}} & ALMA1 & -26.9824 & -0.7504 & 1.20 & 0.53 & 57.1   & 0.80 &  $ 4.23  \pm 0.64 $ & $ 0.07  \pm 0.008 $ & $ 1.9  \pm   0.22 $   & 6.6 \\ 
                                    & ALMA2 & -26.9834 & -0.7510 & 0.82 & 0.53 & 83.4   & 0.66 &  $ 2.01  \pm 0.47 $ & $ 0.051 \pm 0.008 $ & $ 1.4  \pm   0.20  $   & 4.3 \\ 
                                    & ALMA3 & -26.9861 & -0.7511 & 0.92 & 0.52 & 76.1   & 0.69 &  $ 1.98  \pm 0.54 $ & $ 0.054 \pm 0.010 $ & $ 1.5  \pm   0.27 $   & 3.7 \\ 
                                    & ALMA4 & -26.9834 & -0.7493 & 0.86 & 0.31 & 60.0   & 0.52 &  $ 1.66  \pm 0.27 $ & $ 0.064 \pm 0.008 $ & $ 1.7  \pm   0.22 $   & 6.3 \\ 
                                    & ALMA5 & -26.9838 & -0.7502 & 0.47 & 0.38 & 147.8  & 0.42 &  $ 1.09  \pm 0.19 $ & $ 0.053 \pm 0.008 $ & $ 1.4  \pm   0.20  $   & 5.8 \\ 
                                    & ALMA6 & -26.9829 & -0.7489 & 0.45 & 0.37 & 170.8  & 0.41 &  $ 1.08  \pm 0.22 $ & $ 0.054 \pm 0.009 $ & $ 1.4  \pm   0.24 $   & 4.8 \\ 
                                    & ALMA7 & -26.9831 & -0.7522 & 0.66 & 0.45 & 48.0   & 0.55 &  $ 1.07  \pm 0.30 $ & $ 0.037 \pm 0.008 $ &  $ 0.98 \pm   0.22 $  & 3.6 \\ 
                                    & ALMA8 & -26.9821 & -0.7515 & 0.54 & 0.37 & 115.6  & 0.45 &  $ 1.01  \pm 0.23 $ & $ 0.046 \pm 0.008 $ &  $ 1.2  \pm   0.23 $  & 4.4 \\ 
  \hline                                                                                                                                            
  \multirow{7}*{\shortstack{AGAL029.556+00.186 \\ (NA4)}} & ALMA1 & 29.5567  & 0.1826  & 0.83 & 0.61 & 62.7   & 0.71 &  $ 15.34 \pm 1.65 $ & $ 0.49  \pm 0.015 $ & $ 8.9  \pm   0.27 $   & 9.3 \\ 
                                    & ALMA2 & 29.5590  & 0.1847  & 0.84 & 0.50 & 50.4   & 0.65 &  $ 11.48 \pm 1.26 $ & $ 0.36  \pm 0.016 $ & $ 6.5  \pm   0.29 $   & 9.1 \\ 
                                    & ALMA3 & 29.5597  & 0.1839  & 0.77 & 0.32 & 47.7   & 0.50 &  $ 6.13  \pm 0.92 $ & $ 0.25  \pm 0.024 $ & $ 4.5  \pm   0.43 $   & 6.6 \\ 
                                    & ALMA4 & 29.5575  & 0.1870  & 0.69 & 0.48 & 152.9  & 0.57 &  $ 3.91  \pm 0.58 $ & $ 0.13  \pm 0.011 $ & $ 2.3  \pm   0.20 $   & 6.8 \\ 
                                    & ALMA5 & 29.5580  & 0.1856  & 0.56 & 0.30 & 89.8   & 0.41 &  $ 0.97  \pm 0.25 $ & $ 0.053 \pm 0.011 $ & $ 0.96 \pm   0.21 $   & 3.9 \\ 
                                    & ALMA6 & 29.5581  & 0.1872  & 0.33 & 0.29 & 135.1  & 0.31 &  $ 0.94  \pm 0.18 $ & $ 0.078 \pm 0.013 $ & $ 1.4  \pm   0.24 $   & 5.1 \\ 
                                    & ALMA7 & 29.5580  & 0.1850  & 0.38 & 0.30 & 165.0  & 0.34 &  $ 0.71  \pm 0.20 $ & $ 0.049 \pm 0.012 $ & $ 0.88 \pm   0.21 $   & 3.5 \\ 
    \hline

\end{tabular}
      \begin{tablenotes}
      \item [\textit{(a)}] Cores are ranked with their mass, e.g., ALMA1 is the most massive core while ALMA2 is the second massive core.
      \item [\textit{(b)}] Major, minor semi-axes, and equivalent radius, respectively.
      \item [\textit{(c)}] Primary beam corrected integrated flux and its error.
      \item [\textit{(d)}] Primary beam corrected pixel maximum flux and its error. The units are mJy and mJy Beam$^{-1}$ for $F_{1}^{\rm p}$ and $F_{2}^{\rm p}$, respectively.
      \end{tablenotes}
      \end{threeparttable}
\end{table*}

\begin{table*}
\begin{threeparttable}
\setlength{\tabcolsep}{4.pt}
\renewcommand{\arraystretch}{1.}
\centering
\tiny
\caption{Physical parameters of cores.} \label{ALMACoresPhysical} 
\begin{tabular}{rrrrrrrrrr}
\hline
\hline
Clump & Core & $M_{\rm core}$\tnote{\textit{($\alpha$)}}   & $M_{\rm core}^{\rm cold}$\tnote{\textit{($\beta$)}} & $M_{\rm core}^{\rm warm}$\tnote{\textit{($\beta$)}}   & $r$\tnote{\textit{($\gamma$)}} & $\Sigma$ & \nhtcd & \nhtnd  \\
      &      &(\msun)        & (\msun)    & (\msun)    & (AU)       & (g cm$^{-2}$) &  ($10^{23}$ cm$^{-2}$) & ($10^6$ cm$^{-3}$) \\
   \hline                                                                                                                                                                                                                                                      
\multirow{3}*{\shortstack{AGAL010.214-00.306 \\ (AS1, $T_{\rm dust} = 16.6$~K)}}  & ALMA1 & $14.1 \pm 5.4$ & $18.6 \pm 7.2 $ & $11.3 \pm 4.3$ & 3940 &  $2.57 \pm 0.92$ & $5.47 \pm 1.96$ & $6.96 \pm 2.54$\\                                                                              
                                                                                  & ALMA2 & $12.0 \pm 4.5$ & $15.8 \pm 6.0 $ & $9.6  \pm 3.6$ & 3700 &  $2.51 \pm 0.89$ & $5.35 \pm 1.89$ & $7.30 \pm 2.63$\\                                                                              
                                                                                  & ALMA3 & $10.9 \pm 4.5$ & $14.4 \pm 6.0 $ & $8.7  \pm 3.6$ & 4400 & $1.58 \pm 0.63$ & $3.38 \pm 1.33$ & $3.85 \pm 1.54$ \\                                                                              
 \hline                                                                                                                                                                                                                                                        
  \multirow{4}*{\shortstack{AGAL009.951-00.366 \\ (NA1, $T_{\rm dust} = 12.0$~K)}}  & ALMA1 & $2.2 \pm 0.9 $ & $3.6  \pm 1.4 $ & $1.6  \pm 0.7$ & 2800 & $0.81 \pm 0.31$ & $1.73 \pm 0.66$ & $3.10 \pm 1.19$ \\                                                                              
                                                                                    & ALMA2 & $1.8 \pm 0.8 $ & $2.9  \pm 1.2 $ & $1.3  \pm 0.6$ & 2600 & $0.77 \pm 0.30$ & $1.65 \pm 0.65$ & $3.22 \pm 1.29$ \\                                                                             
                                                                                    & ALMA3 & $0.6 \pm 0.3 $ & $1.0  \pm 0.4 $ & $0.4  \pm 0.2$ & 1700 & $0.62 \pm 0.26$ & $1.32 \pm 0.55$ & $3.97 \pm 1.68$ \\                                                                              
                                                                                    & ALMA4 & $0.4 \pm 0.2 $ & $0.7  \pm 0.3 $ & $0.3  \pm 0.1$ & 1100 & $1.07 \pm 0.46$ & $2.28 \pm 0.97$ & $10.58 \pm 4.58$ \\                                                                              
 \hline                                                                                                                                                                                                                                                        
  \multirow{9}*{\shortstack{AGAL018.931-00.029 \\ (AS2, $T_{\rm dust} = 18.2$~K)}}  & ALMA1 & $2.6 \pm 0.9 $ & $3.3  \pm 1.2 $ & $2.1  \pm 0.8$ & 2000  & $1.78 \pm 0.62$ & $3.80 \pm 1.33$ & $9.48 \pm 3.36$ \\                                                                              
                                                                                    & ALMA2 & $1.8 \pm 0.7 $ & $2.3  \pm 0.9 $ & $1.5  \pm 0.6$ & 2700  & $0.73 \pm 0.28$ & $1.56 \pm 0.61$ & $2.94 \pm1.16$ \\                                                                              
                                                                                     & ALMA3 & $0.6 \pm 0.2 $ & $0.7  \pm 0.3 $ & $0.5  \pm 0.2$ & 1000 & $1.53 \pm 0.55$ & $3.26 \pm 1.16$ & $15.98 \pm 5.77$ \\                                                                           
                                                                                     & ALMA4 & $0.5 \pm 0.2 $ & $0.7  \pm 0.3 $ & $0.4  \pm 0.2$ & 1500 & $0.64 \pm 0.25$ & $1.36 \pm 0.54$ & $4.46 \pm 1.77$ \\                                                                             
                                                                                     & ALMA5 & $0.5 \pm 0.2 $ & $0.7  \pm 0.3 $ & $0.4  \pm 0.2$ & 1700 & $0.50 \pm 0.19$ & $1.06 \pm 0.40$ & $3.13 \pm 1.20$ \\                                                                             
                                                                                     & ALMA6 & $0.4 \pm 0.2 $ & $0.6  \pm 0.2 $ & $0.4  \pm 0.2$ & 1400 & $0.59 \pm 0.25$ & $1.25 \pm 0.52$ & $4.34 \pm 1.83$ \\                                                                              
                                                                                     & ALMA7 & $0.4 \pm 0.2 $ & $0.5  \pm 0.2 $ & $0.3  \pm 0.1$ & 1500 & $0.48 \pm 0.19$ & $1.02 \pm 0.40$ & $3.45 \pm 1.36$ \\                                                                             
                                                                                     & ALMA8 & $0.2 \pm 0.1 $ & $0.3  \pm 0.1 $ & $0.2  \pm 0.1$ & 1000 & $0.64 \pm 0.28$ & $1.37 \pm 0.59$ & $6.64 \pm 2.87$ \\                                                                              
                                                                                     & ALMA9 & $0.2 \pm 0.1 $ & $0.2  \pm 0.1 $ & $0.2  \pm 0.1$ & 900  & $0.64 \pm 0.26$ & $1.36 \pm 0.55$ & $7.70 \pm 3.11$ \\                                                                              
 \hline                                                                                                                                                                                                                                                        
  \multirow{3}*{\shortstack{AGAL015.503-00.419 \\ (NA2, $T_{\rm dust} = 14.0$~K)}}   & ALMA1 & $6.0 \pm 2.5 $ & $8.7  \pm 3.6 $ & $4.6  \pm 1.9$ & 2700 & $2.38 \pm 0.79$ & $5.08 \pm 1.68$ & $9.53 \pm 3.36$ \\                                                                              
                                                                                     & ALMA2 & $0.5 \pm 0.2 $ & $0.8  \pm 0.3 $ & $0.4  \pm 0.2$ & 1500 & $0.66 \pm 0.25$ & $1.40 \pm 0.53$ & $4.69 \pm 1.85$ \\                                                                              
                                                                                     & ALMA3 & $0.2 \pm 0.1 $ & $0.3  \pm 0.1 $ & $0.2  \pm 0.1$ & 1000 & $0.53 \pm 0.22$ & $1.13 \pm 0.46$ & $5.57 \pm 2.37$ \\                                                                              
 \hline                                                                                                                                                                                                                                                        
  \multirow{9}*{\shortstack{AGAL022.531-00.192 \\ (AS3, $T_{\rm dust} = 12.4$~K)}}  & ALMA1 & $17.9 \pm 7.1$ & $27.7 \pm 11 $ & $13.1 \pm 5.2$ & 3700 & $3.60 \pm 1.17$ & $7.67 \pm 2.50$ & $10.26 \pm 3.54$ \\                                                                              
                                                                                     & ALMA2 & $3.9  \pm 1.7$ & $6.0  \pm 2.6 $ & $2.8  \pm 1.2$ & 3000 & $1.20 \pm 0.44$ & $2.55 \pm 0.95$ & $4.22 \pm 1.64$ \\                                                                             
                                                                                     & ALMA3 & $2.4  \pm 1.0$ & $3.7  \pm 1.5 $ & $1.8  \pm 0.7$ & 2000 & $1.67 \pm 0.56$ & $3.55 \pm 1.20$ & $8.83 \pm 3.16$ \\                                                                              
                                                                                     & ALMA4 & $1.4  \pm 0.6$ & $2.2  \pm 0.9 $ & $1.0  \pm 0.4$ & 1600 & $1.53 \pm 0.52$ & $3.27 \pm 1.11$ & $10.25 \pm 3.67$ \\                                                                             
                                                                                     & ALMA5 & $1.1  \pm 0.5$ & $1.7  \pm 0.8 $ & $0.8  \pm 0.4$ & 2100 & $0.68 \pm 0.28$ & $1.45 \pm 0.60$ & $3.38 \pm 1.45$ \\                                                                              
                                                                                     & ALMA6 & $0.9  \pm 0.4$ & $1.3  \pm 0.5 $ & $0.6  \pm 0.3$ & 1100 & $2.06 \pm 0.70$ & $4.40 \pm 1.49$ & $20.35 \pm 7.26$ \\                                                                            
                                                                                     & ALMA7 & $0.8  \pm 0.4$ & $1.3  \pm 0.6 $ & $0.6  \pm 0.3$ & 1700 & $0.79 \pm 0.33$ & $1.69 \pm 0.71$ & $4.95 \pm 2.14$ \\                                                                             
                                                                                     & ALMA8 & $0.8  \pm 0.3$ & $1.2  \pm 0.5 $ & $0.6  \pm 0.3$ & 1700 & $0.80 \pm 0.29$ & $1.70 \pm 0.62$ & $5.10 \pm 1.95$ \\                                                                              
                                                                                     & ALMA9 & $0.8  \pm 0.4$ & $1.2  \pm 0.6 $ & $0.6  \pm 0.3$ & 1700 & $0.72 \pm 0.29$ & $1.53 \pm 0.61$ & $4.41 \pm 1.84$ \\                                                                              
 \hline                                                                                                                                                                                                                                                        
  \multirow{8}*{\shortstack{AGAL018.801-00.297 \\ (NA3, $T_{\rm dust} = 12.3$~K)}}  & ALMA1 & $10.3 \pm 4.0$ & $16.1 \pm 6.2 $ & $7.5  \pm 2.9$ & 5000 & $1.19 \pm 0.43$ & $2.53 \pm 0.92$ & $2.56 \pm 0.94$ \\                                                                             
                                                                                    & ALMA2 & $6.9  \pm 2.5$ & $10.7 \pm 3.9 $ & $5.0  \pm 1.8$ & 3600 & $1.52 \pm 0.52$ & $3.25 \pm 1.11$ & $4.55 \pm 1.58$ \\                                                                              
                                                                                    & ALMA3 & $1.7  \pm 0.7$ & $2.6  \pm 1.0 $ & $1.2  \pm 0.5$ & 2200 & $0.95 \pm 0.36$ & $2.03 \pm 0.77$ & $4.57 \pm 1.76$ \\                                                                             
                                                                                    & ALMA4 & $1.5  \pm 0.6$ & $2.4  \pm 1.0 $ & $1.1  \pm 0.5$ & 2100 & $0.95 \pm 0.36$ & $2.02 \pm 0.76$ & $4.71 \pm 1.79$ \\                                                                              
                                                                                    & ALMA5 & $1.3  \pm 0.6$ & $2.1  \pm 0.9 $ & $1.0  \pm 0.4$ & 3600 & $0.30 \pm 0.13$ & $0.63 \pm 0.27$ & $0.88 \pm 0.38$ \\                                                                             
                                                                                    & ALMA6 & $1.3  \pm 0.6$ & $2.1  \pm 0.9 $ & $1.0  \pm 0.4$ & 1800 & $1.13 \pm 0.45$ & $2.40 \pm 0.97$ & $6.60 \pm 2.68$ \\                                                                             
                                                                                    & ALMA7 & $0.9  \pm 0.4$ & $1.3  \pm 0.6 $ & $0.6  \pm 0.3$ & 1400 & $1.18 \pm 0.50$ & $2.53 \pm 1.06$ & $8.84 \pm 3.74$ \\                                                                             
                                                                                    & ALMA8 & $0.5  \pm 0.2$ & $0.8  \pm 0.4 $ & $0.4  \pm 0.2$ & 1700 & $0.50 \pm 0.21$ & $1.06 \pm 0.45$ & $3.17 \pm 1.36$ \\                                                                             
 \hline                                                                                                                                                                                                                                                        
  \multirow{8}*{\shortstack{AGAL333.016-00.751 \\ (AS4, $T_{\rm dust} = 15.6$~K)}} & ALMA1 & $1.6 \pm 0.7 $ & $2.1  \pm 0.9 $ & $1.2  \pm 0.5$ & 2700 & $0.59 \pm 0.21$ & $1.27 \pm 0.45$ & $2.35 \pm 0.88$ \\                                                                              
                                                                                   & ALMA2 & $0.7 \pm 0.3 $ & $1.0  \pm 0.5 $ & $0.6  \pm 0.3$ & 2200 & $0.42 \pm 0.16$ & $0.89 \pm 0.35$ & $1.98 \pm 0.82$ \\                                                                             
                                                                                   & ALMA3 & $0.7 \pm 0.3 $ & $1.0  \pm 0.4 $ & $0.6  \pm 0.3$ & 2400 & $0.37 \pm 0.16$ & $0.79 \pm 0.33$ & $1.68 \pm 0.73$ \\                                                                             
                                                                                   & ALMA4 & $0.6 \pm 0.3 $ & $0.8  \pm 0.4 $ & $0.5  \pm 0.2$ & 1800 & $0.55 \pm 0.20$ & $1.18 \pm 0.42$ & $3.36 \pm 1.27$ \\                                                                              
                                                                                   & ALMA5 & $0.4 \pm 0.2 $ & $0.5  \pm 0.2 $ & $0.3  \pm 0.1$ & 1400 & $0.55 \pm 0.20$ & $1.17 \pm 0.42$ & $4.06 \pm 1.56$ \\                                                                             
                                                                                   & ALMA6 & $0.4 \pm 0.2 $ & $0.5  \pm 0.2 $ & $0.3  \pm 0.1$ & 1400 & $0.58 \pm 0.22$ & $1.24 \pm 0.47$ & $4.47 \pm 1.79$ \\                                                                              
                                                                                   & ALMA7 & $0.4 \pm 0.2 $ & $0.5  \pm 0.3 $ & $0.3  \pm 0.2$ & 1900 & $0.32 \pm 0.14$ & $0.68 \pm 0.29$ & $1.84 \pm 0.81$ \\                                                                             
                                                                                   & ALMA8 & $0.4 \pm 0.2 $ & $0.5  \pm 0.2 $ & $0.3  \pm 0.1$ & 1500 & $0.45 \pm 0.18$ & $0.96 \pm 0.38$ & $3.18 \pm 1.31$ \\                                                                             
 \hline                                                                                                                                                                                                                                                        
  \multirow{7}*{\shortstack{AGAL029.556+00.186 \\ (NA4, $T_{\rm dust} = 12.8$~K)}}  & ALMA1 & $13.1 \pm 4.7$ & $19.8 \pm 7.2 $ & $9.7  \pm 3.5$ & 3200 & $3.60 \pm 1.21$ & $7.67 \pm 2.59$ & $12.01 \pm 4.13$ \\                                                                             
                                                                                    & ALMA2 & $9.8  \pm 3.6$ & $14.8 \pm 5.4 $ & $7.2  \pm 2.6$ & 2900 & $3.24 \pm 1.10$ & $6.91 \pm 2.34$ & $11.87 \pm 4.09$ \\                                                                              
                                                                                    & ALMA3 & $5.2  \pm 2.0$ & $7.9  \pm 3.0 $ & $3.9  \pm 1.5$ & 2200 & $2.97 \pm 1.05$ & $6.34 \pm 2.24$ & $14.25 \pm 5.13$ \\                                                                             
                                                                                     & ALMA4 & $3.3  \pm 1.3$ & $5.1  \pm 1.9 $ & $2.5  \pm 0.9$ & 2600 & $1.42 \pm 0.50$ & $3.03 \pm 1.07$ & $5.91 \pm 2.12$ \\                                                                              
                                                                                    & ALMA5 & $0.8  \pm 0.4$ & $1.3  \pm 0.5 $ & $0.6   \pm 0.3$ & 1800 & $0.69 \pm 0.29$ & $1.48 \pm 0.61$ & $4.06 \pm 1.69$ \\                                                                              
                                                                                    & ALMA6 & $0.8  \pm 0.3$ & $1.2  \pm 0.5 $ & $0.6  \pm 0.2$ & 1400 & $1.17 \pm 0.44$ & $2.50 \pm 0.93$ & $9.00 \pm 3.42$ \\                                                                              
                                                                                    & ALMA7 & $0.6  \pm 0.3$ & $0.9  \pm 0.4 $ & $0.4  \pm 0.2$ & 1500 & $0.74 \pm 0.32$ & $1.59 \pm 0.68$ & $5.27 \pm 2.27$ \\                                                                             
   \hline                                                                                                                                           
   
   \end{tabular}
      \begin{tablenotes}
      \item [\textit{($\alpha$)}] Core mass and its error at clump dust temperature \dustt. 
      \item [\textit{($\beta$)}] Warm (\dustt$+ 3$~K) and cold (\dustt$- 3$~K) core masses and their errors.
      \item [\textit{($\gamma$)}] Core radius.
      \end{tablenotes}
      \end{threeparttable}
\end{table*}

% For the figures of 7~m images
    
%###### Table for ACA leaves

\begin{table*}
\begin{threeparttable}
\centering
\tiny
\setlength{\tabcolsep}{5pt}
\renewcommand{\arraystretch}{1.}
\caption{Astrodendrogram results for 7~m ACA images.} \label{ACALeavesTable} 
\begin{tabular}{ccrrccrcrrccrcc}
\hline
\hline
Clump & Leaves\tnote{\textit{($\alpha$)}} & $l$    & $b$        & major\tnote{\textit{($\beta$)}}    & minor\tnote{\textit{($\beta$)}}    & PA       & $R$\tnote{\textit{($\beta$)}}       & $F$\tnote{\textit{($\gamma$)}}   & $F^p$\tnote{\textit{($\gamma$)}} & S/N & $r$\tnote{\textit{($\beta$)}} & $M$ & $M^{\rm cold}$ & $M^{\rm warm}$ \\
      &       &($\deg$) &   ($\deg$) & (\arcsec)& (\arcsec)& ($\deg$) &(\arcsec) &(mJy) & (mJy)  &     &  pc & \msun &  \msun & \msun \\
    \hline                                                                                                          
AS1 & ACA1 & 10.2151 & -0.3053 & 1.47 & 1.10 & 58.2   & 1.27 & $26.08 \pm 1.44$ & $1.4  \pm 0.068$  & 18.2 & 0.02 & $7.4  \pm 2.6 $ & 9.8  & 5.9 \\   
AS1 & ACA2 & 10.2135 & -0.3049 & 1.39 & 0.99 & 72.1   & 1.17 & $23.49 \pm 1.25$ & $1.5  \pm 0.069$  & 18.8 & 0.02 & $6.6  \pm 2.3 $ & 8.8  & 5.3 \\ 
\hline
NA1 & ACA1 & 9.9499  & -0.3669 & 2.55 & 1.52 & -167.4 & 1.97 & $8.55  \pm 1.76$ & $0.24 \pm 0.033$  & 4.9  & 0.03 & $3.9  \pm 1.6 $ & 6.1  & 2.8 \\   
NA1 & ACA2 & 9.9525  & -0.3661 & 2.34 & 1.41 & -178.1 & 1.81 & $6.91  \pm 1.54$ & $0.22 \pm 0.033$  & 4.5  & 0.03 & $3.1  \pm 1.3 $ & 4.9  & 2.2 \\   
NA1 & ACA3 & 9.9566  & -0.3638 & 1.48 & 1.05 & -144.5 & 1.24 & $4.21  \pm 1.28$ & $0.26 \pm 0.061$  & 3.3  & 0.02 & $1.9  \pm 0.9 $ & 3.0  & 1.4 \\   
NA1 & ACA4 & 9.9485  & -0.3629 & 1.47 & 1.01 & 51.0   & 1.22 & $2.67  \pm 0.83$ & $0.17 \pm 0.041$  & 3.2  & 0.02 & $1.2  \pm 0.6 $ & 1.9  & 0.9 \\   
\hline
AS2 & ACA1 & 18.9316 & -0.0247 & 3.04 & 1.72 & -166.4 & 2.29 & $25.95 \pm 5.73$ & $0.58 \pm 0.073$  & 4.5  & 0.04 & $7.3  \pm 2.9 $ & 9.3  & 6.0 \\   
AS2 & ACA2 & 18.9286 & -0.0320 & 2.55 & 1.98 & 176.8  & 2.25 & $16.47 \pm 4.31$ & $0.36 \pm 0.062$  & 3.8  & 0.04 & $4.6  \pm 2.0 $ & 5.9  & 3.8 \\   
AS2 & ACA3 & 18.9338 & -0.0295 & 3.13 & 1.06 & -159.4 & 1.82 & $7.07  \pm 2.34$ & $0.2  \pm 0.057$  & 3.0  & 0.03 & $2.0  \pm 0.9 $ & 2.5  & 1.6 \\   
\hline
NA2 & ACA1 & 15.5041 & -0.4182 & 2.69 & 2.23 & -157.7 & 2.45 & $21.86 \pm 3.45$ & $0.58 \pm 0.038$  & 6.3  & 0.04 & $7.9  \pm 3.4 $ & 11.4 & 6.1 \\   
NA2 & ACA2 & 15.4999 & -0.4242 & 1.82 & 1.09 & 157.0  & 1.41 & $5.13  \pm 1.66$ & $0.23 \pm 0.061$  & 3.1  & 0.02 & $1.9  \pm 1.0 $ & 2.7  & 1.4 \\   
\hline
AS3 & ACA1 & 22.5322 & -0.1926 & 4.93 & 1.74 & -142.6 & 2.93 & $38.95 \pm 6.26$ & $0.66 \pm 0.046$  & 6.2  & 0.07 & $43.6 \pm 18.5$ & 67.6 & 31.9 \\  
AS3 & ACA2 & 22.5295 & -0.1936 & 3.20 & 1.29 & -143.0 & 2.03 & $7.87  \pm 2.26$ & $0.22 \pm 0.046$  & 3.5  & 0.05 & $8.8  \pm 4.3 $ & 13.7 & 6.4 \\   
\hline
NA3 & ACA1 & 18.8052 & -0.3043 & 2.00 & 1.80 & 52.4   & 1.90 & $28.99 \pm 5.83$ & $0.99 \pm 0.11 $  & 5.0  & 0.05 & $31.0 \pm 12.4$ & 48.4 & 22.6 \\  
NA3 & ACA2 & 18.8004 & -0.2961 & 4.63 & 1.82 & 49.2   & 2.90 & $22.22 \pm 4.85$ & $0.3  \pm 0.038$  & 4.6  & 0.07 & $23.8 \pm 9.7 $ & 37.1 & 17.3 \\  
\hline
AS4 & ACA1 & -26.9832& -0.7496 & 4.30 & 2.67 & -137.0 & 3.39 & $27.63 \pm 5.36$ & $0.27 \pm 0.033$  & 5.2  & 0.06 & $10.1 \pm 4.5 $ & 13.7 & 8.0 \\   
\hline
NA4 & ACA1 & 29.5586 & 0.1853  & 5.42 & 1.59 & 55.4   & 2.94 & $32.93 \pm 5.41$ & $0.48 \pm 0.041$  & 6.1  & 0.06 & $28.0 \pm 10.7$ & 42.5 & 20.7 \\  
NA4 & ACA2 & 29.5566 & 0.1827  & 3.18 & 1.53 & 59.0   & 2.20 & $18.43 \pm 3.19$ & $0.47 \pm 0.04 $  & 5.8  & 0.05 & $15.7 \pm 6.1 $ & 23.8 & 11.6 \\  
    \hline                                                                                                                                          
\end{tabular}
      \begin{tablenotes}
      \item [\textit{($\alpha$)}] Leaves are ranked with their mass, e.g., ACA1 is the most massive leaf while ACA2 is the second massive leaf.
      \item [\textit{($\beta$)}] Major, minor semi-axes, and equivalent radius, respectively.
      \item [\textit{($\gamma$)}] $F$ is the primary beam corrected integrated flux. $F^p$ is the primary beam corrected pixel maximum flux. The unit has been transformed from mJy Beam$^{-1}$ to mJy by dividing the flux in the unit of mJy Beam$^{-1}$ with the beam size in the unit of pixel.
      \end{tablenotes}
      \end{threeparttable}
\end{table*}

\section{Math derivations} \label{Appendix-math}
\subsection{Mean ratio of projected separation to actual separation \citep{sanhueza19}}
Setting $\alpha$ as the angle between actual separation $S_{\rm actu}$ and projected separation $S_{\rm proj}$, we have
\begin{equation*}
S_{\rm actu} \times cos(\alpha) = S_{\rm proj}.
\end{equation*}
The mean ratio is
\begin{equation*}
\overline{S_{\rm proj}/S_{\rm actu}} =  \overline{cos(\alpha)} = \frac{1}{\pi} \int_{-\frac{\pi}{2}}^{\frac{\pi}{2}} cos(x) dx = \frac{2}{\pi}.
\end{equation*}
\subsection{Molecular weight per hydrogen molecule $\mu_{\rm H_2}$ and mean molecular weight per free particle $\mu$ \citep{Kauffmann2008}}
Assuming a mass ratio of $\mathcal{M}(\rm H)/\mathcal{M} \approx 0.71$, $\mathcal{M}(\rm He)/\mathcal{M} \approx 0.27$, $\mathcal{M}(\rm Z)/\mathcal{M} \approx 0.02$ for
hydrogen, helium, and metals, respectively, where $\mathcal{M} = \mathcal{M}(\rm H) + \mathcal{M}(\rm He) + \mathcal{M}(\rm Z)$. The molecular weight per hydrogen molecule is:

\begin{equation*}
\begin{split}
\mu_{\rm H_{2}} & = \frac{\mathcal M}{m_{\rm H}\times \rm N(H_2)} = \frac{\mathcal M}{m_{\rm H}\times \frac{1}{2} \rm N(H)} \\ 
& =  \frac{2 \mathcal M}{m_{\rm H}\times \rm N(H)} = \frac{2\mathcal M}{{\mathcal M}\rm (H)} \\
&\approx 2.8, 
\end{split}
\end{equation*}
where $\rm N(H)$,  $\rm N(H_{2})$, and $m_{\rm H}$ are the numbers of H, \hmole, and the mass of H atoms, respectively.

In the calculation of thermal velocity dispersion, $\sigma_{\rm th}$ is related to the number of free particles, thus, we need to calculate the mean molecular weight per free particle, which is:
\begin{equation*}
\begin{split}
\mu = \frac{\mathcal M}{m_{\rm H}\times \rm N} & \approx \frac{\mathcal M}{m_{\rm H}\times {\rm N(H_{2})} +  m_{\rm H}\times {\rm N(He)}} \\ 
&=  \frac{\mathcal M}{\frac{1}{2} \mathcal M{\rm(H)} + m_{\rm H}\times \rm N(He)} \\
&= \frac{\mathcal M}{\frac{1}{2}{\mathcal M}\rm (H) +  \frac{1}{4}{\mathcal M}\rm (He)} \\ &\approx 2.37.
\end{split}
\end{equation*}

\end{appendix}

\end{document}